\documentclass[a4paper,10pt]{article}
\usepackage{jheppub}

\usepackage[utf8]{inputenc}
\usepackage[english]{babel}
\usepackage{amsmath,amsfonts,amssymb}
\usepackage{fancyhdr}
\usepackage{physics}
\usepackage{graphicx}
\usepackage[framemethod=TikZ]{mdframed}
\usepackage{amsthm}
\usepackage{caption}
\usepackage{subcaption}
\usepackage{array,booktabs}
\usepackage{enumitem}
\usepackage{lipsum}
\usepackage{tensor}
\usepackage{cancel}
\usepackage{mathtools, nccmath}
\usepackage{empheq}
\usepackage{physics}
\usepackage[nottoc]{tocbibind}
\usepackage{lastpage}
\usepackage{hyperref}
\usepackage{tensor}
\usepackage{slashed}
\usepackage{mathrsfs}
\usepackage[vcentermath]{youngtab}
\usepackage{quiver}
\usepackage{cleveref}
\crefname{figure}{figure}{figures}
\Crefname{figure}{Figure}{Figures}
\usepackage{ragged2e}
\usepackage{comment}
\usepackage{float}

\theoremstyle{definition}
\newtheorem{theorem}{Theorem}[section]
\newtheorem{definition}[theorem]{Definition}

\newtheorem{example}[theorem]{Example}

\newcommand{\hypgeo}[2]{%
  {\vphantom{F}}_{#1}\kern-\scriptspace F_{#2}%
}

\hypersetup{
	bookmarksnumbered=true,
	colorlinks=true,
	urlcolor=black,
	linkcolor=teal,
	citecolor=purple,
	linkbordercolor = {white}
}

\newcommand{\be}{\begin{equation}} \newcommand{\ee}{\end{equation}}
\newcommand{\bea}{\begin{equation} \begin{aligned}} \newcommand{\eea}{\end{aligned} \end{equation}}

\newcommand{\Z}{\mathbb{Z}}
\newcommand{\R}{\mathbb{R}}

\newcommand{\N}{\mathcal{N}}

\newcommand{\fh}{\mathfrak{h}}

\newcommand{\cF}{\mathcal{F}}

\newcommand{\cL}{\mathcal{L}}

\newcommand{\cN}{\mathcal{N}}
\newcommand{\cO}{\mathcal{O}}

\newcommand{\cZ}{\mathcal{Z}}

\newcommand{\bC}{\mathbb{C}}

\newcommand{\bR}{\mathbb{R}}

\newcommand{\bZ}{\mathbb{Z}}

\numberwithin{equation}{section}

\title{Global variants of $\mathcal{N}=1^*$ theories and Calogero--Moser systems}

\author[a,b]{Jerem\'ias Aguilera Damia,}
\author[c]{Riccardo Argurio,}
\author[d]{Antoine Bourget,} 
\author[e]{Valdo Tatitscheff,}
\author[c]{\\ Romain Vandepopeliere~}

\affiliation[a]{Institut de Ciencies del Cosmos (ICC)\\
Universitat de Barcelona Facultat de F\'isica, Mart\'i i Franqu\`es, 1, E-08028 Barcelona, Spain}
\affiliation[b]{Instituto de Fisica La Plata,\\ Universidad Nacional de La Plata, C.C. 67, 1900 La Plata, Argentina}
\affiliation[c]{Physique Théorique et Mathématique and International Solvay Institutes\\
Université Libre de Bruxelles, C.P. 231, 1050 Brussels, Belgium}
\affiliation[d]{Institut de Physique Théorique\\ 
Université Paris Saclay, CEA, CNRS, Gif sur Yvette, France}
\affiliation[e]{Institute for Mathematics\\
Ruprecht-Karls-Universität Heidelberg, 69120 Heidelberg, Germany}

\emailAdd{jeremiasad@fqa.ub.edu, riccardo.argurio@ulb.be, antoine.bourget@ipht.fr, valdo.tatitscheff@normalesup.org, romain.vandepopeliere@ulb.be}

\abstract{Global variants of four-dimensional gauge theories are specified by their spectrum of genuine Wilson–'t Hooft line operators.
The choice of global variant has significant consequences when spacetime is taken to be $\mathbb{R}^3 \times S^1$. We focus on $\mathcal N=1^*$ theories, which are closely connected to twisted elliptic Calogero–Moser systems.
We establish, on general grounds, how this gauge-theoretic topological data manifests itself on the integrable system side by introducing a notion of global variants for complex many-body integrable systems associated with Lie algebras. Focusing on $\mathcal N=1^*$ theories of type $A$ and $B_2$, we elucidate the implications for the structure of gapped vacua, the emergent (generalized) symmetries realized in each vacuum, and the action of spontaneously broken modular invariance.} 

\preprint{}
 \clearpage

\begin{document}

\maketitle







\section{Introduction}

Global properties of Lie groups have arisen to recent interest in the context of four-dimensional gauge theories \cite{Aharony:2013hda}. It is well-known that given a compact simple Lie algebra, one can build several different Lie groups out of it through the exponential map. Such groups differ by their center, and by their fundamental homotopy group. The simplest example is the one of the Lie algebra $\mathfrak{su}(2)$, from which one can define two Lie groups: $\mathrm{SU}(2)$, which has $\bZ_2$ as its center and a trivial fundamental group (it is simply connected); and $\mathrm{SO}(3)\cong \mathrm{SU}(2)/\bZ_2$, which has trivial center but fundamental group $\bZ_2$.

When considering a gauge theory, the conventional Lagrangian formulation makes use of the Lie algebra, in which the gauge fields take value. Local physics then depends on the gauge algebra only. This includes important phenomena such as flux confinement, the presence of a mass gap, and the spontaneous breaking of (continuous or discrete) chiral symmetries. A better characterization of confinement in gauge theories however requires the use of extended operators \cite{Wilson:1974sk, tHooft:1977nqb}. It is using the latter that one can enlarge Landau's paradigm to fit also confinement \cite{Gaiotto:2014kfa}. However, when dealing with extended operators, namely lines in four dimensions, one needs to specify not only the gauge {\em algebra} but also the gauge {\em group}. Indeed, such operators probe also global aspects of the gauge group, not only the infinitesimal ones. Both the center and the fundamental group of the gauge group lead to global symmetries of the theory, under which different line operators are charged.

It turns out that four-dimensional gauge theories lead to a finer classification of global forms of a gauge group, that are usually referred to as {\em global variants}. They are essentially related to the following fact. When the gauge group is not simply-connected, instantons may exist while having a fractional number with respect to the normalization fixed for the simply connected variant. As a consequence, the $\theta$ angle of the theory can have a periodicity which is not $2\pi$ but a multiple of it. In some cases then, one observes that after a $2\pi$ shift of $\theta$ one goes from one global variant to another, which shares the same global form of the gauge group. In other words, the Witten effect \cite{Witten:1979ey} changes the charges of the line operators that are the observables of a given theory. 

At the same time, gauge theories, particularly supersymmetric ones, are deeply intertwined with a vast array of physical and mathematical structures. For instance, four-dimensional $\mathcal{N}=2$ gauge theories are naturally linked to complex integrable systems through Seiberg–Witten theory \cite{Donagi:1995cf}. This raises the question of whether the concept of global variants can also be extended to these structures and whether such an extension leads to meaningful refinements of conventional perspectives. In this paper, we focus more specifically on twisted elliptic Calogero–Moser (CM) systems (see e.g.~\cite{Olshanetsky:1981dk, Olshanetsky:1983wh, DHoker:1999yni, Kumar:2001iu}). CM systems constitute a fundamental class of integrable systems traditionally associated with simple Lie algebras, and their twisted elliptic versions are related to a particular class of gauge theories---namely, the $\mathcal{N}=2^*$ and $\mathcal{N}=1^*$ massive deformations of $\mathcal{N}=4$ Super-Yang-Mills (SYM) theories.

The connection between CM systems and $\mathcal{N}=1^*$ Super-Yang-Mills (SYM) theories \cite{Donagi:1995cf, Dorey:1999sj} is most transparent when the latter are formulated on $\mathbb{R}^3 \times S^1$. In this setting, the correspondence establishes a bijection between the extrema of a given CM system and the vacua of the associated $\mathcal{N}=1^*$ theory, which further specializes to a one-to-one matching between isolated extrema and gapped vacua.
However, the structure of the set of gapped vacua of the gauge theory on $\mathbb{R}^3 \times S^1$ depends on the global variant under consideration.\footnote{Recall that on $\bR^4$, the number of vacua of the gauge theory is determined by local dynamics, hence it is the same for any global variant. On $\mathbb{R}^3\times S^1$ instead, the non-trivial cycle in spacetime allows the global features of the gauge group to play a role, leading possibly to a different number of vacua.} Consequently, for the correspondence to hold in all cases, the CM system itself must be associated with a specific global variant. One of the main objectives of this paper is to define the notion of global variants for CM systems and to verify that they indeed preserve the expected relation to all global variants of $\mathcal{N}=1^*$ theories.

A prominent role in the classification of both the global variants, and the vacua/extrema, will be played by the duality transformations inherited from $\cN=4$ SYM on the gauge theory side, and the modular properties of the CM system on the other side. Both can be seen as transformations in $\mathrm{SL}(2,\bZ)$, and they are indeed mapped to each other. The duality transformations relating different vacua of the gauge theory can actually be promoted to duality symmetries, at some specific values of the original $\cN=4$ YM coupling. They are then generally non-invertible symmetries \cite{Choi:2021kmx, Kaidi:2021xfk}, in the sense that they can relate vacua displaying different physical properties, i.e.~confinement vs.~screening of probe charges.

While we define the notion of global variants of CM systems to align with their expected correspondence with supersymmetric gauge theories, we believe that this notion could also be derived more directly. Specifically, following an approach similar to \cite{Dorey:1999sj, Kumar:2001iu}, one could derive the potential of CM systems as a resummation of instanton contributions, carefully incorporating the topological subtleties associated with global variants of gauge theories. We anticipate that this more direct approach would lead to an equivalent definition of global variants of CM systems. Further in this spirit, it would be interesting to see whether it is possible to perform a topological manipulation on the CM system itself to go from one global variant to another one.

The number of vacua of $\cN=1^*$ SYM on $\mathbb{R}^3\times S^1$ has been related to the superconformal index of $\cN=4$ SYM via the Bethe Ansatz Equation (BAE) approach \cite{ArabiArdehali:2019orz, Benini:2021ano}, which expresses the index as a sum over solutions to BAE equations \cite{Closset:2017bse}. This conjectured relation naturally fits within the broader framework of the Bethe/Gauge correspondence \cite{Nekrasov:2009uh}, and the role of global variants in modifying the sum over Bethe solutions has recently been explored in \cite{Closset:2024sle}. Given the eventual relevance of this relation to black hole microstate counting, it is natural to ask whether the choice of global form in this context can also lead to physically distinct results. It should be emphasized, however, that the present paper focuses exclusively on the gapped vacua of $\cN=1^*$ SYM. In contrast, the correspondence with index computations involves contributions from both gapped and gapless vacua. Establishing a concrete connection between these two approaches remains an interesting direction for future work.

The plan of the paper is as follows. In \cref{sec: review} we review global variants of gauge theories, and the dualities relating them for $\cN=4$ SYM. The main players will be the gauge algebras $\mathfrak{su}(N)$ and $\mathfrak{so}(5)$, the latter being our main example for a non-simply laced algebra. In \cref{Sec:N=1*andIRphases} we recall what is known about vacua of $\cN=1^*$ SYM, with particular emphasis on gapped vacua. We also introduce the relation with Calogero-Moser systems, which are the focus of \cref{sec:GlobvarCM}, where we slowly build evidence towards a proposal for a definition of a global variant of such a model. In \cref{Sec:typeA} we expand and test our proposal, mostly on the algebras $\mathfrak{su}(6)$ and  $\mathfrak{su}(4)$ ($\mathfrak{su}(3)$ is worked out in the preceding section). Finally, in \cref{sec: so(5)} we apply our proposal to the example of $\mathfrak{so}(5)$, which turns out to be a rather non-trivial generalization of the previous cases.

\vspace{0.5cm}

\subsection*{Summary of the results}

We now give a bird's eye view of our results. 
In order to present our results, we need to give a more precise definition of the starting point, namely the global variants of a gauge theory, and the Calogero-Moser systems. For notations and further precisions we refer to the main text. 

\paragraph{Global variants}
Global variants of gauge theories on $\mathbb{R}^4$ with gauge algebra $\mathfrak{g}$ are in one-to-one correspondence with Lagrangian subgroups $\mathcal{L} \subset \widehat{\mathcal Z} \times \mathcal Z$, where $\mathcal Z$ denotes the center of the connected and simply-connected compact Lie group with Lie algebra $\mathfrak g$. This amounts to a choice of mutually local line operators.

\paragraph{Calogero-Moser systems}
Twisted elliptic Calogero-Moser (CM) systems appear 
by identifying their phase space to the Coulomb branch of $4d$ $\mathcal N=2^*$ theories on $\mathbb{R}^3\times S^1$ \cite{Donagi:1995cf, Gorsky:1995zq, Martinec:1995by, Seiberg:1996nz, Kapustin:1998xn, DHoker:1998zuv, DHoker:1998xad, Kumar:2001iu}. For a given gauge theory, the CM system depends only on the Lie algebra $\mathfrak{g}$ and the complexified coupling constant $\tau$, and it is formally defined by a potential $V_{\mathfrak{g}}^{\mathrm{tw}}(Z ; \tau )$ which is written in terms of Weierstrass functions. The potential depends on $Z = \sigma + \tau \, a\in \mathfrak{h}^{\ast} \oplus \tau \mathfrak{h}$,
with the variable $\sigma$ identified with the scalar dual to the $3d$ unbroken abelian gauge field and the variable $a$ with the holonomy of the gauge field along the circle $S^1$.
Importantly, the potential of the CM system is symmetric under the Weyl group $W_{\mathfrak{g}}$ and outer automorphisms of $\mathfrak{g}$, and it naturally has the following periodicities:
\begin{equation}
V_{\mathfrak{g}}^{\mathrm{tw}} (Z + 
(\Lambda_w \oplus \tau \Gamma_w )) = V_{\mathfrak{g}}^{\mathrm{tw}} (Z )\ . 
\end{equation}
Finally, the $\mathcal N=1^*$ mass deformation lifts the $\mathcal N=2^*$ Coulomb branch entirely, except for the singular loci at which BPS particles become massless. In terms of the CM system, the singular loci on the Coulomb branch of $\mathcal N=2^*$ theories compactified on a circle $S^1$ are identified with the extrema of the potential.\\

\noindent We now state our main result.

\paragraph{Global variant of a Calogero-Moser system}
The \emph{global variant $\mathcal{L}$ of the twisted elliptic Calogero--Moser system} corresponding to $\mathfrak g$, or \emph{$(\mathfrak{g} , \mathcal{L})$ CM} for short, is the Calogero--Moser system where we gauge the translation symmetries in
\begin{equation}
        \left\{ \lambda+\chi\tau | \lambda\in\Lambda_w \, , \,\chi\in\Gamma_w \, , \, \left(\phi_\Lambda(\lambda),\phi_\Gamma(\chi)\right)\in\mathcal L \right\} \ , 
\end{equation}
where $\phi_\Lambda$ and $\phi_\Gamma$ are the abelian group morphisms from $\Lambda_w$ and $\Gamma_w$ to $\widehat{\mathcal Z}$ and $\mathcal Z$ respectively, that induce the isomorphisms $\Lambda_w/\Lambda_r\cong\widehat{\mathcal Z}$ and $\Gamma_w/\Gamma_r\cong \mathcal Z$.\footnote{Presumably, this notion of global variant should extend to many other families of integrable systems associated with Lie algebras, particularly those governing the low-energy dynamics on the Coulomb branch of $4d$ $\mathcal{N}=2$ theories. These include periodic Toda systems (pure SYM theories), spin generalizations of elliptic Calogero–Moser systems (elliptic models, i.e.~necklace quiver theories), and, more broadly, all Hitchin systems (theories in class $\mathcal S$), as well as certain integrable systems of non-Hitchin type ($4d$ $\mathcal N=2$ theories not in class $\mathcal S$).}

Configurations of points in the complex plane related by gauged translations and symmetries of the root system $\Delta(\mathfrak g)$ are considered equivalent. Configurations which cannot be related by such operations are considered distinct.

Our main claim is that there is a one-to-one correspondence
\begin{center}
 \begin{tikzpicture}
 \node[align=center, draw, text width = 5cm] (1) at (0,0) {Isolated extrema of the $(\mathfrak{g} , \mathcal{L})$ CM};
 \node[align=center, draw, text width = 5cm] (2) at (8,0) {Gapped vacua of the $(\mathfrak{g} , \mathcal{L})$ $\mathcal N=1^*$ theory on $\mathbb{R}^3\times S^1$};
 \draw[<->] (3,0)--(5,0) node[midway,above] {$1:1$};
 \end{tikzpicture}
\end{center} 

Given this correspondence, it is interesting to investigate how the gapped vacua of the different global variants of $\mathcal N=1^*$ are permuted by modular transformations. In particular, we initiate the study of how many different orbits there are under modular transformations, cf.~\cref{eq:totalCountVacua}, and how many vacua are contained in each orbit. Already in the case where $\mathfrak{g} = \mathfrak{su}(N)$, we observe a fascinating non-trivial dependence on the factorization of $N$. We hope to return to this issue in the future.

\paragraph{Low energy effective description of $\cN=1^*$ vacua} Our main application is the study of gapped vacua of $\cN=1^*$ SYM theories. A complete characterization of these ground states requires determining the pattern of line operators which are condensed on each vacua. This in turn prescribes the realization of global symmetries at low energies. We leverage the analysis of the extrema of the twisted CM potential to obtain this information. For instance, the occurrence of additional degeneracies when the theory is compactified on $\bR^3\times S^1$ signals the presence of topological order, with the spontaneous breaking of a 1-form symmetry. Furthermore, when the theory enjoys (non-invertible) self-duality symmetries, the modular transformations permuting the extrema of the CM potential instruct us on how the gapped ground states organize into representations of these symmetries. Most notably, 
this structure depends on the global variant associated to the gauge theory and the CM system. Finally, all such insights enable an explicit uplift to four dimensions, hence leading to a complete characterization of these gapped vacua in $\bR^4$.

\begin{figure}
\begin{minipage}{.5\linewidth}
\centering
\subfloat[][$\rho = \mathrm{Spin}(5)$]{\label{fig:spin5SUMMARY}\includegraphics[width=0.9\linewidth]{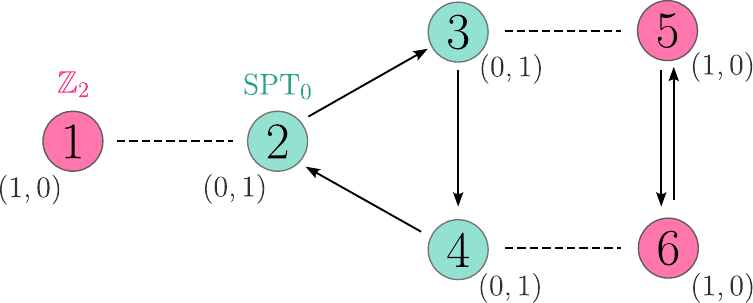}}
\end{minipage}%
\begin{minipage}{.5\linewidth}
\centering
\subfloat[][$\rho = \mathrm{SO}(5)_+$]{\label{fig:so5+SUMMARY}\includegraphics[width=0.9\linewidth]{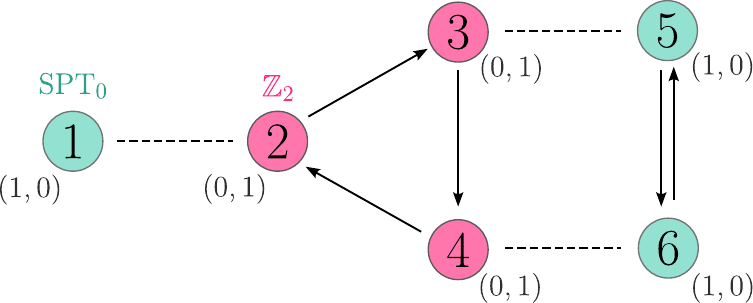}}
\end{minipage}\par\medskip \medskip
\caption{Vacuum structure of ${\rm Spin}(5)$ and $\mathrm{SO}(5)_+$ $\mathcal{N}=1^*$ theories on $\mathbb{R}^4$. Nodes colored in pink and teal respectively correspond to gapped vacua hosting a $\bZ_2$ gauge theory and a $\bZ_2$ SPT. }
\label{fig:so5VariantVacSUMMARY}
\end{figure}

The approach just described becomes extremely useful when there is no available analytical expressions for the expectation value of the superpotential on these ground states. As an illustrative example of this kind, in \cref{fig:so5VariantVacSUMMARY} we display the vacuum structure corresponding to $\cN=1^*$ with gauge groups ${\rm Spin}(5)$ and $\mathrm{SO}(5)_+$, both corresponding to the gauge algebra $\mathfrak{so}(5)$. The equivalence classes of line operators for this theory are labeled by a pair of integers $(n,m)\in \bZ_2\times \bZ_2$. 
Contrarily to the $\mathfrak{su}(N)$ case, here there is no possible bijection between the gapped vacua (of which there are six) and the non-trivial line operators (of which there are three).
The nodes in these graphs represent the gapped vacua and we also prescribe the condensed line operator, together with the corresponding topological field theory taking place on each of them. 
This structure is completely determined by consistency with the realization of the global symmetries, in particular the orbits under the action of the non-invertible $S_2$-duality symmetry at the self-dual coupling $\tau^*=i/\sqrt{2}$, and the action of the modular $T$-transformation, which is in turn related to the emergence of discrete symmetries that are spontaneously broken at low energies.            

\section{Global variants of gauge theories and duality}
\label{sec: review}

The notion of global variant can be regarded as a piece of the defining data concerning a generic gauge theory. For concreteness, let us consider a four dimensional gauge theory. The latter is characterized by specifying a Lie algebra $\mathfrak{g}$, a set of (typically complex) bare couplings,\footnote{At this stage, we are not interested in whether the theory undergoes renormalization, for which the specified coupling plays the role of a microscopic (UV) parameter, or alternatively describes a fixed point with the coupling being a coordinate on a conformal manifold (if there is such). The latter scenario corresponds to the case of $\cN=4$ SYM, whereas the former applies to $\cN=1^*$ SYM theory.  } and a choice of global variant $\rho$. The local dynamics associated to a putative gauge theory is usually determined by the first two items, namely the Lie algebra and the couplings. On the other hand, the global variant $\rho$ corresponds to specifying a consistent set of line operators, as we will briefly review below following \cite{Aharony:2013hda} (see also \cite{Kapustin:2005py, Gaiotto:2010be}). 

A gauge theory based on a Lie algebra $\mathfrak{g}$ admits a set of line operators falling in conjugacy classes labeled by pairs 
\be\label{alllines}
(a_{\rm e},b_{\rm m})\in \frac{\Lambda_w\times \Gamma_w}{\Lambda_r \times \Gamma_r} \cong \widehat{\mathcal{Z}} \times \mathcal{Z}\ ,
\ee
where $\Lambda_w$, respectively $\Lambda_r$, denotes the weight and root lattices, while $\Gamma_w$, respectively $\Gamma_r$, are the coweight and coroot lattices. In addition, $\cZ$ corresponds to the center of the universal covering group $\widetilde G$ associated to $\mathfrak{g}$ and $\widehat{\mathcal{Z}}\equiv {\rm Hom}(\mathcal{Z},U(1))$ is its Pontryagin dual. We provide a description of some relevant aspects about Lie algebras in \cref{App:Lie}.
Physically, non-trivial conjugacy classes in $\Lambda_w/\Lambda_r$ comprise external electric charges, {\it i.e.} Wilson lines, which cannot be completely screened by dynamical gluons. Conversely, the quotient $\Gamma_w/\Gamma_r$ describes analogous objects for the Langlands (or electric-magnetic) dual algebra $\mathfrak{g}^L$. From the perspective of the theory with Lie algebra $\mathfrak{g}$, $\Gamma_w/\Gamma_r$ classifies external magnetic charges. Throughout this article we will mostly focus on theories based on self-dual algebras, namely $\mathfrak{g}\cong \mathfrak{g}^L$.\footnote{This includes simply-laced algebras along with the self-dual non simply-laced ones $B_2$, $G_2$ and $F_4$.} Moreover, the center $\mathcal{Z}$ will be described by a finite Abelian group, hence $\widehat{\mathcal{Z}}\cong \mathcal{Z}$. 

For the sake of brevity, we will illustrate the case of $\mathcal{Z}=\widehat{\mathcal{Z}}=\bZ_N$, therefore $(a_{\rm e},b_{\rm m})=(n,m)\in \bZ_N\times \bZ_N$.\footnote{The generalization of this analysis to the case of $\mathcal{Z}$ being a product group, such as $\mathcal{Z}=\bZ_k\times\bZ_{k'}$, is technically more cumbersome, but it does not add any new insight at the conceptual level.} Within $\widehat{\mathcal{Z}}\times\mathcal{Z}$, there is a natural bilinear action given by the Dirac pairing: 
\be
\langle (n,m),(n',m')\rangle=n m'-n' m \quad {\rm mod} \; N\ .
\ee
A consistent spectrum of line operators, namely a global variant $\rho$, is determined by a Lagrangian subgroup $\cL\subset \widehat{\mathcal{Z}}\times\mathcal{Z}$. This corresponds to a maximal set of mutually local line operators:
\bea
(n,m) \, , \, (n',m') \, \in \cL \,\,\,\, \Rightarrow \,\,\,\, \langle (n,m),(n'm')\rangle=0 \,\,\, {\rm mod} \,\,\, N\ , \\
\langle (n,m),(n'm')\rangle=0 \quad {\rm mod} \; N \qquad \forall (n',m')\in \cL \quad \Rightarrow \quad (n,m)\in \cL \ .
\eea
For $\mathcal{Z}=\bZ_N$, the counting of all possible inequivalent choices is given by:
\be\label{eq:sigma1N}
\sigma_1(N) = \sum_{d \,| N} \, d \ .
\ee
In the absence of matter fields transforming in representations charged under the center $\mathcal{Z}$, the choice made for the spectrum of line operators completely determines the 1-form global symmetry featured by the gauge theory. 

As a central piece of the global data pertaining to gauge theories, the global variant plays important roles in several aspects concerning their dynamics. As mentioned, it prescribes the genuine line operators and potential 1-form global symmetries, therefore having direct impact in the characterization of infrared phases for which line operators act typically as order parameters. 

A further important aspect concerns the interplay between the global variant and duality, as we will briefly review now. For future convenience, we will focus on the case of $\cN=4$ SYM theory, namely the UV fixed point of the $\cN=1^*$ flow which will be the main subject of study in the remaining sections. The $\cN=4$ SYM theory belongs to a conformal manifold spanned by the exactly marginal coupling
\be\label{eq: comp coupling}
\tau=\frac{\theta}{2\pi} + \frac{4\pi i}{g_\mathrm{YM}^2}\ .
\ee 
Given a choice of Lie algebra and global variant $\rho$, namely a Lagrangian subgroup as described above, we denote the corresponding theory by $\mathsf{T}_\rho(\tau)$. There is a well defined action of the $\mathrm{SL}(2,\bZ)$ duality group on the landscape of $\cN=4$ SYM theories \cite{Kapustin:2006pk}.  
A duality transformation $I_g$ associated to a given element 
\begin{equation}
    g=\begin{pmatrix}
    a & b \\ c & d
\end{pmatrix}\in \mathrm{SL}(2,\bZ)\ ,
\end{equation} 
defines an isomorphism 
\be
I_g \colon \,\, \mathsf{T}_\rho(\tau) \, \longrightarrow \, \mathsf{T}_{\rho^g}(\tau^g)\ , \quad \tau^g =\frac{a\tau+b}{c\tau+d}\ ,
\ee
with $a,b,c,d \in \bZ$ and $ad-bc=1$.
The resulting global variant $\rho^g$ is given by the action on the conjugacy classes of line operators, with the pair $(n,m)$ transforming as a vector under left multiplication
\be
I_g\colon \,\, \cL_\rho \, \longrightarrow \,  \cL_{\rho^g}\equiv \left\lbrace g.(n,m) \, | \, (n,m)\in\cL_\rho \right\rbrace\, .
\ee
Let us emphasize that the two $\cN=4$ SYM theories $\mathsf{T}_\rho(\tau)$ and $\mathsf{T}_{\rho^g}(\tau^g)$ stand as equivalent representations of the same physical system, though described in terms of a different set of variables. 

Importantly, for a fixed $\mathcal{Z}$, the global variants $\rho$ do not furnish a faithful representation of the full $\mathrm{SL}(2,\bZ)$, but rather a (typically projective) representation of a finite group $\cF_\mathcal{Z}$. For the prototypical example of $\mathcal{Z}=\bZ_N$, $\cF_{\bZ_N}=\mathrm{PSL}(2,\bZ_N)$ is generated by elements $S$ and $T$ satisfying\footnote{Regarded as elements of $\mathrm{SL}(2,\bZ)$, these are conventionally assigned the following matrix representations
\be
S=\begin{pmatrix}
0 & -1 \\
1 & 0
\end{pmatrix} \ , \qquad 
T=\begin{pmatrix}
1 & 1 \\
0 & 1
\end{pmatrix}\ .
\ee
In presence of fermions, one actually needs to consider the metaplectic group ${\rm Mp}(2,\bZ_N)$ obtained through the extension by fermion number $(-1)^F$. This subtlety is not going to play any significant role for the purposes of this work and we will henceforth ignore it.} 
\be
S^2=(ST)^3=1 \ , \quad T^N=1\ , 
\ee
where in so doing we have modded out by charge conjugation. 

Let us list the duality orbits concerning the Lie algebras that we will focus on as examples in the rest of this article:   
\paragraph{$\mathbf{\mathfrak{su}(3)}$:} There are four global variants associated to the Lie algebra $\mathfrak{su}(3)$. Under the action of the duality group, these global variants fit within the following orbits
\be\label{eq:dualityorbitsu(3)}
\begin{tikzcd}[column sep=-10pt]
& \mathrm{SU}(3) \arrow[loop above, "T"] \arrow[d,dashed, leftrightarrow, "S"] & \\
& \mathrm{PSU}(3)_0 \arrow[rd, "T"] & \\
\mathrm{PSU}(3)_2 \arrow[ur, "T"] & & \mathrm{PSU}(3)_1 \arrow[ll, "T"] \arrow[ll,dashed, leftrightarrow, bend left = 50, "S"]
\end{tikzcd}
\ee
where $\mathrm{PSU}(N)_k$ denotes the group obtained by gauging the whole center $\bZ_N$ with discrete torsion $k\in \bZ_N$. 

\paragraph{$\mathbf{\mathfrak{su}(4)}$:} By including the global variants obtained by gauging a $\bZ_2$ subgroup of the $\bZ_4$ 1-form symmetry, one is left with 7 possible global variants transforming as follows under duality:  
\bea
\begin{tikzcd}
\raisebox{0pt}[0.5em][0pt]{$(\mathrm{SU}(4)/\bZ_2)_0 \quad\;\;$} \arrow[loop right, "{S,T}"]
\end{tikzcd} \\[-1.5em]
\begin{tikzcd}[row sep=tiny]
& & \mathrm{PSU}(4)_1 \arrow[rd, "T"] \arrow[dd, leftrightarrow, "S", dashed] \\
\mathrm{SU}(4) \arrow[loop below, "T"] \arrow[r, leftrightarrow, "S", dashed] & \mathrm{PSU}(4)_0 \arrow[ru, "T"]  & & \mathrm{PSU}(4)_2 \arrow[ld,"T"] \arrow[r, leftrightarrow, "S", dashed] & {(\mathrm{SU}(4)/\bZ_2)_1} \arrow[loop below, "T"] \\
& & \mathrm{PSU}(4)_3 \arrow[lu,"T"]
\end{tikzcd}
\eea

\paragraph{$\mathbf{\mathfrak{su}(6)}$:} 
For this Lie algebra, there are in total 12 global variants, and they are all interconnected by a chain of $S$ and $T$ dualities: 
\bea
\begin{tikzcd}[column sep=-15pt]
& & & \mathrm{SU}(6) \arrow[loop above, "T"] \arrow[d,leftrightarrow,"S", dashed] & & & \\
& & & \mathrm{PSU}(6)_0 \arrow[drrr, "T"] & & & \\
\mathrm{PSU}(6)_5 \arrow[urrr, "T"] \arrow[rrrrrr,leftrightarrow,"S", dashed] & & & & & & \mathrm{PSU}(6)_5 \arrow[lddddd,"T"] \\
& & (\mathrm{SU}(6)/\mathbb{Z}_3)_1 \arrow[rd, "T"] & & (\mathrm{SU}(6)/\mathbb{Z}_3)_2 \arrow[ll,"T"] & & \\
& & & (\mathrm{SU}(6)/\mathbb{Z}_3)_0 \arrow[ru, "T"] \arrow[d,leftrightarrow,"S", dashed] & & & \\
& & & (\mathrm{SU}(6)/\mathbb{Z}_2)_0 \arrow[d,leftrightarrow,"T"] & & & \\
& & & (\mathrm{SU}(6)/\mathbb{Z}_2)_1 \arrow[d,leftrightarrow,"S",dashed] & & & \\
& \mathrm{PSU}(6)_4 \arrow[luuuuu,"T"] \arrow[uuuur,leftrightarrow,"S",dashed] & & \mathrm{PSU}(6)_3 \arrow[ll,"T"] & & \mathrm{PSU}(6)_2 \arrow[ll,"T"] \arrow[uuuul,leftrightarrow,"S",dashed] &
\end{tikzcd}
\eea

\paragraph{$\mathbf{\mathfrak{so}(5)}$:} The three global variants associated to the algebra $\mathfrak{so}(5)$ are $\mathrm{Spin}(5)$, $\mathrm{SO}(5)_+$ and $\mathrm{SO}(5)_-$. The 1-form symmetry is $\bZ_2^{(1)}$ and there is no non-trivial Witten effect for this class of theories, implying a trivial action for $T$. We therefore have:
\be\label{eq: so(5) S-duality}
\begin{tikzcd}[row sep=tiny]
 & \mathrm{Spin}(5) \arrow[loop below, "T"] \arrow[r, leftrightarrow, "S"] & \mathrm{SO}(5)_+  \arrow[loop below, "T"]  & \mathrm{SO}(5)_-  \arrow[loop below, "{S,T}"]
\end{tikzcd}
\ee

As we have seen in the above examples, the spectrum of lines associated to a given global variant $\rho$ can be associated to gauge theories with gauge group of the form $G=\widetilde G/\Pi$, for $\Pi\subset\mathcal{Z}$. This is related to the fact that there is yet another way to map gauge theories with different global variants which does not rely on the notion of duality. This is achieved by means of topological manipulations $\varphi$, generated by partial or complete gauging of the 1-form symmetry, possibly after stacking the theory with a symmetry-protected topological (SPT) phase (namely, an invertible phase for the background gauge field of the 1-form symmetry). Any two global variants can be reached by such topological manipulations. An in depth description of this matter is not essential for our purposes and we refer to \cite{Kaidi:2022uux} for more exhaustive studies. 
We will limit ourselves to state that for a given duality denoted by the element $g$, there exists a topological manipulation $\varphi_g$ with action
$
\varphi_g \,\, : \,\,  \mathsf{T}_\rho(\tau) \, \longrightarrow \, \mathsf{T}_{\rho^g}(\tau)
$. 
This operation can be interpreted as a map between the partition functions of $\mathsf{T}_{\rho}$ and $\mathsf{T}_{\rho^g}$ at fixed $\tau$. Alternatively, acting with $\varphi_g$ on half of spacetime, one may label by $\varphi_g$ a (topological) interface between two $\cN=4$ SYM theories specified by different choices of line operators, hence not physically equivalent. However, when the coupling satisfies $g\cdot \tau^*=\tau^*$ for some $g\in \mathrm{SL}(2,\bZ)$, it becomes a topological interface of the theory with itself (up to duality), hence a global symmetry. Equivalently, one can define a topological symmetry defect by the composition $\varphi_g^\dagger \circ I_g \,:  \mathsf{T}_\rho(\tau^*) \, \to \, \mathsf{T}_{\rho}(\tau^*)$. Self-duality symmetries of this type have been studied in the context of various four dimensional gauge theories, including $\cN=4$ SYM, in e.g.~\cite{Kaidi:2021xfk, Choi:2022zal}.

\section{The \texorpdfstring{$\cN=1^*$}{N=1*} SYM theory and its infrared phases}\label{Sec:N=1*andIRphases}

The $\cN=1^*$ theory is most conveniently studied as a massive deformation of $\cN=4$ SYM theory. 
More precisely, denoting $\Phi_i$ ($i=1,2,3$) the adjoint $\cN=1$ chiral multiplets contained in the $\cN=4$ vector multiplet, the $\cN=1^*$ SYM theory corresponds to the following deformation of the $\cN=4$ superpotential:
\be\label{eq: W}
W={\rm Tr}\left[\Phi_1[\Phi_2,\Phi_3]+m\sum_i \Phi^2_i\right]\ , \quad \Phi_i \in \mathfrak{g}\ .
\ee 
with $m$ a mass parameter. Consequently, supersymmetry is reduced to $\cN=1$, with moreover a $U(1)_R$ symmetry which is explicitly broken.\footnote{More generally, one may assign a different mass $m_i$ to the adjoint chirals, hence breaking also the flavor $\mathrm{SU}(3)$ symmetry. This distinction is not going to play any role in the following.}
The deformation proportional to $m$ is relevant and the system generically flows to strong coupling. In this context, the complexified coupling \cref{eq: comp coupling} should be regarded as a microscopic parameter, i.e.~as the UV initial conditions for the RG flow.

The duality group of the UV $\cN=4$ SYM theory has a precise realization on the infrared dynamics of the $\cN=1^*$ theory \cite{Argyres:1999xu}. Furthermore, when the flow starts at the specific loci $\tau^*$, the self-duality symmetries naturally extend to the $\cN=1^*$ theory \cite{Damia:2023ses}.\footnote{There is a subtlety that stems from the fact that dualities have a non-trivial action on the superspace coordinates and, as such, on the fermionic components of the chiral multiplet \cite{Kapustin:2006pk} (see also \cite{Intriligator:1998ig}).
To define a preserved duality transformation one must then compose it with an element of the broken $U(1)_R$ symmetry.} These self-duality global symmetries are preserved by the RG flow triggered by the deformed superpotential \cref{eq: W}. As such, they are realized (either preserved or spontaneously broken) within the intricate IR structure of $\cN=1^*$ SYM theory, that we now review. Explicit examples of this phenomenon have been studied in detail in \cite{Damia:2023ses}, focusing on Lie algebras of type $A$.

The classification of the infrared phases of $\cN=1^*$ theory dates back to \cite{Donagi:1995cf, Dorey:1999sj, Polchinski:2000uf}. With the aid of supersymmetry, this rich structure of gapped and gapless vacua can be accessed without any need of non-perturbative techniques. The starting point amounts to considering the $F$-term equations arising from varying the deformed superpotential:
\be\label{eq: F term}
[\Phi_i , \Phi_j] =-2m \epsilon_{ijk} \; \Phi_k \ .
\ee 
Since $\Phi_i$ is valued in the Lie algebra $\mathfrak{g}$, the solutions to the above equations are completely determined by reducible embeddings (of appropriate dimensionality)\footnote{For the case of $\mathfrak{g}=\mathfrak{su}(N)$, the solutions are determined by $N$-dimensional representations of $\mathfrak{su}(2)$.} of $\mathfrak{sl}(2,\bC)$ into the (complexification of) $\mathfrak{g}$.

Clearly, for any $\mathfrak{g}$ there will be a trivial solution $\Phi_i=0$. Physically, within this locus one can reliably integrate out the massive chiral multiplets, hence obtaining a pure $\cN=1$ SYM theory at intermediate energies. In turn, the dynamics of $\cN=1$ SYM will lead to various confining vacua in the infrared, characterized by certain condensation patterns of magnetically charged line operators.    
On the other hand, any non-trivial solution to \cref{eq: F term} will partially Higgs the gauge group. Since $\Phi_i\in \mathfrak{g}$, the center of the gauge group (which depends on the global variant under consideration) will always remain as part of the gauge symmetry. A more refined determination of the Higgsing patterns corresponding to each solution requires the characterization of nilpotent orbits \cite{Bourget:2015lua}. This procedure will be detailed in section \ref{subsec:unbrokengaugegroups}, here we review the standard results for the Lie algebras of interest in the rest of the paper. 

\paragraph{Type $A$ Lie algebras:} 
For gauge algebras of the form $\mathfrak{g}=\mathfrak{su}(N)$, the solutions are counted by integer partitions of $N$. More precisely, for a generic partition of the form:
\begin{equation}
    N=\sum_{i=1}^k \mu_i \, \lambda_i\ , \quad \lambda_1\geq\dots \geq \lambda_k \in\mathbb{N}_{>0}\ ,\quad \mu_i\in\mathbb{N}_{>0}\ ,
\end{equation}
there is a unique solution to \cref{eq: F term} (up to isomorphism). In terms of the embedding, each integer $\lambda_i$ in the above partition encodes the dimension of the corresponding $\mathfrak{sl}(2,\bC)$ representation, which appears with multiplicity $\mu_i$. For later convenience, let us introduce the following compact notation:
\begin{equation}
    N=\sum_{i=1}^k \mu_i \, \lambda_i \; \Longleftrightarrow \; \underline{\lambda} = \{\lambda_1^{\mu_1}\ldots \lambda_k^{\mu_k}\}\ .
\end{equation}
For a given partition $\underline{\lambda}$, the unbroken gauge algebra takes the form
\be\label{eq:classicalHiggsing}
\bigoplus_{i} \mathfrak{su}(\mu_i) \oplus \mathfrak{u}(1)^{\oplus (k-1)}\ , 
\ee
while the global form of the unbroken gauge group will be determined by the global form of the UV theory. 

In general, we will be interested in studying gapped vacua, hence partitions with only one non-vanishing term ($k=1$). We will nevertheless make some comments on the gapless vacua at the end of this section. The trivial solution corresponds to $\underline{\lambda} = \{1^N\}$. Another special solution which is always present for any $N$ corresponds to $\underline{\lambda} = \{N\}$, leading to a complete Higgsing of the gauge bosons. This will be called a {\it Higgs vacuum} and, depending on the global form of the gauge group, it may be either trivially gapped or host a discrete gauge theory. For gauge algebras of the form $\mathfrak{su}(N)$, there is always one instance of this vacuum while it may appear more than once for other types of Lie algebras. 

According to the standard classification \cite{tHooft:1977nqb}, all gapped phases of $\cN=1^*$ SYM can be described by the condensation of a certain genuine or non-genuine line operator. For instance, on the Higgs phase alluded before the fundamental Wilson line (which is genuine in the $\mathrm{SU}(N)$ global variant) acquires an expectation value. On the contrary, the solution with $\mathfrak{su}(N)$ preserved gauge algebra gives rise to $N$ vacua characterized by the condensation of monopole/dyonic lines. 
A peculiar feature of $\cN=1^*$ theories with type $A$ algebras is that the number of gapped vacua is also given by \cref{eq:sigma1N}, and indeed each such vacuum can be associated to the condensation of one of the lines defining a global variant, in a one-to-one fashion.
This property does not hold for theories with other types of gauge algebras.

\paragraph{Type $B$ and $D$ Lie algebras:} The case of algebras of type $B$ and $D$ has been studied in \cite{Bourget:2015lua} by exploiting results from the theory of nilpotent orbits. 

For $\mathfrak{so}(2N+1)$, the solutions to the F-term equations \ref{eq: F term} are in one-to-one correspondence with partitions of $2N+1$ such that even integers occur with even multiplicities
\be\label{eq: type B partitions}
2N+1= \sum_{i=1}^{k} \mu_i \lambda_i\ , \quad \mu_i\in 2\bZ\; {\rm for }\; \lambda_i\in 2\bZ\ ,
\ee
whereas multiplicities for odd integers are not constrained. The preserved gauge algebra for a given partition of the form \cref{eq: type B partitions} reads:
\be\label{eq: preserved B}
\left(\bigoplus_{i \in D_e} \mathfrak{sp}(\mu_i) \right) \oplus \left(\bigoplus_{i\in D_o} \mathfrak{so}(\mu_i)\right) \ ,
\ee
where, for a given partition $\{\lambda_1^{\mu_1}\dots \lambda_k^{\mu_k}\}$, the set $D_e$ (resp. $D_o$) comprises the indices $i$ for which $\lambda_i$ is even (resp. odd). Again, the global aspects of the preserved gauge group will depend on the global variant $\rho$ under consideration. More precisely, the global form of \cref{eq: preserved B} will be determined by the centralizer group of the nilpotent orbit associated to the partition. Once more, the subset of four-dimensional gapped vacua corresponds to solutions without $\mathfrak{u}(1)=\mathfrak{so}(2)$ factors.
In section \ref{sec: so(5)} we will study the first non-trivial case $\mathfrak{g}=\mathfrak{so}(5)$ in great detail. 

For gauge algebras of type $D$, namely $\mathfrak{g}=\mathfrak{so}(2N)$, the story is almost the same as for type $B$: one has to consider partitions
\be
2N= \sum_{i=1}^{k} \mu_i \lambda_i\ , \quad \mu_i\in 2\bZ \; {\rm for }\; \lambda_i\in 2\bZ\ .
\ee
In particular, there may exist partitions of the above form for which $\mu_i=0$ whenever $\lambda_i\in 2\bZ+1$. This are called very even partitions and give rise to two different vacua, related by the $\bZ_2$ automorphism of type $D$ algebras \cite{Bourget:2015lua}. The preserved gauge algebra takes the same form \cref{eq: preserved B} as for type $B$. Due to certain subtleties that arise when dealing with theories based on this type of algebras, we will present a detailed analysis elsewhere.

Contrary to what happens for type $A$ gauge algebras, for the other classical algebra the map between line operators defining global variants and gapped vacua is no longer injective, the latter being more and more numerous than the former as the rank grows. In particular, the same line may acquire a non-zero expectation value in several massive vacua.

We now introduce the crucial relation between vacua in $d=4$ and $d=3$, upon a circle reduction.
Remarkably, the IR vacua of the $\cN=1^*$ theory can be accessed through the exact solution of the model compactified in $\bR^3\times S^1$ with supersymmetric periodicity conditions \cite{Donagi:1995cf, Dorey:1999sj}. The three dimensional theory is governed by an effective action for a set of chiral superfields $\{z_a\}$ ($a=1,\ldots, n$). The number $n$ generically depends on the rank of the particular Lie algebra $\mathfrak{g}$. 
In turn, the effective superpotential for this theory has been determined exactly as a function of the bare coupling $\tau$ and is given by the potential of the $n$-particle integrable Calogero--Moser (CM) Hamiltonian. The problem of finding the vacua of the three dimensional model is then reduced to finding complex extrema of the CM potential. Throughout this work, we will be mainly interested in the subset of gapped vacua of the $\cN=1^*$ theory, as opposed to vacua containing massless photons, hence we will focus only on isolated minima of the potential. 

The integrable Calogero-Moser system associated to a Lie algebra $\mathfrak{g}$ has been the focus of investigation in several contexts (see e.g.~\cite{DHoker:1999yni}). However, the role played by the global variant $\rho$ still remains elusive. It is the main goal of this work to bridge this gap by developing a precise framework in which the choice of global variant becomes manifest. This will amount to carefully analyzing the symmetries of a twisted version of the CM potential, together with a detailed account of several examples. As it turns out, the global variant $\rho$ has direct consequences on the characterization of the solutions and their degeneracies. Along the way, we will also describe how the action of duality in $\cN=4$ SYM descends in the IR to certain modular transformations acting on the coordinates of the CM system. These modular transformations have been previously studied, see for instance \cite{Bourget:2015cza}. For self-dual values $\tau^*$ of the modular parameter, where the system enjoys additional self-duality symmetries as described before, the solutions of the CM system accommodate into modules under the action of such (typically non-invertible) symmetries. This manifests itself into degeneracies of the exact values of the superpotential corresponding to each vacua. We will show explicit realizations of this phenomenon in several examples.

As it turns out, the uplift of the vacuum solutions to four dimensions hides a series of subtleties. Physically, there are two sources of discrepancies among the finite discrete sets of massive vacua in three and four dimensions respectively, which are rooted in the compactification procedure: 

\paragraph{I)} Given a gapped vacuum in four dimensions, the associated centralizer group may contain discrete abelian factors, leading to topological order with a spontaneously broken 1-form symmetry. A prototypical example is the Higgs vacuum for a global form given by the simply-connected group ({\it e.g.} $\rho=\mathrm{SU}(N)$ for $\mathfrak{g}=\mathfrak{su}(N)$). Upon compactification, it costs no energy to wrap non-trivial deconfined line operators along the $S^1$, leading to a series of gapped degenerate vacua distinguished by the expectation values of such line operators. Physically, the four dimensional spontaneously broken 1-form symmetry gives rise to a broken 0-form symmetry when compactified to three dimensions, hence the additional degeneracy. Consequently, many gapped vacua in three dimensions may map to a single gapped vacuum in four dimensions.

\paragraph{II)} As previously explained, there are solutions to the F-term equation leading to gapless vacua in four dimensions. However, some of them may acquire a gap upon compactification. This is the case, for example, when the centralizer group contains discrete factors acting non-trivially on the continuous abelian ones, such as $\mathrm{O}(2)=\mathrm{SO}(2)\rtimes \bZ_2$. In fact, even if the there is a massless photon on $\bR^4$ (see \cite{Heidenreich:2021xpr, Antinucci:2022eat} for a recent discussion of Maxwell theory based on $\mathrm{O}(2)$ gauge group), when compactifying in presence of a non-trivial Wilson line for $\bZ_2$, the photon acquires a mass as a consequence of the semi-direct product. Indeed, the non-trivial $\bZ_2$ Wilson line effectively imposes anti-periodic boundary conditions on the circle for all the fields charged under it, namely all the fields in the $\cN=1$ vector multiplet. As a result, the KK tower has masses given by $m_k=\frac{2\pi}{\beta}\left|k+\frac12\right|$, removing the presence of all zero modes. Note that this leads to a single gapped vacuum, since reducing without the $\bZ_2$ Wilson line yields a gapless vacuum, which in 3d has also a continuous degeneracy parameterized by the $\mathrm{SO}(2)$ Wilson line wrapped on the $S^1$ (see \cite{Damia:2023gtc} for a discussion of such moduli space).

\section{Global variants of elliptic Calogero--Moser systems}\label{sec:GlobvarCM}

Having just motivated their crucial interest for the characterization of the vacua of $\cN=1^*$ SYM, we turn in this section to the study of elliptic Calogero-Moser integrable systems. We start by reviewing their properties, in particular their symmetries, their twisted generalization, and their relation to both $\cN=2^*$ and $\cN=1^*$ SYM theories. We then put forward our proposal to endow such Calogero-Moser systems with data that specifies the global variant of the gauge group of $\cN=1^*$ SYM. We illustrate it with the $\mathfrak{su}(3)$ theories, before plunging into more involved examples in the next two sections.

\subsection{Elliptic Calogero--Moser systems}

The elliptic Calogero--Moser system of Dynkin type $A_{N-1}$ describes the system of $N$ particles on a torus, i.e.~an elliptic curve $E_\tau$, interacting through the pairwise ``repulsive'' potential:
\begin{equation}
    V_{A_{N-1}}(Z ; \tau ) = \frac{g}{2}\sum_{i\neq j} \wp (z_i -z_j ; \tau) = g \sum_{i < j} \wp (z_i -z_j ; \tau)\ ,
\end{equation}
where the indices $i$ and $j$ appearing in the sums run from $1$ to $N$, and where $g$ is a coupling constant. For each $i=1,\dots,N$, $z_i\in E_\tau$ denotes the position of the $i$-th particle, the $z_i$'s satisfy the constraint $\sum z_i=0$.\footnote{Root systems of types $A$, $E$, and $G$ are typically represented in an $(r+1)$-dimensional vector space, where $r$ is the rank. For instance, in type $A_{N-1}$, the roots are usually given by $e_i - e_j$ for all $i \neq j = 1, \dots, N$. Consequently, the configuration of points has an additional translation degree of freedom:
\begin{equation}\label{Eq:shiftsymm}
    (z_1,\dots,z_N)\rightarrow (z_1+\omega,\dots,z_N+\omega)\ , \quad \omega\in\mathbb{C}\ .
\end{equation}
There are two natural ways to fix this freedom: either by imposing $\sum z_i = 0$ or by freezing one of the coordinates, e.g. $z_N = 0$. We will always use the former, but we will make a brief comment on the latter later on.} 
The function
\begin{equation}
    \wp(z;\tau) = \frac{1}{z^2} + \sum_{(m,n)\neq(0,0)} \left[\frac{1}{(z+m+n\tau)^2} - \frac{1}{(m+n\tau)^2}\right]\ ,
\end{equation}
is the Weierstrass $\wp$-function on $E_\tau$. This dynamical system is completely integrable.

The relationship with the root system $\Delta_{A_{N-1}}$ of type $A_{N-1}$ follows from rewriting:
\begin{equation}
    V_{A_{N-1}}(Z ; \tau ) = \frac{g}{2} \sum_{\alpha \in \Delta_{A_{N-1}}} \wp ( \alpha(Z) ; \tau) = g\sum_{\alpha \in \Delta^+_{A_{N-1}}} \wp ( \alpha(Z) ; \tau)\ ,
\end{equation}
where $\Delta^+_{A_{N-1}}\subset \Delta_{A_{N-1}}$ stands for the subset of positive roots, and $Z = (z_1,\dots,z_N)\in E_\tau^N$. The relation between this more abstract notation and the previous one is simply
\begin{equation}
    (e_i-e_j)(Z) = z_i-z_j\ .
\end{equation}
The periodicity of $\wp$ implies that the potential does not depend on the lift of $Z$. Moreover, $V_{A_{N-1}}(Z ; \tau)$ is invariant under the action of the Weyl group $W_{A_{N-1}}$ on $\mathfrak h$.

Similarly, there exists an elliptic Calogero--Moser system for every finite-dimensional compact simple Lie algebra $\mathfrak g$, which is also completely integrable (see e.g. \cite{Olshanetsky:1981dk} and references therein).\footnote{In fact, there exist completely integrable Calogero--Moser systems for every finite Coxeter group.} The potential reads: 
\begin{equation}
    V_{\mathfrak{g}}(Z ; \tau )  = \frac{1}{2} \sum_{\alpha \in \Delta(\mathfrak{g})} g_{\nu(\alpha)} \wp( \alpha (Z) ; \tau)= \sum_{\alpha \in \Delta^+(\mathfrak{g})} g_{\nu(\alpha)} \wp( \alpha (Z) ; \tau)\ ,
\end{equation}
where $\Delta(\mathfrak{g})\subset \mathfrak{h}^*$ is the root system of $\mathfrak{g}$, and $\Delta^+(\mathfrak{g})\subset \Delta(\mathfrak{g})$ a chosen subset of positive roots, $\mathfrak h$ being a Cartan subalgebra of $\mathfrak g$. Note that for Lie algebras of type $B$, $C$, $D$ and $F_4$, the vector $Z\in\mathfrak h$ encodes the position of $r$ particles on $E_\tau$, where $r = \mathrm{rank}(\mathfrak g)$. For Lie algebras of type $A$, $E$ and $G_2$, we already mentioned that one rather takes $Z$ to encode the position of $r+1$ particles on $E_\tau$, with a suitable constraint.

The index $\nu(\alpha)$ is defined as follows:
\begin{equation}\label{eq:indexCM}
    \nu(\alpha) = \frac{\vert\alpha_\mathrm{long}\vert^2}{\vert \alpha\vert^2} = \frac{2}{\vert \alpha\vert^2}\ .
\end{equation} 
The norm $\vert\cdot\vert$ is induced by the Killing form, normalized such that long roots have squared length two. Since the couplings $g_{\nu(\alpha)}$ remain constant along the orbits of roots under the Weyl group $W_{\mathfrak g}$, the potential $V_{\mathfrak{g}}(Z ; \tau )$ is invariant under the action of $W_{\mathfrak g}$ on $\mathfrak h$. 

When $\mathfrak g$ is simply-laced, all the roots have the same length and there is a single $W_{\mathfrak g}$-orbit, hence $g_{\nu(\alpha)} = g_1$ for all roots $\alpha$. In contrast, when $\mathfrak g$ is not simply-laced, there is more than one $W_{\mathfrak g}$-orbit of roots---in fact, exactly two. The orbit of long roots is assigned the coupling $g_1$, while the orbit of short roots is assigned the coupling $g_s$. One has $\nu(\alpha_s)\equiv s=2$ for types $B$, $C$ and $F_4$, and $s=3$ for type $G_2$.

\paragraph{Twisted elliptic Calogero--Moser systems}

When the Lie algebra $\mathfrak g$ is not simply-laced, implying that there are roots of different lengths in $\Delta(\mathfrak{g})$, one can define a mild modification of the elliptic Calogero--Moser system known as twisted elliptic Calogero--Moser system. It is characterized by the following potential:
\begin{equation}
    V_{\mathfrak{g}}^{\mathrm{tw}}(Z ; \tau ) := \sum_{\alpha \in \Delta^+(\mathfrak{g})} g_{\nu(\alpha)} \wp_{\nu(\alpha)}\left( \alpha (Z) ; \tau\right)\ ,
\end{equation}
where $\nu(\alpha)$ is defined in \cref{eq:indexCM}, and $\wp_{\nu(\alpha)}$ is the $\nu(\alpha)$-twisted Weierstrass $\wp$-function: 
\begin{equation}
    \wp_{n}(z;\tau) := \sum_{k=0}^{n-1} \wp\left(z+\frac{k}{n};\tau\right)\ .
\end{equation}
Twisted $\wp$-functions have smaller periods along the ``horizontal'' direction:
\begin{equation}
    \wp_n(z;\tau) = \wp_n(z+\frac{1}{n};\tau) = \wp_n(z+\tau;\tau)\ .
\end{equation}
The twisted potential rewrites:
\begin{equation}\label{eq:twistedCMpotential}
    V_{\mathfrak{g}}^{\mathrm{tw}}(Z ; \tau ) := g_1\sum_{\alpha_l \in \Delta_l^+(\mathfrak{g})} \wp\left( \alpha_l (Z) ; \tau\right)+g_s\sum_{\alpha_s \in \Delta_s^+(\mathfrak{g})} \wp_{s}\left( \alpha_s (Z) ; \tau\right)\ ,
\end{equation}
where $\Delta_l^+(\mathfrak g)$ and $\Delta_s^+(\mathfrak g)$ are the sets of long and short positive roots of $\mathfrak g$, respectively.

\subsection{Symmetries of the potential}\label{subsec:symmetriesCMpotential}

Now we look at the symmetries of the potentials introduced in the previous subsection, in particular the periodicities in $Z$ induced by the periodicities of the Weierstrass $\wp$-function on $E_\tau$.

The potentials $V_{\mathfrak g}(Z;\tau)$ and $V_{\mathfrak g}^{\mathrm{tw}}(Z;\tau)$ are invariant under all symmetries of the root system $\Delta(\mathfrak g)$, including the Weyl group $W_\mathfrak{g}$ discussed above, as well as outer automorphisms of $\mathfrak g$. Let us now concentrate on the periodicities in $Z$.

By definition, the coweight lattice $\Gamma_w\subset \mathfrak h$ of $\mathfrak g$ is dual to the root lattice $\Lambda_r\subset\mathfrak{h}^*$, i.e. for any coweight $\chi\in\Gamma_w$ and any root $\alpha\in\Delta(\mathfrak{g})$ one has $\alpha(\chi)\in\mathbb{Z}$. Therefore, any shift of $Z$ by coweights leaves $V_{\mathfrak g}(Z;\tau)$ invariant: 
\begin{equation}
  \forall \, \chi_H,\chi_V\in\Gamma_w \ , \quad   V_{\mathfrak g}(Z + \chi_H + \chi_V\tau ;\tau)=V_{\mathfrak g}(Z;\tau) \ . 
\end{equation}
Here and below the subscripts $H$ and $V$ refer to the ``horizontal'' (real) and ``vertical'' (i.e., that of $\tau$) directions.

Now, when $\mathfrak g$ is simply-laced and the Killing form is normalized such that roots have squared length two, the isomorphism $\phi : \mathfrak h \rightarrow \mathfrak h^*$ induced by the Killing form identifies coroots with roots and coweights with weights. Let us decompose each $z_i$ as $z_i = \widetilde{\sigma}_i+a_i\tau$, with $\widetilde{\sigma} = (\widetilde{\sigma}_1,\dots,\widetilde{\sigma}_N)\in\mathfrak h$ and $a = (a_1,\dots,a_N) \in\mathfrak{h}$. Let also $\sigma = \phi(\widetilde{\sigma}) = (\sigma_1,\dots,\sigma_N)\in\mathfrak{h^*}$. Then, the invariance of $V_{\mathfrak g}$ under shifts of $\widetilde{\sigma}$ and $a$ by coweights can be reformulated as invariance under shifts of $\sigma$ by weights and of $a$ by coweights. We will see in the next subsection that from a physical viewpoint, it is more natural to consider the variables $\sigma_i$ instead of $\widetilde{\sigma}_i$, so that we can rewrite:
\begin{equation}
    \alpha(Z) = \alpha(\widetilde{\sigma})+\alpha(a)\tau = (\alpha,\sigma)+\alpha(a)\tau\ ,
\end{equation}
where $(\cdot,\cdot)$ denotes the Killing form on $\mathfrak h^*$.

Let us now consider a non simply-laced algebra $\mathfrak g$. In this case, the isomorphism induced by the Killing form does not map weights to coweights; rather, the coweight lattice is a strict sublattice of the (dual) weight lattice. By definition, every weight $\lambda$ of $\mathfrak g$ satisfies:
\begin{equation}
    \forall \alpha\in\Delta(\mathfrak g)\ ,\quad \left(\frac{2 \alpha}{(\alpha,\alpha)},\lambda\right) \in\mathbb{Z}\ .
\end{equation}
When $\alpha=\alpha_l$ is a long root, i.e. $(\alpha_l,\alpha_l)=2$, one has $(\alpha_l,\lambda)\in\mathbb{Z}$. In contrast, for $\alpha=\alpha_s$ a short root, one has $(\alpha_s,\lambda)\in \frac{1}{\nu(\alpha_s)}\mathbb{Z}$. Therefore, it is the twisted potential $V_{\mathfrak{g}}^{\mathrm{tw}}$ which is invariant under any shift of $\sigma$ by weights. Moreover, $V_{\mathfrak{g}}^{\mathrm{tw}}$ is also invariant under any shift of $a$ by coweights, just as the non-twisted potentials discussed above.

Setting $V_{\mathfrak{g}}^{\mathrm{tw}} := V_{\mathfrak g}$ when $\mathfrak g$ is simply-laced, one can uniformly describe the group of symmetries of $V_{\mathfrak{g}}^{\mathrm{tw}}$ for any simple algebra $\mathfrak g$: \textit{it is the semi-direct product of the group of symmetries of the root system with the group of horizontal translations by weights and vertical translations by coweights}.

\subsection{Calogero--Moser systems as Seiberg--Witten integrable systems}\label{Subsec:CMasSW}

As promised, here we review the physical interpretation of the variables $Z$ appearing in the potentials above. Elliptic Calogero–Moser systems appear in Seiberg–Witten theory as they encode the low-energy effective theory on the Coulomb branch of $\mathcal N=2^*$ theories \cite{Donagi:1995cf, Gorsky:1995zq, Martinec:1995by, DHoker:1998xad}. More precisely, the phase space of the (twisted) elliptic Calogero--Moser corresponding to a compact simple Lie algebra $\mathfrak g$ can be naturally identified with the Coulomb branch of the $\mathcal N=2^*$ theory with gauge algebra $\mathfrak g$ compactified on a circle \cite{Seiberg:1996nz,Kapustin:1998xn,DHoker:1998zuv,Kumar:2001iu}. 

Under this correspondence, the ``vertical'' variable $a$ in elliptic Calogero--Moser systems is identified with the holonomy of the gauge field along the circle $S^1$ on which the $\mathcal N=2^*$ theory is compactified:
\begin{equation}
    a = \frac{1}{2\pi}\oint_{S^1} A\ .
\end{equation}
Up to a global gauge transformation, one can assume that $a$ belongs to the Cartan subalgebra $\mathfrak h$ of $\mathfrak g$. 
Now, the periodicity of $a$ depends on the global variant of the gauge group. Let us consider first the simply connected universal covering group $\tilde G$ built out of $\mathfrak g$. The gauge field $A$ being in the adjoint representation of $\mathfrak g$, the variable $a$ undergoes additive shifts by coroots under gauge transformations along the circle $S^1$. On the other hand, the finer additive shifts of $a$ by coweights are global symmetry transformations, related to transformations in the center $\cZ$ of $\tilde G$, and eventually to the 1-form symmetry of the gauge theory \cite{Gaiotto:2014kfa}. Going then to a non-simply connected global form $G=\tilde G/\Pi$ with $\Pi\subset \cZ$ amounts to reducing the periodicity of $a$ to (a subset of) shifts by coweights.\footnote{Moreover, generic values of $a$ break the gauge group $G$ to the semi-direct product of its Cartan subgroup by its Weyl group. This implies that $a$ is also only defined up to Weyl transformations. The same is true for $\sigma$ defined hereafter.}

On the other hand, the ``horizontal'' variable $\sigma$  corresponds to the scalar dual to the $3d$ unbroken abelian gauge field. This implies that $\sigma$ naturally lives in the dual $\mathfrak h^*$ of the Cartan subalgebra, and that its finest periodicity naturally corresponds to shifts by weights of $\mathfrak g$. We will shortly establish how the exact periodicity depends on the global variant of the gauge group. 

We can then summarize the above by stating that, before specifying a global variant, for a (relative) $\mathcal N=2^*$ theory characterized by some gauge algebra $\mathfrak g$, the natural periodicities for $a$ and $\sigma$ correspond to translations by coweights and weights of $\mathfrak g$, respectively. These periodicities align precisely with those of the twisted Calogero--Moser systems discussed in the previous subsection, thereby explaining the emergence of twisted elliptic Calogero--Moser systems in the context of $\mathcal{N}=2^*$  theories rather than genuine elliptic Calogero--Moser systems \cite{DHoker:1998zuv,Kumar:2001iu}.

Gauge theory fixes the ratio of $g_l=g_1$ and $g_s$ to be $g_s = g_1/\nu(s)$ for twisted elliptic Calogero--Moser systems corresponding to $\mathcal N=2^*$ theories, where $\nu(s)$ is the index of \cref{eq:indexCM} for short roots \cite{Kumar:2001iu}.

\subsection{Calogero--Moser systems and \texorpdfstring{$\mathcal N=1^*$}{N=1*} theories}\label{sec: N=1* and CM}

As we have seen, $\mathcal N=1^*$ theories are mass deformations of $\mathcal N=4$ SYM theories where the three $\mathcal N=1$ chiral superfields in the $\mathcal N=4$ vector superfield are taken to be massive. In fact, 
$\mathcal N=1^*$ theories can also be seen as mass deformations of `intermediate' $\mathcal N=2^*$ theories, in which the adjoint chiral superfield in the $\mathcal N=2$ vector superfield is assigned a mass. 
This fits in the more general framework initiated in \cite{Seiberg:1994rs}, where one softly breaks $\mathcal N=2$ to $\mathcal{N}=1$ in a Lagrangian $\mathcal N=2$ theory by turning on a mass term for the adjoint chiral superfield in each $\mathcal N=2$ vector superfield.

Such mass deformation lifts the $\mathcal N=2^*$ Coulomb branch entirely, expect for the singular locus at which BPS particles become massless. In terms of (twisted) Calogero--Moser systems, the singular locus on the Coulomb branch of $\mathcal N=2^*$ theories compactified on a circle $S^1$ is identified with the extrema of the Calogero--Moser potential (\cref{eq:twistedCMpotential}). Massive vacua of the $\mathcal N=1^*$ theory are identified with isolated extrema of the corresponding Calogero--Moser integrable system. 

Isolated extrema of (twisted) Calogero--Moser systems are solutions $Z^*$ of:
\begin{align}
\frac{\partial V (Z ; \tau)}{\partial Z}\bigg\rvert_{Z = Z^* (\tau)}  &= 0\ , \label{eq:extrema} \\
\bigg\rvert \frac{\partial^2 V (Z ; \tau)}{\partial Z_a \partial Z_b} \bigg \rvert_{Z = Z^* (\tau)}  &> 0\ , \label{eq:isolated}
\end{align}
where here (and in the sequel) $V$ refers to the twisted Calogero--Moser potential corresponding to some simple compact Lie algebra $\mathfrak g$, with the convention spelled out at the end of \cref{subsec:symmetriesCMpotential} for simply-laced $\mathfrak{g}$.

It turns out that it is very interesting to consider extrema as functions of $\tau$. 
At any value of $\tau$, we have a finite number $s$ of solutions to \cref{eq:extrema}-\cref{eq:isolated}. We can then pick a numbering $Z_i^* (\tau)$ of the solutions, with an index $i=1 , \dots , s$. In order for this numbering to be consistent as $\tau$ is varied, one might attempt to require that the functions $\tau \mapsto Z_i^* (\tau)$ be continuous for each $i$, but this inevitably leads to multivalued functions on any given fundamental domain. The best we can do then is to require these functions to be continuous on the upper-half plane minus branch cuts, which are a priori arbitrary. When $\tau$ crosses one of these branch cuts, the index $i$ can jump to another index. 
With $Z_i^* (\tau)$ now well-defined, we introduce for $i=1 , \dots , s$ 
\begin{equation}
V_i (\tau) = V \left( Z_i^* ( \tau) ; \tau \right)\ ,
\end{equation}
the value of the potential on the $i$-th isolated extremum. These functions display a surprisingly rich behavior, depending on the algebra $\mathfrak{g}$. 

Let us first discuss the simplest case, in which the potential is untwisted, so that it straightforwardly satisfies 
\begin{equation}
        V (Z ; \tau +1) = V (Z ; \tau ) \qquad \textrm{and} \qquad \frac{1}{\tau^2} V \left( \frac{Z}{\tau} ; \frac{-1}{\tau} \right) = V (Z ; \tau ) \ , \label{eq:transformPot}
\end{equation} 
and furthermore, we assume that it is possible to make a choice of the $Z_i^* (\tau)$-defining branch cuts along the boundary of the fundamental domain of the modular group. These two assumptions are satisfied for $\mathfrak{g} = \mathfrak{su}(n)$. It is then easy to check that 
\begin{equation}
    \mathcal{V}  (\tau) \equiv \begin{pmatrix}
V_1(\tau)  \\
V_2 (\tau) \\
... \\
V_{s}(\tau)
\end{pmatrix} 
\end{equation}
is a rank $s$ vector-valued modular form of weight 2, characterized by the permutations $\sigma_S$, $\sigma_T$ and $\sigma_U$ corresponding to the action of $S$, $T$ and $U=ST$ on the potentials in an orbit $I$: 
\begin{align}
\label{eq:SpotTransform}
V_i & \xmapsto{\, \, \, \,  S \, \, \, \,} V_{\sigma_S (i)}  \Longleftrightarrow V_i \left(  \frac{-1}{\tau} \right) = \tau^2 V_{\sigma_S (i)} (\tau)\ , \\ 
V_i & \xmapsto{\, \, \, \,  U \, \, \, \,} V_{\sigma_U (i)}  \Longleftrightarrow V_i \left(  \frac{-1}{\tau + 1} \right) = (\tau+1)^2 V_{\sigma_U (i)} (\tau)\ , \\
V_i & \xmapsto{\, \, \, \,  T \, \, \, \,} V_{\sigma_T (i)}  \Longleftrightarrow V_i \left(  \tau + 1 \right) = V_{\sigma_T (i)} (\tau)\ . 
\end{align}
Equivalently, the individual components $V_i(\tau)$ are weight 2 modular forms of a certain congruence subgroup of the modular group. This implies in particular that:
\begin{align} \label{eq:SUTtransf}
S \textrm{ at } & \tau = i \, : \qquad \, \, \, \, V_i(i) = - V_{\sigma_S (i)}(i)\ , \\
U \textrm{ at }&  \tau = e^{2\pi i / 3} \, : \, \, \,  \, V_i\left(e^{2\pi i / 3}\right) = e^{2\pi i / 3} \,  V_{\sigma_U (i)}\left(e^{2\pi i / 3}\right)\ , \\
T \textrm{ at } & \tau = i\infty \, : \, \, \, \, \,  \, \, \, \, \, V_i(i\infty) =  V_{\sigma_T (i)}(i\infty)\ . 
\end{align}
In words, potentials exchanged under $S$ have opposite values at $ \tau = i$, while an $S$-singlet must vanish at this point. Potentials that are exchanged under $U$ are arranged as an equilateral triangle centered at the origin of the complex plane at $\tau = e^{2\pi i / 3}$, while a $U$-singlet must vanish at this point. Lastly, multiplets under $T$ can contain arbitrarily many potentials; however they must all take the same value at $\tau = i \infty$. 

More generally, the relations of \cref{eq:transformPot} are not satisfied when the potential is the twisted one, namely for non-simply laced algebras, also because Langlands duality implies that one should send $\tau$ to $-\frac{1}{s\tau}$. However, in the self-dual cases $B_2$, $G_2$ and $F_4$, exceptional isomorphisms can be used to restore \cref{eq:transformPot} by including a shift (see \cite[(2.10)]{Bourget:2015cza}, \cite[(5.22)]{Bourget:2015upj}). We will discuss such an example in section \ref{sec: so(5)}.

More crucial is the fact that the assumption concerning the branch cuts is conjectured to be satisfied only by $\mathfrak{su}(n)$ and $\mathfrak{so}(5) = \mathfrak{sp}(2)$. This means that in all the other cases, it is necessary to include branch cuts in the interior of the fundamental domain, ending on monodromy points \cite{Bourget:2015cza}. 
In view of these complications, we focus in this paper on the cases $\mathfrak{g} = \mathfrak{su}(n)$ and $\mathfrak{g} = \mathfrak{so}(5)$. Other cases are left for future work.

One last remark is that when \cref{eq:transformPot} holds, 
one can build a complex characteristic polynomial of degree $s$ whose coefficients are in the space of modular forms of the modular group \cite{Bourget:2017goy}:
\begin{equation}
P (w) \equiv \prod_{i = 1}^{s} ( w - V_i (\tau)) = \sum_{i=0}^{n_I} c_i(\tau)  w^i\ , 
\end{equation}
The coefficients $c_i (\tau)$ are modular forms of weight $2 n_I-2i$ for $\mathrm{PSL}(2,\Z)$, and can be expressed as combinations of the Eisenstein series. 
In particular, $c_{n_I -1} = 0$, i.e. the sum of all potentials within a given orbit is identically zero.

\subsection{Global variants of \texorpdfstring{$\mathcal N=1^*$}{N=1*} theories and massive vacua on \texorpdfstring{$\mathbb{R}^3\times S^1$}{R3xS1}}

We now start to consider more closely the distinction between global variants of $\mathcal N=1^*$ theories. 

\subsubsection{Pure \texorpdfstring{$\mathcal N=1$}{N=1} SYM theories} 

Let $\mathfrak g$ be a compact simple gauge algebra, $\widetilde{G}$ the unique connected and simply-connected Lie group with Lie algebra $\mathfrak g$, and $\mathcal{Z}$ its center. 

As reviewed in section \ref{sec: review}, a global variant $\rho$ of a $\mathfrak{g}$-gauge theory is defined to be a maximal set of mutually local Wilson--'t~Hooft line operators. The choice of global form constrains the matter content of the theory. For pure SYM or $\mathcal N=1^*$ theories, which are the focus of our discussion, the matter only transforms in the adjoint representation of the gauge group, hence suitable for any global variant. In such a theory $\mathcal T$, line operators fit in equivalence classes labeled by elements of $\widehat{\mathcal{Z}}\times \mathcal{Z}\cong \mathcal{Z}^2$, where $\widehat{\mathcal{Z}}$ is the Pontryagin dual of $\mathcal{Z}$. The global variant $\rho$ determines the gauge group $\widetilde{G}/\Pi$ of the theory, where $\Pi$ is a subgroup of $\mathcal{Z}$. Note that in general there are several global forms of the gauge group corresponding to the same gauge algebra, and several global variants (i.e.~non-equivalent choices of lines) corresponding to the same global form of the gauge group. 

On flat four-dimensional spacetime $\mathbb{R}^4$, distinct global variants of pure $\mathcal N=1$ SYM theories sharing the same Lie algebra $\mathfrak g$ have the same number of vacua, but the dynamics in these vacua depend on the specific global variant. For instance, pure $\mathcal N=1$ SYM theory with gauge group $\mathrm{SU}(N)$ admits $N$ vacua on $\mathbb{R}^4$, which are all trivially gapped and equivalent, as they are permuted by the spontaneously broken elements of the invertible $\mathbb{Z}_{2N}$ R-symmetry. In contrast, the invertible R-symmetry of pure $\mathrm{PSU}(N)$ $\mathcal N=1$ SYM theories is $\mathbb{Z}_2$ and it is unbroken, hence even though these theories also display $N$ gapped vacua on $\mathbb{R}^4$, they do not necessarily display equivalent physics.

More precisely, local dynamics of $\mathfrak{su}(N)$ theories is believed to yield $N$ vacua characterized by the condensation of the line of electric-magnetic charge $(l,1)$ in the $l^\textrm{th}$ vacuum, where $l=0,1, \dots , N-1$. The physics in the vacua crucially depend on the global variant one considers, and this even if all the vacua are gapped. The $\mathrm{SU}(N)$ theory displays trivially gapped vacua only.\footnote{In fact, the vacua are distinguished by an SPT phase for the unbroken $\Z_N^{(1)}$ electric symmetry. We use here the terminology `trivially gapped' for all these vacua, though sometimes it is reserved to vacua where this SPT phase is trivial.} For $N$ prime, all $\mathrm{PSU}(N)_i$ theories have exactly one vacuum (the $i^\textrm{th}$ one) exhibiting a $\Z_N$ gauge theory, whereas the $N-1$ other vacua are trivially gapped.

Upon compactification on a circle $S^1$, a gapped vacuum on $\mathbb{R}^4$ supporting a discrete $\mathbb{Z}_k$ gauge theory yields $k$ gapped vacua distinguished by the value of the VEV of the line charged under $\mathbb{Z}_k$ and wrapping $S^1$. This implies for instance that pure $\mathrm{SU}(N)$ $\mathcal N=1$ SYM admits $N$ gapped vacua on $\mathbb{R}^3\times S^1$, while global variants of pure $\mathrm{PSU}(N)$ $\mathcal N=1$ SYM admit
\begin{equation}\label{eq:NumberGappedVacPureN=1}
   I^{\mathcal{N}=1} \left[ (\mathrm{PSU}(N))_{s}  \right] =   \sum\limits_{\ell = 1}^N \gcd \left( N ,  \ell \right) 
\end{equation}
gapped vacua on $\mathbb{R}^3\times S^1$. Global variants of $\mathrm{SU}(N)/\Z_k$ theories, where $k$ is a divisor of $N$, can be studied in a similar way, giving a number of vacua \cite[(2.3)]{Bourget:2016yhy}:
\begin{equation}
    I^{\mathcal{N}=1} \left[ (\mathrm{SU}(N)/\Z_k)_{s}  \right] =  \frac{N}{k} \sum\limits_{\ell = 1}^k \gcd \left( k , s + \ell \frac{N}{k} \right) \ . 
\end{equation}

\subsubsection{\texorpdfstring{$\mathcal N=1^*$}{N=1*} theories} 

Isolated extrema of (twisted) Calogero--Moser systems are expected to correspond to gapped vacua of $\mathcal N=1^*$ theories on $\mathbb{R}^3\times S^1$. Therefore, the choice of global variant for $\mathcal N=1^*$ theories should reflect in some way into the set of isolated extrema of (twisted) Calogero--Moser systems. The example of $\mathfrak{so}(5)$ $\mathcal N=1^*$ theories was worked out in detail in \cite{Bourget:2015upj}; it was found there that the periodicities of twisted $\mathfrak{so}(5)$ Calogero--Moser systems can be refined so as to accommodate the distinction between the three global variants $\mathrm{Spin}(5)$, $\mathrm{SO}(5)_+$ and $\mathrm{SO}(5)_-$ of $\mathfrak{so}(5)$ $\mathcal N=1^*$ theories. In particular, one finds that the integrable systems corresponding to the global variants $\mathrm{Spin}(5)$ and $\mathrm{SO}(5)_+$ admit 10 isolated extrema, while the one corresponding to the global variant $\mathrm{SO}(5)_-$ admits 7 isolated extrema. Accordingly, the global variants $\mathrm{Spin}(5)$ and $\mathrm{SO}(5)_+$ have 10 massive vacua of $\mathbb{R}^3\times S^1$, while $\mathrm{SO}(5)_-$ has only 7. We will discuss $\mathfrak{so}(5)$ theories further in \cref{sec: so(5)}.

As seen in section \ref{Sec:N=1*andIRphases}, solutions to the F-term equations in $\mathcal N=1^*$ theories are in one-to-one correspondence with nilpotent orbits in the gauge group $G$ of the theory. The VEVs of the scalar fields solving the F-term equations and corresponding to some nilpotent orbit $\mathcal O$ effectively reduce the dynamics to that of pure $\mathcal N=1$ SYM theory with gauge group $\mathrm{Cent}_G(\mathcal O)$, the centralizer of $\mathcal O$ in $G$. When $\mathrm{Cent}_G(\mathcal O)$ does not admit abelian factors, pure $\mathcal N=1$ SYM theory with gauge group $\mathrm{Cent}_G(\mathcal O)$ in turn confines, leading to a finite number of gapped vacua for each such nilpotent orbit in $G$. The dynamics in these vacua strongly depends on the specific global variant at hand, potentially leading to different numbers of gapped vacua on $\mathbb{R}^3\times S^1$, as explained above.

A last subtlety is that some gapless vacua on $\mathbb{R}^4$ can lead to additional gapped vacua on $\mathbb{R}^3\times S^1$. This issue does not arise in type $A$---this follows from the Bala--Carter--Sommers theorem \cite{Sommers:1998ago}. We will come back to it in \cref{sec: so(5)}, as this phenomenon does occur for $\mathfrak{so}(5)$ $\mathcal N=1^*$ theories.

Somewhat surprisingly, the number of isolated extrema of $\mathrm{su}(N)$ elliptic Calogero--Moser systems is precisely the sum $\sigma_1(N)$ of divisors of $N$ (cf.~\cref{eq:sigma1N}), provided one identifies solutions that are mapped to each other through translations by weights and coweights (horizontally and vertically, respectively). This number is also the number of massive $\mathcal N=1^*$ vacua on $\mathbb{R}^4$. However, as already emphasized, one rather expects a one-to-one correspondence between isolated extrema of (twisted) elliptic Calogero--Moser systems and massive vacua of $\mathcal N=1^*$ theories on $\mathbb{R}^3\times S^1$, which are more numerous than those on $\mathbb{R}^4$. We will show that this apparent mismatch is resolved by considering refined periodicities in Calogero--Moser diagrams, implementing global variants. Building on the analysis of $\mathfrak{su}(3)$ theories in \cref{subsec:exsu3}, we will propose a general definition of global variants of (twisted) elliptic Calogero–Moser systems in \cref{subsec:genproposal}. In \cref{Sec:typeA}, we will further explore this framework in type $A$ and test it against examples of increasing complexity, specifically the $\mathfrak{su}(6)$ and $\mathfrak{su}(4)$ $\mathcal N=1^*$ theories.

Before diving into the detailed analysis of
$\mathfrak{su}(3)$ theories, let us first provide some general intuition about the expected structure of global variants of twisted elliptic Calogero–Moser systems. Let $\rho$ be a specific global variant of $\mathcal N=1^*$ theories with compact simple gauge algebra $\mathfrak g$. 
As reviewed in \cref{sec: review}, it is determined by a consistent maximal set of mutually local lines, which in turn selects a Lagrangian subset of the set of all lines \cref{alllines}. Let us focus for the moment on such a set generated by (purely electric) Wilson lines and (purely magnetic) 't Hooft lines. The generalization to any global variant will be the core of our proposal in \cref{subsec:genproposal}. Let $\Lambda_w^G$ and $\Gamma_w^G$ be the character and cocharacter lattices of $G=\tilde{G}/\Pi$  to which these mutually local Wilson and 't Hooft lines belong, respectively. Let also $\Lambda_r$, $\Lambda_w$, $\Gamma_r$ and $\Gamma_w$ be the root, weight, coroot and coweight lattices of $\mathfrak g$, with $\Pi=\Gamma_w^G/\Gamma_r=\Lambda_w/\Lambda_w^G$, see \cref{App:Lie}.

The gauge-theoretic origin of the complex variables $z_i$ appearing in twisted elliptic Calogero--Moser systems, discussed in \cref{Subsec:CMasSW}, motivates the following ``gauged'' periodicities in this kind of global variant, rather than those discussed previously:
\begin{itemize}
    \item translations by $G$-cocharacters of $a$ (``vertical'' direction), i.e.~if two tuples $a$ and $a'$ are related as $a' = a+\chi$ where $\chi\in\Gamma_w^G$, they must be considered equivalent.
    \item translations by $G$-characters of $\sigma$ (``horizontal'' direction), i.e.~if two tuples $\sigma$ and $\sigma'$ are related as $\sigma' = \sigma+\lambda$ where $\lambda\in\Lambda_w^G$, they must be considered equivalent.
\end{itemize}

A key point in our reasoning concerns the symmetry of the potential $V_{\mathfrak g}^\mathrm{tw}$. We have seen that it is symmetric under horizontal translations by elements of $\Lambda_w$, and vertical translations by elements of $\Gamma_w$. In equations, if: 
\begin{equation}
    Z'=Z+\chi\tau+\lambda \in\mathbb{C}^N\ ,
\end{equation} 
with $\chi\in\Gamma_w$ and $\lambda\in\Lambda_w$, then: 
\begin{equation}\label{eq:equalpot}
    V_{\mathfrak g}^\mathrm{tw}(Z',\tau)=V_{\mathfrak g}^\mathrm{tw}(Z,\tau)\ .
\end{equation} 
However, \cref{eq:equalpot} does not imply that $Z'$ and $Z$ are physically equivalent configurations. Moreover, it may happen that $Z'$ and $Z$ project to the same configuration on the torus (modulo Weyl) even if they are not equivalent, which can obscure the multiplicity of the solutions we seek (cf.~\cref{App:othergaugefixtypeA} for a general discussion of this phenomenon in type $A$). In what follows, we will refer to translations by $\Gamma_w^G\tau+\Lambda_w^G$ as \textit{gauged}, while the coset 
\begin{equation}
    (\Gamma_w\tau+ \Lambda_w) / (\Gamma_w^G\tau+\Lambda_w^G)    
\end{equation}
will be interpreted as generating global symmetries.

\subsection{\texorpdfstring{$\mathfrak{su}(3)$}{su(3)} theories}\label{subsec:exsu3}

Let $G$ be a compact connected Lie group with Lie algebra $\mathrm{Lie}(G) = \mathfrak{su}(3)$, i.e. either $G=\widetilde
G = \mathrm{SU}(3)$ or $G=G_\mathrm{ad}=\mathrm{PSU}(3)$. Solutions to the F-term equations correspond to nilpotent orbits in $\mathfrak{su}(3)$, or equivalently, to partitions of $3$. There are three such partitions: $[3]$, $[2,1]$ and $[1^3]$, among which only $[3]$ and $[1^3]$ lead to pure $\mathcal N=1$ SYM theories without abelian gauge factors, hence to massive vacua.

\begin{itemize}
    \item For the partition $[3]$, the gauge group breaks spontaneously to its center $\mathcal{Z}(G)$ (since the center always acts trivially in the adjoint representation), which is $\mathcal{Z}(G)=\mathbb{Z}_3$ for $G=\mathrm{SU}(3)$ and trivial for $G=\mathrm{PSU}(3)$.
    \item The partition $[1^3]$ corresponds to zero vacuum expectation values for the three adjoint scalar in the theory. The gauge group $G$ is unbroken, and the dynamics reduces to pure $\mathcal N=1$ SYM with gauge group $G$.
\end{itemize}

Upon compactification on $S^1$, the partition $[3]$ leads to three gapped vacua when $G=\mathrm{SU}(3)$, since the unbroken gauge group is $\mathcal Z(\mathrm{SU}(3))=\mathbb{Z}_3$, yielding 3 massive vacua on $\mathbb{R}^3\times S^1$. In contrast, for $G=\mathrm{PSU}(3)$, the group is entirely spontaneously broken in the partition $[3]$, resulting in a single massive $3d$ vacuum associated with it.

Considering the partition $[1^3]$ instead, pure $\mathcal{N}=1$ SYM with gauge group $\mathrm{SU}(3)$ gives rise to $3$ massive vacua on $\mathbb{R}^3\times S^1$, whereas any global variant of pure $\mathcal N=1$ SYM with gauge group $\mathrm{PSU}(3)$ admits 5 massive vacua on $\mathbb{R}^3\times S^1$, according to \cref{eq:NumberGappedVacPureN=1}. In conclusion, $\mathcal N=1^*$ theories with either $\mathrm{SU}(3)$ or $\mathrm{PSU}(3)$ gauge group always admit 6 massive vacua in $3d$, though they do not arise in the same way. In hindsight, these theories must admit the same number of vacua in $3d$, because the $\mathrm{SU}(3)$ theory and all global variants corresponding to the group $\mathrm{PSU}(3)$ are in the same duality orbit (cf. \cref{eq:dualityorbitsu(3)}).

Let $ \mathfrak h$ be the usual Cartan subalgebra of $\mathfrak{su}(3)$, i.e.~$\mathfrak h$ is the algebra of traceless real diagonal matrices, and let $\mathfrak h^*$ be its dual. We refer to \cref{App:Lie} for all the definitions and our conventions concerning algebras. The coroots corresponding to the standard choice of simple roots $\alpha_1$ and $\alpha_2$ are:
\begin{equation}\label{eq:corootssu3}
    H_{\alpha_1}=\left(\begin{array}{ccc}
    1 & 0 & 0 \\ 0 & -1 & 0 \\ 0 & 0 & 0 \end{array}\right)\ , \quad H_{\alpha_2}=\left(\begin{array}{ccc}
    0 & 0 & 0 \\ 0 & 1 & 0 \\ 0 & 0 & -1 \end{array}\right)\ .
\end{equation}
As for the fundamental coweights, we take:
\begin{equation}\label{eq:coweigthssu3}
    \chi_1=\left(\begin{array}{ccc}
    \frac{2}{3} & 0 & 0 \\ 0 & -\frac{1}{3} & 0 \\ 0 & 0 & -\frac{1}{3} \end{array}\right)\ , \quad \chi_2=\left(\begin{array}{ccc}
    \frac{1}{3} & 0 & 0 \\ 0 & \frac{1}{3} & 0 \\ 0 & 0 & -\frac{2}{3} \end{array}\right)\ .
\end{equation}

\begin{figure}
    \centering
    \includegraphics[width=\textwidth]{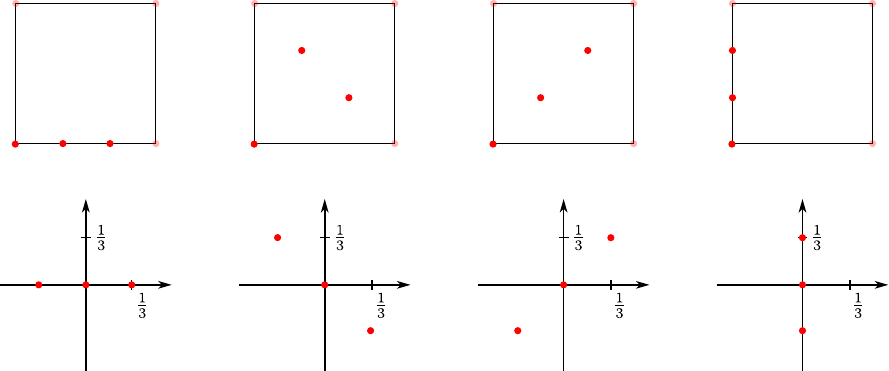}
    \caption{Extrema of the $A_2$ Calogero--Moser system in a fundamental cell for $E_{\tau=i}$ (top) and in the complex plane (bottom).}
    \label{fig:extremaCMA2}
\end{figure}

Let $Z=(z_1,z_2,z_3)\in\mathbb{C}^3$ be a configuration of three points in the complex plane. The Weyl group permutes these points arbitrarily. The Calogero--Moser potential $V_{\mathfrak{su}(3)}(Z;\tau)$ only depends on the projection 
\begin{equation}
    \widetilde{Z} = (\widetilde{z}_1,\widetilde{z}_2,\widetilde{z}_3)    
\end{equation} 
of $Z$ on the elliptic curve $E_\tau$, i.e. $\widetilde{z}_i\sim \widetilde{z}_i+1\sim \widetilde{z}_i+\tau$. The extrema of the CM potential for such configurations of $\widetilde{z_i}$'s are known, see for instance \cite{Dorey:1999sj, Dorey:2001qj}. In \cref{fig:extremaCMA2} they are represented as points in a fundamental cell for the elliptic curve. The different equivalent choices of such a fundamental cell boil down to permuting the four extrema. 
Our reasoning is independent of the precise value of $\tau$; therefore, in the figures, we set $\tau = i$.

Below each such extrema, we also display a configuration $Z \subset \mathbb{C}^3$ corresponding to it and satisfying the condition $\sum z_i = 0$, which we will use as a basis for the manipulations. For future reference, note that each configuration in $\mathbb{C}$ shown in \cref{fig:extremaCMA2} contains a point located at the origin of the plane $\mathbb{C}$; hence, it can also be considered as satisfying $z_3 = 0$ (modulo Weyl).\footnote{This is a property of $\mathfrak{su}(3)$ that does not apply in general, e.g.~it is already not true for  $\mathfrak{su}(2)$ and $\mathfrak{su}(4)$.}
    
Note that in fact, the extrema displayed in \cref{fig:extremaCMA2} are the ones that remain after modding by translations by the coweight lattice, both vertically and horizontally. 

\subsubsection{\texorpdfstring{$\mathrm{SU}(3)$}{SU(3)}}

When $G=\widetilde G=\mathrm{SU}(3)$, one has $\Lambda_w^G=\Lambda_w$ and $\Gamma_w^G=\Gamma_r$, meaning that the horizontal periodicities are given by the weight lattice, mapped by the Killing form to the coweight lattice 
\begin{equation}
    \Gamma_w = \mathbb{Z}\langle \chi_1,\chi_2\rangle \ , 
\end{equation}
whereas the vertical periodicities are given by the coroot lattice 
\begin{equation}
    \Gamma_r = \mathbb{Z}\langle H_{\alpha_1},H_{\alpha_2}\rangle \ .
\end{equation}

\begin{figure}
    \centering
    \includegraphics[width=\textwidth]{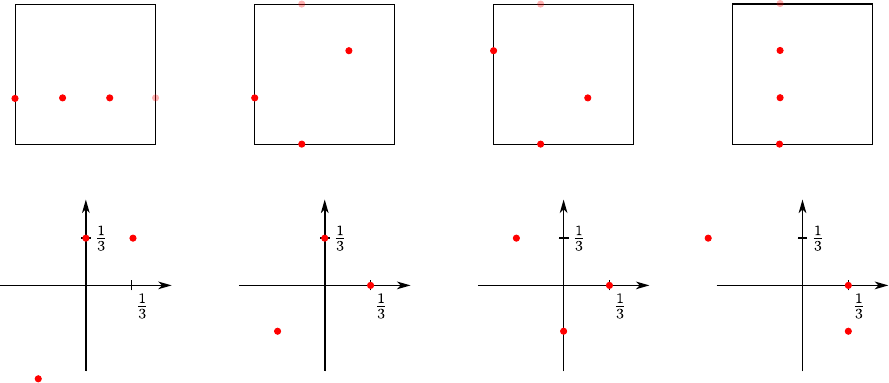}
    \caption{Shifted extrema of the $A_2$ Calogero--Moser system.}
    \label{fig:extremaCMA2bis}
\end{figure}

In \cref{fig:extremaCMA2bis} we show four (of the eight) configurations obtained from those of \cref{fig:extremaCMA2} by vertical or horizontal translations either by $\pm\chi_1$ or by $\pm\chi_2$. 
However, they can be either equivalent or distinct from the diagrams of \cref{fig:extremaCMA2}, once we impose the physical periodicities specific to the present global variant, namely vertical translations by $\Gamma_r$ and horizontal translations by $\Gamma_w$.
Vertical shifts by elements of $\Gamma_w$ are in general not gauged. However, it can be the case that they can be compensated by horizontal translations by elements of $\Gamma_w$, which are gauged. Therefore, not all vertical shifts by coweights are necessarily generating inequivalent Calogero--Moser extrema.

Consider for example the rightmost configuration (in the complex plane $\mathbb{C}$) in \cref{fig:extremaCMA2bis}. It is obtained from the one on the right of \cref{fig:extremaCMA2} as 
\begin{equation}
    (z_1,z_2,z_3)\rightarrow (z_1-2/3,z_2+1/3,z_3+1/3) = (z_1,z_2,z_3)-\chi_1\ ,
\end{equation}
where the $z_i$'s are ordered from top to bottom (this can always be assumed up to Weyl transformations). The shift by $-\chi_1$ in the horizontal direction is gauged for the $\mathrm{SU}(3)$ global variant, hence the rightmost extrema in \cref{fig:extremaCMA2bis} is to be identified with the rightmost one in \cref{fig:extremaCMA2}. Similarly, the two extrema in the middle of \cref{fig:extremaCMA2bis} must be identified with the two middle ones in \cref{fig:extremaCMA2bis}, as they can be obtained by horizontal translations by $\chi_2$ and $-\chi_1$, respectively. Actually, they can also be obtained as vertical translations by $-\chi_1$ and $-\chi_2$, respectively; thus, such vertical shifts can in this case be compensated by horizontal ones.

In contrast, the leftmost configuration in \cref{fig:extremaCMA2bis} is obtained from the leftmost one in \cref{fig:extremaCMA2} by a vertical shift by $-\chi_1$, with the $z_i$'s ordered from left to right. This shift is not gauged when $G=\mathrm{SU}(3)$. Moreover, this translation cannot be undone by a horizontal shift. Therefore, the leftmost extremum in \cref{fig:extremaCMA2bis} is physically distinct from the leftmost one in \cref{fig:extremaCMA2}. Similarly, the extremum obtained from the leftmost one in \cref{fig:extremaCMA2} after a vertical shift by $-2\chi_1$ is physically distinct from the leftmost extrema in \cref{fig:extremaCMA2} and \cref{fig:extremaCMA2bis}. Any further vertical shift by coweights is gauge-equivalent to one of these three vacua, as $\Gamma_w/\Gamma_r\cong \mathbb{Z}_3$.

\begin{figure}
    \centering
    \includegraphics[width=0.75\textwidth]{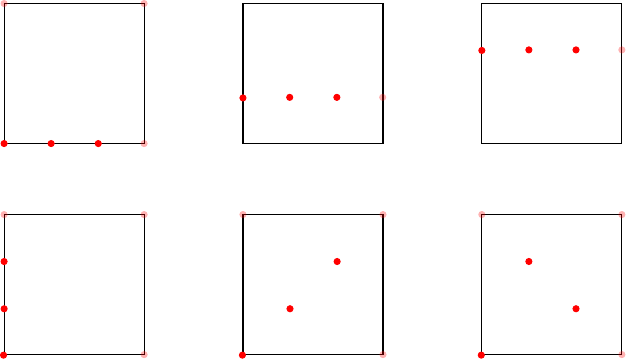}
    \caption{Extrema of the $\mathrm{SU}(3)$ Calogero--Moser system}
    \label{fig:extremaCMA2totalSU3}
\end{figure}

In conclusion, we have found six distinct isolated extrema for the global variant $\mathrm{SU}(3)$, shown in \cref{fig:extremaCMA2totalSU3}. The three extrema in the first row of \cref{fig:extremaCMA2totalSU3} are related by a symmetry (a vertical shift by a coweight modulo coroots); in particular they share the same value of the potential $V_{\mathfrak{su}(3)}$. They correspond to the three vacua of the $\mathcal N=1^*$ $\mathrm{SU}(3)$ theory in the generic nilpotent orbit, i.e.~the partition $[3]$, where the gauge group breaks spontaneously to $\mathbb{Z}_3$, leading to three vacua on $\mathbb{R}^3\times S^1$. In contrast, the three extrema on the second row of \cref{fig:extremaCMA2totalSU3} correspond to the three gapped vacua of $\mathcal N=1^*$ $\mathrm{SU}(3)$ theory in the nilpotent orbit corresponding to the partition $[1^3]$, i.e.~to the three confining vacua of pure $\mathrm{SU}(3)$ $\mathcal N=1$ SYM.

The computation of the multiplicity of each extremum $E$ in \cref{fig:extremaCMA2} can be formalized as follows. One generates all the alternative configurations by shifting $E$ both horizontally and vertically by the finest elements at our disposal, namely the coweights. Then, the choice of gauge group, or rather, of global variant, determines whether a given configuration $E'$ is equivalent to the starting one $E$. For the global variant $\mathrm{SU}(3)$, we have to solve the equation 
\begin{equation}\label{eq:inequivalentCMSU3}
    E' + h_1 \, \chi_1 + h_2\, \chi_2 + v_1\, \tau H_{\alpha_1} + v_2 \,\tau H_{\alpha_2}  \approx E \subset \mathbb{C}\ , 
\end{equation}
where $h_a, v_a\in\mathbb{Z}$, for $a = 1,2$, and $\approx$ stands for equivalence up to Weyl transformations, i.e.~in this case permutations of the $z_i$. If it admits solutions, then $E'$ and $E$ must be identified, otherwise they are inequivalent isolated extrema of the $\mathrm{SU}(3)$ Calogero--Moser system. Such computations can easily be implemented on formal computational software.

Let us make a brief comment on the other gauge fixing for the translation symmetry, namely imposing $z_3=0$ instead of $\sum z_i=0$. In this case, the translations by the coweights need to be composed with an overall shift in order to reinstate the gauge condition $z_3=0$. This has the effect of blurring the distinction between equivalent and inequivalent vacua. More details about this are given in \cref{App:othergaugefixtypeA}. Therefore, in what follows, we will consider configurations satisfying $\sum z_i = 0$, unless explicitly stated otherwise.

\subsubsection{\texorpdfstring{$\mathrm{PSU}(3)_0$}{PSU(3)0}} 

\begin{figure}
    \centering  \includegraphics[width=\textwidth]{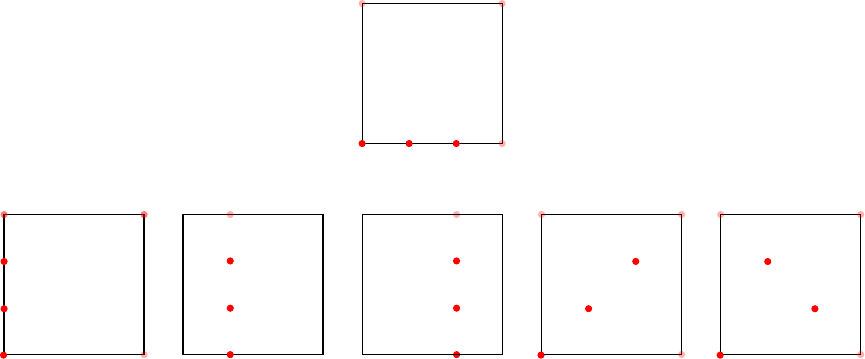}
    \caption{Inequivalent Calogero--Moser diagrams for the global variant $\mathrm{PSU}(3)_0$.}
    \label{fig:extremaCMA2totalPSU30}
\end{figure}

Let us now turn to the global variant $\mathrm{PSU}(3)_0$ of the $\mathfrak{su}(3)$ $\mathcal N=1^*$ theory. One has $\Lambda_w^G=\Lambda_r$, and $\Gamma_w^G=\Gamma_w$, hence the horizontal periodicities for Calogero--Moser diagrams are translations by coroots, and the vertical ones, translations by coweights. Therefore, the study of isolated extrema of Calogero--Moser diagrams closely follows that for the global variant $\mathrm{SU}(3)$, with the horizontal and vertical directions exchanged. 
In the convention $\sum z_i=0$, one finds the six inequivalent extrema shown in \cref{fig:extremaCMA2totalPSU30}.

The equation corresponding to \cref{eq:inequivalentCMSU3} for the global variant $\mathrm{PSU}(4)_0$ reads:
\begin{equation}
    E' + h_1 \, H_{\alpha_1} + h_2\,H_{\alpha_2}  + v_1\, \tau \chi_1  + v_2 \,\tau  \chi_2  \approx E\ , 
\end{equation}
with $h_{1,2}, v_{1,2}\in\mathbb{Z}$.

These isolated extrema of the $\mathrm{PSU}(3)_0$ global variant of the $A_2$ Calogero--Moser systems correspond to the gapped vacua of the $\mathrm{PSU}(3)_0$ $\mathcal N=1^*$ vacua on $\mathbb{R}^3\times S^1$, in the following way. The extremum on the first row of \cref{fig:extremaCMA2totalPSU30} corresponds to the generic nilpotent orbit $[3]$, where the gauge group fully breaks spontaneously, leading to a single vacuum on $\mathbb{R}^3\times S^1$. 

The extrema on the second row of \cref{fig:extremaCMA2totalPSU30} rather correspond to the nilpotent orbit $[1^3]$, where the dynamics reduces to that of $\mathrm{PSU}(3)_0$ pure $\mathcal N=1$ SYM. On $\mathbb{R}^4$, this theory admits three confining vacua, two of which are trivially gapped, and one supports a discrete magnetic $\mathbb{Z}_3$ gauge symmetry. On $\mathbb{R}^3\times S^1$, these vacua lead respectively to two and three vacua, for a total of five vacua. They correspond to the five Calogero--Moser diagrams on the second row of \cref{fig:extremaCMA2totalPSU30}. In particular, the three leftmost ones are related by a global symmetry, hence they correspond to the three gapped vacua on $\mathbb{R}^3\times S^1$ obtained from the single vacuum on $\mathbb{R}^4$ supporting a discrete $\mathbb{Z}_3$ gauge symmetry.

\subsubsection{\texorpdfstring{$\mathrm{PSU}(3)_{1,2}$}{PSU(3)1,2}}

\begin{figure}
    \centering  \includegraphics[width=\textwidth]{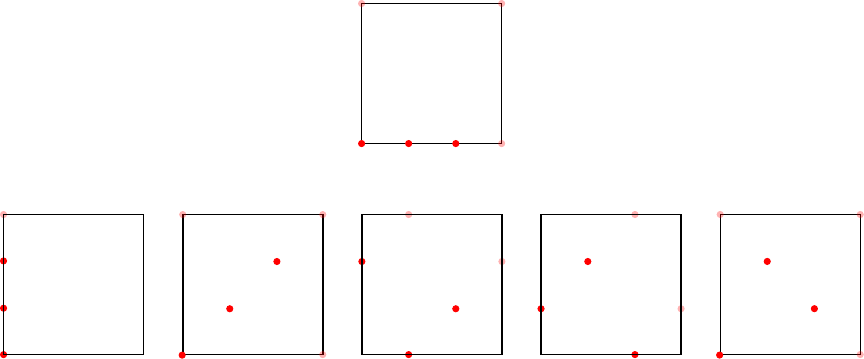}
    \caption{Inequivalent Calogero--Moser diagrams for the global variant $\mathrm{PSU}(3)_1$.}
    \label{fig:extremaCMA2totalPSU31}
\end{figure}

\begin{figure}
    \centering  \includegraphics[width=\textwidth]{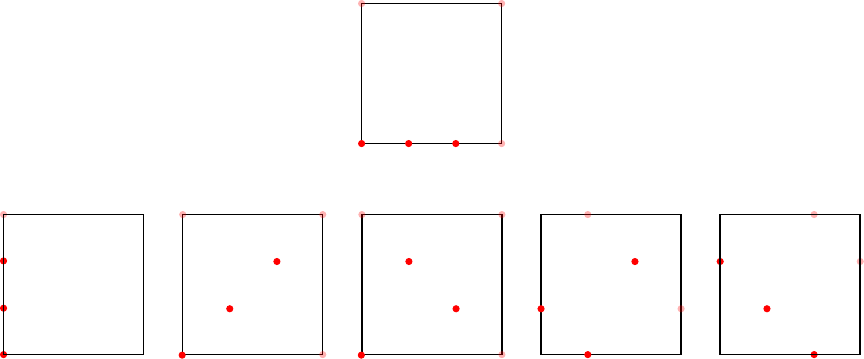}
    \caption{Inequivalent Calogero--Moser diagrams for the global variant $\mathrm{PSU}(3)_2$.}
    \label{fig:extremaCMA2totalPSU32}
\end{figure}

In order to analyze the vacua of $\mathcal N=1^*$ $\mathrm{PSU}(3)_1$ and $\mathrm{PSU}(3)_2$ theories on $\mathbb{R}^3\times S^1$ and the corresponding extrema of the Calogero--Moser systems, one needs to consider the following periodicities. The gauge group being $\mathrm{PSU}(3)$ for both global variants $\mathrm{PSU}(3)_1$ and $\mathrm{PSU}(3)_2$, we take the horizontal periodicity to be translations by $\Lambda_w^{\mathrm{PSU}(3)}\cong\Lambda_r\cong\Gamma_r$, i.e. coroots. The distinction between the global variants $\mathrm{PSU}(3)_{0,1,2}$ is then determined by the direction along which one can translate by coweights: in the case of $\mathrm{PSU}(3)_0$, this is the vertical ($\tau$) direction. In the case of $\mathrm{PSU}(3)_1$, one rather gauges translations by coweights in the diagonal $(1,1)$ direction, i.e. along $1+\tau$.
Lastly, for the global variant $\mathrm{PSU}(3)_2$, one gauges translations by coweights in the $(2,1)$ direction, i.e. along $2+\tau$. In other words, the equation corresponding to \cref{eq:inequivalentCMSU3} for the global variant $\mathrm{PSU}(3)_k$ ($k=0,1,2$) is 
\begin{equation}
    E' + h_1 \, H_{\alpha_1} + h_2\,H_{\alpha_2}  + d_1\, (k+\tau) \chi_1  + d_2 \,(k+\tau)  \chi_2  \approx E\ .
\end{equation}
The isolated Calogero--Moser extrema are shown in \cref{fig:extremaCMA2totalPSU31} for the global variant $\mathrm{PSU}(3)_1$, and in \cref{fig:extremaCMA2totalPSU32} for the global variant $\mathrm{PSU}(3)_2$.

The identification between these isolated extrema and the gapped $\mathcal N=1^*$ vacua on $\mathbb{R}^3\times S^1$ can be done as before.

We will now explain how this generalizes to arbitrary gauge groups and global variants.

\subsection{General proposal}\label{subsec:genproposal}

We build upon the analysis of $\mathfrak{su}(3)$ theories to conjecture a general framework for the periodicities in twisted elliptic Calogero--Moser systems, thereby defining the notion of global variants of twisted elliptic Calogero--Moser systems.  In the next sections, we will provide evidence to support our proposal, in type $A$ (\cref{Sec:typeA}) and $\mathfrak{so}(5)$ theories (\cref{sec: so(5)}).

In essence, the proposal can be very intuitively understood from the previous example. The gauge symmetry that we have to use as an equivalence relation between extrema for a given global variant is uniquely determined by the Lagrangian subgroup that defines the latter.

Let us formalize this intuition, by introducing first, for a given compact simple Lie algebra $\mathfrak g$,
the following abelian group morphisms:
\begin{align}
	\phi_\Lambda\colon \Lambda_w &\longrightarrow \widehat{\mathcal Z}\ ,\\
	\phi_\Gamma\colon \Gamma_w &\longrightarrow \mathcal Z\ ,
\end{align}
such that the kernel of these morphisms is respectively $\Lambda_r$ and $\Gamma_r$, hence $\phi_\Lambda$ and $\phi_\Gamma$ induce the isomorphisms $\Lambda_w/\Lambda_r\cong\widehat{\mathcal Z}$ and $\Gamma_w/\Gamma_r\cong \mathcal Z$.

Recall from \cref{sec: review} that a global variant $\rho$ of an $\mathcal N=1^*$ theory with gauge algebra  $\mathfrak g$ is the data of the equivalence classes of Wilson--'t~Hooft line operators present in the theory, which form a Lagrangian subgroup $\mathcal L$ of $\widehat{\mathcal Z}\times \mathcal Z$. We define the global variant $\rho$ of the (twisted) $\mathfrak g$ elliptic Calogero--Moser system as follows:

\begin{definition}\label{def:globvarCM}
	The global variant $\rho$ of the twisted elliptic Calogero--Moser system corresponding to $\mathfrak g$ is the Calogero--Moser system on $E_\tau$ for which the gauged translations are those by elements of the form 
	\begin{equation}
		\lambda+\chi\tau\ ,
	\end{equation}
	where $\lambda\in\Lambda_w$, $\chi\in\Gamma_w$, and $\left(\phi_\Lambda(\lambda),\phi_\Gamma(\chi)\right)\in\mathcal L$. This condition defines a sublattice of $\Lambda_w\times\Gamma_w$. Configurations of points in the complex plane related by gauged translations and symmetries of the root system $\Delta(\mathfrak g)$, which includes both the Weyl group $W(\mathfrak g)$ and outer automorphisms of $\mathfrak g$, are considered equivalent. Configurations which cannot be related by such operations are considered distinct.
\end{definition}

This notion of global variant of twisted elliptic Calogero--Moser systems is such that the (isolated) extrema of some global variant $\rho$ correspond bijectively to the (gapped) vacua of the same global variant $\rho$ of $\mathcal N=1^*$ theories on $\mathbb{R}^3\times S^1$.

\subsection{Generalizations to other integrable systems}\label{subsec:generalization}

Presumably, \cref{def:globvarCM} should extend to many other families of integrable systems associated to Lie algebras \cite{Olshanetsky:1981dk,Olshanetsky:1983wh}, particularly those that encode the low-energy dynamics on the Coulomb branch of $\mathcal N=2$ theories, such as periodic Toda systems---which correspond to pure $\N=2$ SYM theories \cite{Martinec:1995by}---and spin generalizations of elliptic Calogero–Moser systems, which are associated with necklace $\N=2$ quivers \cite{Dorey:2001qj,Argurio:2024kdr}. 

A large class of integrable systems related to $4d$ $\mathcal N=2$ theories is the class of Hitchin systems, which corresponds to $4d$ $\mathcal N=2$ theories in class $\mathcal S$ \cite{Gaiotto:2009we}. In this case, part of the discussion, particularly the study of dualities of Hitchin integrable systems, should match the analysis of \cite{Kapustin:2006pk}. 

Global variants of such integrable systems will differ from their standard definition when the corresponding $\N=2$ theory possesses a non-trivial one-form symmetry. In essence, our proposed notion extends the family of integrable systems associated with Lie algebras to integrable systems associated with Lie groups---or, more specifically, their global variants. 

Evidence supporting this perspective for any $4d$ $\mathcal{N}=2$ theory comes from the analysis in \cite{Gaiotto:2010be}, particularly Section 6.3, from which we adopt the notation for the remainder of this section. Let $\Gamma$ denote the lattice of possible low-energy charges of ``vanilla'' BPS states in a given $4d$ $\mathcal{N}=2$ theory on its Coulomb branch, and let $\rho_1, \dots, \rho_n$ represent the distinct global variants of that theory. Each global variant $\rho_i$ defines a charge lattice $\Gamma_{\rho_i}$, which contains $\Gamma$ as a sublattice and corresponds to the possible low-energy charges of BPS particles in the presence of line operators in $\rho_i$. Defining $\Gamma_{\mathrm{rel}}$ as the union of all $\Gamma_{\rho_i}$, one has the following inclusions for each $i$:
\begin{equation} 
\Gamma \subset \Gamma_{\rho_i} \subset \Gamma_{\mathrm{rel}}\ . 
\end{equation}

Upon compactification of the global variant $\rho_i$ of the $4d$ $\mathcal{N}=2$ theory on a circle $S^1$, the resulting low-energy $3d$ $\mathcal{N}=4$ theory is a sigma model into a hyperkähler manifold $\mathcal{M}_{\rho_i}$, which depends on $\rho_i$. This manifold is a torus fibration over the Coulomb branch $\mathcal{B}$, with fiber
\begin{equation} 
(\mathcal{M}_{\rho_i})_u = \mathrm{Hom}((\Gamma_{\rho_i})_u, \mathbb{R}/\mathbb{Z})
\end{equation}
at a generic point $u \in \mathcal{B}$. The authors of \cite{Gaiotto:2010be} introduce analogous hyperkähler spaces $\mathcal{M}$ and $\mathcal{M}_{\mathrm{rel}}$, corresponding to $\Gamma$ and $\Gamma_{\mathrm{rel}}$, respectively, though these do not correspond to genuine $4d$ $\mathcal{N}=2$ theories ($\mathcal{M}_\mathrm{rel}$ corresponds to a \textit{relative} $4d$ $\mathcal N=2$ theory). For each $i$, one then obtains the following finite coverings:
\begin{equation} 
\mathcal{M}_{\mathrm{rel}} \rightarrow \mathcal{M}_{\rho_i} \rightarrow \mathcal{M}\ . 
\end{equation}

The global variant $\rho_i$ of the Calogero–Moser integrable systems we define has phase space $\mathcal{M}_{\rho_i}$, with periodicities determined by the global variant $\rho_i$, as described in \cref{def:globvarCM}. In contrast, the hyperkähler manifold $\mathcal{M}_{\mathrm{rel}}$ serves as the phase space for the standard notion of twisted Calogero–Moser systems, where horizontal periodicities are gauged weight translations and vertical periodicities are gauged coweight translations. Finally, $\mathcal{M}$ corresponds to a related notion of twisted Calogero–Moser systems, with horizontal gauge periodicities given by root translations and vertical ones by coroot translations.

\section{Type \texorpdfstring{$A$}{A}: general theory and examples}\label{Sec:typeA}

In this section, we test our proposal in two examples of increasing intricacy, namely based on gauge algebras $\mathfrak{su}(6)$ and $\mathfrak{su}(4)$. The first case is such that the order of the center $\cZ$ has a pair of coprime divisors, $6=2\times 3$, while in the second case it has a square divisor, $4=2^2$. Before attacking the two examples, we discuss some general features of all models with $A_{N-1}$ gauge algebras, including subtleties concerning the (classical) unbroken gauge groups in the gapped vacua, the matching of the R-symmetry between UV and IR, and how the vacua are mapped under duality transformations.

Since simple Lie algebras of type $A$ are simply-laced, the isomorphism given by the Killing form, normalized so that roots have squared length two, identifies the root lattice $\Lambda_r \subset \mathfrak{h}^*$ with the coroot lattice $\Gamma_r \subset \mathfrak{h}$, and the weight lattice $\Lambda_w \subset \mathfrak{h}^*$ with the coweight lattice $\Gamma_w \subset \mathfrak{h}$ (the notations are set in \cref{App:Lie}). Therefore, throughout this section, we will consider both horizontal and vertical periodicities to lie in $\mathfrak{h}$. In other words, in the notation of \cref{sec:GlobvarCM}, we use the horizontal variables $\widetilde{\sigma}$ instead of $\sigma$.

\subsection{Orbits of \texorpdfstring{$\mathfrak{sl}_2$}{sl2}-triples and unbroken gauge groups}\label{subsec:unbrokengaugegroups}

Recall from \cref{sec: review} that the supersymmetric vacua of $\mathcal N=1^*$ theories correspond to gauge orbits of $\mathfrak{sl}(2,\bC)$-triples in the complexified gauge algebra $\mathfrak{g}_\mathbb{C}$ (cf. \cref{eq: F term}). In such an orbit $\mathcal O$, the gauge group $G$ of the theory breaks spontaneously to $\mathrm{Cent}_G(\mathcal O)$, the centralizer of $\mathcal O$ in $G$. Since we are specifically interested in type $A$ theories in this section, we can simply think of the elements in the gauge algebra $\mathfrak{sl}(N,\bC)$ to furnish a $N$-dimensional representation of $\mathfrak{sl}(2,\bC)$. Elements of the theory of $\mathfrak{sl}(2)$-triples and nilpotent orbits in complex simple Lie algebras are given in \cref{App:Lie}; we refer to \cite{collingwood1993nilpotent} for a more thorough discussion.

Now, a generic $N$-dimensional representation of $\mathfrak{sl}(2)$ is always reducible to a direct sum of irreducible representations such that the sum of their dimensions is exactly $N$. As a consequence, $N$-dimensional representations of $\mathfrak{sl}(2)$, and hence orbits of $\mathfrak{sl}(2)$-triples in $\mathfrak{sl}(N)$, are in one-to-one correspondence with partitions of $N$. 

A partition of $N$ is a non-increasing sequence of positive integers 
    \begin{equation}
        \underline{\lambda} =(\lambda_1,\dots,\lambda_1,\lambda_2,\dots,\lambda_2,\lambda_3,\dots,\lambda_k)\ ,
    \end{equation} 
where each $\lambda_i$ appears $\mu_i\in\mathbb{N}_{>0}$ times, $\lambda_{i}>\lambda_{i+1}$, and $k\in\mathbb{N}_{>0}$, such that:
    \begin{equation}
        N = \mu_1\lambda_1+\dots+\mu_k\lambda_k\ .   
    \end{equation}
We denote such partitions $\underline{\lambda}=\{\lambda_1^{\mu_1}\dots\lambda_k^{\mu_k}\}$.

In such a representation, the $N\times N$ matrices defining the triple are block diagonal, with each $\lambda_i\times\lambda_i$ irreducible block appearing $\mu_i$ times:
\begin{equation}
    X_{\underline{\lambda}} = \mathrm{Diag}(X_{\lambda_1},\dots,X_{\lambda_1},X_{\lambda_2},\dots,X_{\lambda_n})\ , 
\end{equation}
with $X$ any matrix in the triple.

The centralizer $\mathrm{Cent}_{\mathfrak{su}(N)}(\cO_{\underline{\lambda}})$ consists of all matrices $M$ in $\mathfrak{su}(N)$ that commute with all the elements in $\cO_{\underline{\lambda}}$, namely matrices in the irreducible block-diagonal form given above. It is the unbroken gauge algebra $\mathfrak{g}_\mathrm{ub}$ in the classical vacuum of the $\N=1^*$ theory corresponding to $\underline{\lambda}$.
Schur's Lemma implies that such an $M$ must take the form:
\begin{equation}\label{eq:embeddingGubinGUV}
        M = \mathrm{Diag}\left(\widetilde{M}^{(1)}\otimes\mathrm{Id}_{\lambda_1},\dots,\widetilde{M}^{(k)}\otimes\mathrm{Id}_{\lambda_k}\right)\ ,
\end{equation}
where for each $i=1,\dots,k$, $\widetilde{M}^{(i)}$ is a $\mu_i\times\mu_i$ matrix.

For instance, for $N=12$ and $\underline{\lambda}=\{3^22^3\}$, the unbroken gauge algebra $\mathfrak{g}_\mathrm{ub}$ consists of the matrices $M$ of the form:
\begin{equation}
    M = \begin{pmatrix}
        \widetilde{M}^{(1)}\otimes\mathrm{Id}_3 & 0_{3\times 2} \\
        0_{2\times 3} & \widetilde{M}^{(2)}\otimes\mathrm{Id}_2
    \end{pmatrix}\in\mathfrak{su}(12)\ ,
\end{equation}
where $\widetilde{M}^{(1)}$ and $\widetilde{M}^{(2)}$ are respectively $2\times 2$ and $3\times 3$. In order for $M$ to be in $\mathfrak{su}(12)$ rather than $\mathfrak{u}(12)$, it must satisfy
\begin{equation}
    \tr(M) = 3\tr(\widetilde{M}^{(1)})+2\tr(\widetilde{M}^{(2)}) = 0\ .
\end{equation}
This implies that $\mathfrak{g}_\mathrm{ub}$ contains a copy of $\mathfrak{u}(1)$. More generally, the centralizer of the orbit corresponding to $\underline{\lambda}=\{\lambda_1^{\mu_1}\dots\lambda_k^{\mu_k}\}$ contains a copy of $\mathfrak{u}(1)^{k-1}$. Therefore, $\mathfrak{g}_\mathrm{ub}$ always contains an abelian factor, unless $k=1$. Since we are primarily interested in gapped vacua of $\mathcal N=1^*$ theories, we henceforth assume that $k=1$.

Then, when $N=\mu\lambda$ and $\underline{\lambda}=\{\lambda^\mu\}$, the centralizer of any such orbit in $\mathfrak{su}(N)$ is of the form:
\begin{equation}
    \mathrm{Cent}_{\mathfrak{su}(N)}( \cO_{\underline{\lambda}}) = \left\{\widetilde{M}\otimes \mathrm{Id}_\lambda\right\}\subset \mathfrak{su}(N)\ ,
\end{equation}
where $\widetilde{M}\in\mathfrak{su}(\mu)$.

Let us now consider the centralizer in the group, instead of the algebra. For any $G$ a compact connected simple Lie group with Lie algebra $\mathfrak{su}(N)$, the centralizer $\mathrm{Cent}_G( \cO_{\underline{\lambda}})$ of the orbit corresponding to $\underline{\lambda}$, which is the unbroken gauge group $G_{\mathrm{ub}}$ in an $\N=1^*$ theory with gauge group $G$ in the classical vacuum corresponding to $\underline{\lambda}$, is a subgroup of $G$ with Lie algebra $\mathfrak{su}(\mu)$.

Consider first the case where the UV gauge group is simply-connected, $G=\mathrm{SU}(N)$:
\begin{equation}\label{eq:unbrokengaugegroup}
    G_{\mathrm{ub}} = \left\{\widetilde{m}\otimes \mathrm{Id}_\lambda\right\} < \mathrm{SU}(N)\ .
\end{equation}
Note that $\widetilde{m}$ is straightforwardly in $\mathrm{U}(\mu)$, but $\widetilde{m}\otimes \mathrm{Id}_\lambda\in\mathrm{SU}(N)$ requires $\det(\widetilde{m}\otimes \mathrm{Id}_\lambda) = \det(\widetilde{m})^\lambda=1$. Reciprocally, every unitary matrix $\widetilde{m}$ such that $\det(\widetilde{m})^\lambda=1$ is in $G_\mathrm{ub}$. Therefore: 
\begin{equation}
    G_\mathrm{ub} = \left\{g\in\mathrm{U}(\mu) \;\mid \; \det(g)^\lambda=1 \right\} < \mathrm{U}(\mu)\ .
\end{equation}

Equivalently, the subgroup $G_\mathrm{ub}$ of $\mathrm{U}(\mu)$ is the preimage of $\mathbb{Z}_\lambda < \mathrm{U}(1)$ under the determinant morphism:
\begin{equation}
    \det\ \colon\ \mathrm{U}(\mu) \longrightarrow \mathrm{U}(1)\ ,
\end{equation}
hence $G_\mathrm{ub}$ fits into the following short exact sequence:
\begin{equation}\label{eq:SESGub}
    1 \longrightarrow \mathrm{SU}(\mu) \longrightarrow G_\mathrm{ub} \longrightarrow \mathbb{Z}_\lambda \longrightarrow 1\ .
\end{equation}

This short exact sequence splits, much like the one for unitary groups: one can take as section the group morphism $\mathbb{Z}_\lambda \longrightarrow G_\mathrm{ub}$ sending $\exp(2ik\pi/\lambda)$, for $k=0,\dots,\lambda-1$, to the diagonal matrix $\mathrm{Diag}(\exp(2ik\pi/\lambda),1,\dots,1)\in G_\mathrm{ub}<\mathrm{U}(\mu)$. We conclude that when $G=\mathrm{SU}(N)$, the unbroken gauge group in the partition $\{\lambda^\mu\}$ is a semi-direct product: 
\begin{equation}
G_\mathrm{ub} = \mathrm{SU}(\mu)\rtimes\mathbb{Z}_\lambda \ .
\end{equation}

It is instructive to consider the restriction of \cref{eq:SESGub} to the centers. One obtains another short exact sequence:
\begin{equation}\label{eq:SEScenterA}
    1 \longrightarrow \mathbb{Z}_\mu \longrightarrow \mathbb{Z}_N \longrightarrow \mathbb{Z}_\lambda \longrightarrow 1\ ,
\end{equation}
where $\mathbb{Z}_\mu \longrightarrow \mathbb{Z}_N$ is multiplication by $\lambda$, whereas $\mathbb{Z}_N \longrightarrow \mathbb{Z}_\lambda$ is reduction modulo $\mu$. 

Let us emphasize that this short exact sequence does \textit{not} necessarily split. It does split when $\lambda$ and $\mu$ are coprime, in which case $\mathbb{Z}_N \cong \mathbb{Z}_\lambda\times\mathbb{Z}_\mu$, and $G_\mathrm{ub} \cong \mathrm{SU}(\mu)\times\mathbb{Z}_\lambda$. In contrast, when $\lambda$ and $\mu$ are not coprime, \cref{eq:SEScenterA} does not split, and $\mathrm{G}_\mathrm{ub}$ is a non-trivial semi-direct product.

In terms of \cref{eq:unbrokengaugegroup}, the group $\mathbb{Z}_N$ in \cref{eq:SEScenterA} corresponds to the diagonal matrices $\zeta_N\cdot\mathrm{Id}_N\in\mathrm{SU}(N)$, where $\zeta_N$ is an $N$-th root of 1, forming the center of $\mathrm{SU}(N)$. These matrices must be in $G_\mathrm{ub}$ since the center acts trivially in the adjoint representation. In words, $G_\mathrm{ub}$ can be described as the result of subdividing the center $\mathbb{Z}_\mu$ of $\mathrm{SU}(\mu)$ in such a way that the center of $G_\mathrm{ub}$ becomes $\mathbb{Z}_N$.

\begin{example}
    For $G=\mathrm{SU}(N)$ and the partition $\{N^1\}$, the centralizer $G_{\mathrm{ub}}$ consists of all matrices of the form $M = \widetilde{m} \otimes \mathrm{Id}_{N} = \widetilde{m} \cdot \mathrm{Id}_{N} \in  G$, where $\widetilde{m}$ is a complex phase such that $\widetilde{m}^N = 1$. Therefore, $G_\mathrm{ub}= \Z_N = \mathcal{Z}$. The low-energy dynamics in the corresponding $4d$ ``Higgs'' vacuum of the $\mathrm{SU}(N)$ $ \N = 1^*$ is a discrete $\Z_N$ gauge theory, leading to $N$ distinct Higgs vacua upon compactification on a circle. 
\end{example}

\begin{example}
    For $G=\mathrm{SU}(6)$ and the partition $\{3^2\}$, $G_\mathrm{ub}$ is the product $G_\mathrm{ub}\cong \mathbb{Z}_3\times\mathrm{SU}(2)$.
\end{example} 

\begin{example}
    For $G=\mathrm{SU}(4)$ and the partition $\{2^2\}$, $G_\mathrm{ub}$ is the semi-direct product $\mathrm{SU}(2)\rtimes\mathbb{Z}_2$ such that $Z(G_\mathrm{ub})=\mathbb{Z}_4$. The center $\mathbb{Z}_2$ of $\mathrm{SU}(2)$ is the subgroup $\mathbb{Z}_2$ of $\mathbb{Z}_4$. Since $\mathbb{Z}_4\ncong\mathbb{Z}_2\times\mathbb{Z}_2$, $G_{\mathrm{ub}}$ cannot be the direct product $\mathrm{SU}(2)\times\mathbb{Z}_2$. 
\end{example}

Let us now address the general case of $G = \mathrm{SU}(N)/\mathbb{Z}_k$, where $k\vert N$; its center is $\mathbb{Z}_{N/k}$. The structure of $G_\mathrm{ub}$ can be deduced from the short exact sequences of \cref{eq:SESGub,eq:SEScenterA}. 
First of all, notice that $N=\mu\lambda=k\frac Nk$ implies that
\be
\lambda = \frac{k}{\gcd(\mu,k)}\gcd\left(\lambda,\frac Nk\right)\ , \qquad \frac Nk = \frac{\mu}{\gcd(\mu,k)}\gcd\left(\lambda,\frac Nk\right)\ .
\ee
The latter relation can also be written as
\be
\frac Nk = \frac{\mu}{\gcd(\mu,k)} 
\frac{\lambda}{k/\gcd(\mu,k)}\ ,
\ee
so that the new short exact sequence, which can be thought of as a quotient of \cref{eq:SEScenterA} by $\mathbb{Z}_k$,  is
\begin{equation}
    1\longrightarrow \mathbb{Z}_{\mu'} \longrightarrow \mathbb{Z}_{N/k} \longrightarrow \mathbb{Z}_{\lambda'} \longrightarrow 1\ ,
\end{equation}
where $\mu' = \mu/\gcd(\mu,k)$ and $\lambda'=\lambda\gcd(\mu,k)/k=\gcd(\lambda,N/k)$. The value of $\mu'$ is the order of $N/\mu$ modulo $N/k$, while that of $\lambda'$ in turn follows from the requirement that $\mu'\lambda'=N/k$.
This determines the unbroken gauge group $G_\mathrm{ub}$ in any given massive partition of $N$, for any UV connected compact simple gauge group $G$ such that $\mathrm{Lie}(G)=\mathfrak{su}(N)$. In general, one has
\begin{equation}
    G_\mathrm{ub} \cong \left(\mathrm{SU}(\mu)/\mathbb{Z}_{\gcd(\mu,k)}\right)\rtimes\mathbb{Z}_{\gcd(\lambda,N/k)}\ ,
\end{equation}
and the semi-direct product is in fact direct when $\mu'$ and $\lambda'$ are coprime.

\begin{example}
    For $G$ such that $\mathrm{Lie}(G)=\mathfrak{su}(6)$ and partitions of $6$ of the form $\lambda^\mu$, one finds the following unbroken gauge groups:
    \begin{table}
        \centering
        \begin{tabular}{c|c|c|c|c}
            & $\mathrm{SU}(6)$ & $\mathrm{SU}(6)/\mathbb{Z}_2$ & $\mathrm{SU}(6)/\mathbb{Z}_3$ & $\mathrm{PSU}(6)$\\[3pt]
            \hline
            $6^1$ & $\mathbb{Z}_6$ & $\mathbb{Z}_3$ & $\mathbb{Z}_2$ & $\{\mathrm{id}\}$ \\[3pt]
            $3^2$ & $\mathbb{Z}_3\times\mathrm{SU}(2)$ & $\mathbb{Z}_3\times\mathrm{PSU}(2)$ & $\mathrm{SU}(2)$ & $\mathrm{PSU}(2)$ \\[3pt]
            $2^3$ & $\mathbb{Z}_2\times\mathrm{SU}(3)$ & $\mathrm{SU}(3)$ & $\mathbb{Z}_2\times\mathrm{PSU}(3)$ & $\mathrm{PSU}(3)$ \\[3pt]
            $1^6$ & $\mathrm{SU}(6)$ & $\mathrm{SU}(6)/\mathbb{Z}_2$ & $\mathrm{SU}(6)/\mathbb{Z}_3$ & $\mathrm{PSU}(6)$
    \end{tabular}
    \end{table}
\end{example}

\begin{example}
    We have seen that for $G=\mathrm{SU}(4)$ and the partition $\{2^2\}$, $G_\mathrm{ub}$ is isomorphic to the semi-direct product $\mathrm{SU}(2)\rtimes\mathbb{Z}_2$ such that $Z(G_\mathrm{ub}) = \mathbb{Z}_4$. For $G=\mathrm{SU}(4)/\mathbb{Z}_2$ and the same partition $\{2^2\}$, $G_\mathrm{ub}$ rather fits into the short exact sequence
    \begin{equation}
        1\longrightarrow \mathrm{PSU}(2) \longrightarrow G_\mathrm{ub} \longrightarrow \mathbb{Z}_2 \longrightarrow 1\ .
    \end{equation}
    Hence, in this case, $G_\mathrm{ub} \cong \mathbb{Z}_2\times\mathrm{PSU}(2)$.
\end{example}

\begin{example}
    For $G=\mathrm{SU}(8)$ and the partition $\{4^2\}$, $G_\mathrm{ub}\cong\mathrm{SU}(2)\rtimes\mathbb{Z}_4$ such that $Z(G_\mathrm{ub}) = \mathbb{Z}_8$. Since $\Z_8$ is not isomorphic to $\mathbb{Z}_2\times\mathbb{Z}_4$, $G_\mathrm{ub}$ is \textit{not} $\mathrm{SU}(2)\times\mathbb{Z}_4$. Considering instead $G=\mathrm{SU}(8)/\mathbb{Z}_2$ and the same partition $\{4^2\}$, the unbroken gauge fits into
    \begin{equation}
        1 \longrightarrow \mathrm{PSU}(2) \longrightarrow G_\mathrm{ub} \longrightarrow \mathbb{Z}_4 \longrightarrow 1\ ,
    \end{equation}
    hence in that case $G_\mathrm{ub}=\mathrm{PSU}(2)\times\mathbb{Z}_4$. 
    
    For $G = \mathrm{SU}(8)/\mathbb{Z}_4$ and the same partition $\{4^2\}$, $G_\mathrm{ub}=\mathrm{PSU}(2)\times\mathbb{Z}_2$.
\end{example}

\begin{example}
    For $G$ such that $\mathrm{Lie}(G) = \mathfrak{su}(8)$ and the partition $\{2^4\}$:
    \begin{itemize}
        \item If $G=\mathrm{SU}(8)$, then $G_\mathrm{ub}\cong\mathrm{SU}(4)\rtimes\mathbb{Z}_2$ such that $Z(G_\mathrm{ub})\cong\mathbb{Z}_8$.
        \item If $G=\mathrm{SU}(8)/\mathbb{Z}_2$, then $G_\mathrm{ub}\cong\left(\mathrm{SU}(4)/\mathbb{Z}_2\right)\rtimes\mathbb{Z}_2$ such that $Z(G_\mathrm{ub})\cong\mathbb{Z}_4$.
        \item If $G=\mathrm{SU}(8)/\mathbb{Z}_4$, then $G_\mathrm{ub}\cong\mathrm{PSU}(4)\times\mathbb{Z}_2$.
        \item If $G=\mathrm{PSU}(8)$, then $G_\mathrm{ub}\cong\mathrm{PSU}(4)$.
    \end{itemize}
\end{example}

\subsection{Modular \texorpdfstring{$T$}{T} transformation and emergent \texorpdfstring{$R$}{R}-symmetry}\label{subsec:TUVandTIR}

Another important aspect of the correspondence between vacua of $\mathcal N=1^*$ theories on $\mathbb{R}^3\times S^1$ and extrema of elliptic Calogero--Moser systems, is the link between the $\mathrm{SL}(2,\mathbb{Z})$ dualities shuffling the vacua on the gauge theory side, and the modular transformations shuffling the extrema on the integrable system side.

The modular parameter $\tau$ of the elliptic curve $E_\tau$ supporting the Calogero--Moser integrable system can be identified with the UV gauge coupling 
\begin{equation}
    \tau_\mathrm{UV} = \frac{4\pi i}{g_\mathrm{UV}^2} + \frac{\theta_\mathrm{UV}}{2\pi}
\end{equation} 
of the $\mathcal N=1^*$ theory. This follows from the identification between $\tau$ and the UV gauge coupling of the corresponding $\N=2^*$ theory. 

The theta term in the Lagrangian of a $\mathfrak g$ Yang--Mills theory, where $\mathfrak g$ is any compact simple Lie algebra, reads (see e.g.~\cite{Nekrasov:2004vw}):
\begin{equation}\label{eq:thetatermYM}
    \frac{\theta}{16\pi^2 h^\vee} \int \tr_\mathrm{adj}(F\wedge F) = \theta k\ ,
\end{equation}
where $\tr_\mathrm{adj}$ refers to the trace in the adjoint representation of $\mathfrak g$, $h^\vee$ is the dual Coxeter number of $\mathfrak g$, and $k$ is the instanton number. Note that \cref{eq:thetatermYM} does not depend on a choice of generators for $\mathfrak g$. The bilinear form:
\begin{equation}\label{eq:normKilling}
    (X,Y) = \frac{1}{2h^\vee}\tr_\mathrm{adj}(XY)\ ,\quad X,Y\in\mathfrak{g}\ ,
\end{equation}
is the Killing form on $\mathfrak g$ for which long roots have squared length two. The normalization of \cref{eq:thetatermYM} ensures that the periodicity of $\theta$ is $2\pi$ when the gauge group $G$ is simply-connected \cite{Kapustin:2005py}. When $G$ is not simply-connected, $k$ can be fractional. For instance, $\theta$ is only $2\pi N$ periodic in $\mathrm{PSU}(N)$ theories.

Letting $\mathcal R$ be any finite-dimensional highest-weight representation of $\mathfrak g$, one has:
\begin{equation}
    \frac{\tr_{\mathrm{adj}}(\cdot)}{\tr_{\mathcal R}(\cdot)} = \frac{I(\mathrm{adj})}{I(\mathcal R)} = \frac{h^{\vee}(\mathfrak g)}{I(\mathcal R)}\ ,
\end{equation}
where $I(\cdot)$ is the Dynkin index, and where we have used the fact that the Dynkin index of the adjoint representation is the dual Coxeter number $h^{\vee}(\mathfrak g)$. For instance, when $\mathfrak{g} = \mathfrak{su}(N)$ and for $\mathcal R=\mathcal{F}$ the fundamental representation, one has 
\begin{equation}
    \tr_{\mathrm{adj}}(\cdot) = 2h^\vee \tr_{\mathcal{F}}(\cdot)\ ,
\end{equation}
hence \cref{eq:thetatermYM} rewrites:
\begin{equation}
    \frac{\theta}{8\pi^2} \int \tr_{\mathcal{F}}(F\wedge F) = \theta k\ .
\end{equation}

Consider now an $\mathcal N=1^*$ theory with gauge algebra $\mathfrak g$, and a solution to the F-term equations, i.e.~an orbit $\cO$ in $\mathfrak g_\mathbb{C}$. In the classical vacuum of the $\mathcal N=1^*$ theory corresponding to $\cO$, the dynamics reduces to that of some global variant of pure $\mathcal N=1$ SYM with gauge algebra $\mathfrak{g}_\mathrm{ub} = \mathrm{Cent}_{\mathfrak g}(\cO)$. The natural embedding $\mathfrak{g}_\mathrm{ub} \subset \mathfrak{g}$ can be non-trivial, implying that the theta term of the pure $\mathcal N=1$ SYM to which the dynamics reduces is not necessarily canonically normalized. This reasoning is valid at least as long as the Lagrangian description of the theories makes sense, i.e.~when $\mathrm{Im}(\tau)$ is large.

For concreteness, let us consider first in detail the example of $\mathfrak{g}=\mathfrak{su}(6)$ and the orbit $\cO$ corresponding to the partition $\underline{\lambda}=\{3^2\}$, hence $\mathfrak{g}_\mathrm{ub}\cong\mathfrak{su}(2)$. Since $\mathfrak{g}_\mathrm{ub}$ is defined as the centralizer of $\cO$ in $\mathfrak g$, one has a natural embedding $\mathfrak{g}_\mathrm{ub}\subset \mathfrak g$ corresponding to (cf.~\cref{eq:embeddingGubinGUV}):
\begin{equation}
    \mathfrak{g}_\mathrm{ub} = \left\{\left. \begin{pmatrix}
        a & b \\
        c & d
    \end{pmatrix}\otimes\mathbf{1}_3 \in\mathfrak{su}(6)\right\vert \begin{pmatrix}
        a & b \\ c & d
        \end{pmatrix} \in \mathfrak{su}(2) \right\}\ .
\end{equation}

In the UV $\mathcal N=1^*$ theory, the theta term writes:
\begin{equation}\label{eq:thetatermsu6}
    \frac{\theta_\mathrm{UV}}{8\pi^2}\int \mathrm{tr}_\mathcal{F}\left(F\wedge F\right)\ .
\end{equation}
Let $H,X,Y$ be the standard Cartan--Weyl generators of $\mathfrak{g}_\mathrm{ub} = \mathfrak{su}(2)$, i.e.
\begin{equation}
    [H,X] = 2X\ , \quad [H,Y] = -2Y\ , \quad [X,Y] = H\ .
\end{equation}
In the fundamental representation, one has:
\begin{equation}
    H = \begin{pmatrix}
        1 & 0 \\ 0 & -1
    \end{pmatrix}\ , \quad E = \begin{pmatrix}
        0 & 1 \\ 0 & 0
    \end{pmatrix}\ , \quad F = \begin{pmatrix}
        0 & 0 \\ 1 & 0
    \end{pmatrix}\ ,
\end{equation}
and $\tr_{\mathcal{F}(\mathfrak{su}(2))}(H^2) = 2$ while $\tr_{\mathcal{F}(\mathfrak{su}(2))}(EF) = \tr_{\mathcal{F}(\mathfrak{su}(2))}(FE) = 1$.

In the representation of $\mathfrak{g}_\mathrm{ub}$ induced by the embedding $\mathfrak{g}_\mathrm{ub} \subset \mathfrak g$ and the fundamental representation of $\mathfrak g = \mathrm{su}(6)$, one rather has $\tr_{\mathcal{F}(\mathfrak{su}(6))}(H^2) = 6$ while $\tr_{\mathcal{F}(\mathfrak{su}(6))}(EF) = \tr_{\mathcal{F}(\mathfrak{su}(6))}(FE) = 3$. In other words, the index of the embedding $\mathfrak{g}_\mathrm{ub} \subset \mathfrak g$ is 3. Therefore, when restricting the theta term of \cref{eq:thetatermsu6} to the generators of $\mathfrak{g}_\mathrm{ub}$, one obtains
\begin{equation}
    \frac{3\theta_\mathrm{UV}}{8\pi^2} \int \mathrm{tr}_{\mathcal{F}(\mathfrak{su}(2))}\left(f\wedge f\right)\ ,
\end{equation}
where $f$ denotes the field strength of the resulting pure $\mathcal N=1$ theory. This theta term is not canonically normalized; denoting $\theta_\mathrm{IR}$ the standard theta angle of pure $\mathcal N=1$ $\mathfrak{su}(2)$ SYM, one has $\theta_\mathrm{IR} = 3\theta_\mathrm{UV}$.

The modular $T$ transformation on the Calogero--Moser side naturally corresponds to the shift $\theta_\mathrm{UV} \rightarrow \theta_\mathrm{UV} + 2\pi$, hence to the shift $\theta_\mathrm{IR} \rightarrow \theta_\mathrm{IR} + 6\pi$. The shift $\theta_\mathrm{IR}\rightarrow \theta_\mathrm{IR}+2\pi$ instead corresponds to $\theta_\mathrm{UV}\rightarrow \theta_\mathrm{UV}+2\pi/3$, which is not a symmetry of the $\mathcal N=1^*$ theory. This is because $\mathcal N=1^*$ theories have no anomalous continuous chiral symmetry, hence $\theta_\mathrm{UV}$ is physical in the UV. Rather, when the UV gauge group is simply-connected i.e. $G=\mathrm{SU}(6)$, $\theta_\mathrm{IR}\rightarrow \theta_\mathrm{IR}+2\pi$ corresponds to the $\mathbb{Z}_4$ R-symmetry of pure $\mathcal N=1$ $\mathbb{Z}_3\times\mathrm{SU}(2)$ SYM, which is emergent. The transformation $\theta_\mathrm{IR}\rightarrow \theta_\mathrm{IR}+2\pi$ exchanges the two confining vacua of pure $\mathcal N=1$ $\mathrm{SU}(2)$ theory. 

Since the modular transformation $T$ corresponds to $\theta_\mathrm{IR}\rightarrow \theta_\mathrm{IR}+6\pi$, then in fact it also exchanges the two isolated extrema of the $\mathrm{SU}(6)$ Calogero--Moser system corresponding to the confining vacua of pure $\mathcal N=1$ $\mathrm{SU}(2)$ SYM. 

In the other extreme, i.e.~when $G=\mathrm{PSU}(6)$, the results of \cref{subsec:unbrokengaugegroups} show that the unbroken gauge group in the partition $\underline{\lambda}=3^2$ is $\mathrm{PSU}(2)$. The UV theta angle is now $12\pi$-periodic, whereas the IR one is $4\pi$-periodic. Therefore, the transformation $T$, which again corresponds to the shift $\theta_\mathrm{IR} \rightarrow \theta_\mathrm{IR}+6\pi$ is of order $2$ on the corresponding vacua on $\mathbb{R}^3\times S^1$, or, equivalently, on the corresponding Calogero--Moser extrema. Moreover, $T$ cyclically permutes the global variants $\mathrm{PSU}(6)_i$, $i=0,\dots,5$. In terms of the vacua of the intermediate pure SYM theory in the partition $3^2$, it must map the vacua of pure $\mathcal N=1$ $\mathrm{PSU}(2)_0$ SYM to vacua of pure $\mathcal N=1$ $\mathrm{PSU}(2)_1$ SYM, and vice-versa.

The reasoning straightforwardly generalizes to the case of the partition $\underline{\lambda} = \lambda^\mu$ in $\mathfrak{su}(\lambda\mu)$ $\mathcal N=1^*$ theories, where one finds that the embedding index of $\mathfrak{g}_\mathrm{ub}\cong\mathfrak{su}(\mu)$ in $\mathfrak{su}(\lambda\mu)$ is $\lambda$, or, in other words, $\theta_\mathrm{IR} = \lambda\theta_\mathrm{UV}$. The transformation $T$ acts on $\theta_\mathrm{IR}$ as:
\begin{equation}
    T\colon\theta_\mathrm{IR} \longmapsto  \theta_\mathrm{IR}+2\lambda\pi\ .
\end{equation}
The relevant quantity, as far as the action of $T$ on the isolated extrema of Calogero--Moser systems is concerned, is $\lambda$ modulo $\mu$. In particular, since $\theta_\mathrm{IR} \rightarrow  \theta_\mathrm{IR}+2\mu\pi$ corresponds to a trivial R-symmetry transformation in pure $\mathcal N=1$ $\mathrm{SU}(\mu)$ SYM, the transformation $T$ will act trivially on the extrema of Calogero--Moser systems whenever $\mu$ divides $\lambda$. This can occur only when $\lambda\mu$ has square factors, the simplest case being $\lambda=\mu=2$.

\subsection{Vacua and dualities} 

Thanks to the results of the previous section, one can now study the action of the duality group $\mathrm{SL}(2,\mathbb{Z})$ on gapped vacua of $\mathfrak{g}$ $\mathcal N=1^*$ theories. We will illustrate the general method by focusing on the concrete case of $\mathfrak g=\mathfrak{su}(3)$ $\mathcal N=1^*$ theories. Recall that the global variants of such theories are shuffled by $\mathrm{SL}(2,\mathbb{Z})$ dualities as shown in \cref{eq:dualityorbitsu(3)2occ}. 

\be\label{eq:dualityorbitsu(3)2occ}
\begin{tikzcd}[column sep=-10pt]
& \mathrm{SU}(3) \arrow[loop above, "T"] \arrow[d,dashed, leftrightarrow, "S"] & \\
& \mathrm{PSU}(3)_0 \arrow[rd, "T"] & \\
\mathrm{PSU}(3)_2 \arrow[ur, "T"] & & \mathrm{PSU}(3)_1 \arrow[ll, "T"] \arrow[ll,dashed, leftrightarrow, bend left = 50, "S"]
\end{tikzcd}
\ee

\paragraph{Action of $T$.} In \cref{eq:dualityorbitsu(3)2occ}, $T$ refers to the shift $\theta_\mathrm{UV} \rightarrow \theta_\mathrm{UV}+2\pi$ of the UV theta angle. This transformation $T$ is related to the $2\pi$ shift $T_\mathrm{IR}$ by $T = T_\mathrm{IR}^I$, where $I$ is the embedding index of $\mathfrak{g}_\mathrm{ub}$ in $\mathfrak{g}$. From this, we can infer that $T$ acts on the gapped vacua of $\mathfrak{su}(3)$ $\mathcal N=1^*$ theories on $\mathbb{R}^3\times S^1$ as follows.

\begin{itemize}
    \item The $\mathrm{SU}(3)$ global variant is fixed by $T$, hence $T$ preserves the set of $3d$ gapped vacua of this global variant. 

    \begin{itemize}
        \item The Higgs vacua (there are three of them on $\mathbb{R}^3\times S^1$) correspond to the partition $\underline{\lambda} = 3^1$ of $3$. They are characterized by the condensation of electrically-charged excitations and by the VEV of the $\mathbb{Z}_3$ Wilson line wrapping $S^1$. Both of these are preserved by the Witten effect; therefore, each Higgs vacuum is fixed by $T$. 
    
        \item The dynamics reduces to that of pure $\mathcal N=1$ $\mathrm{SU}(3)$ SYM in the partition $\underline{\lambda} = 1^3$. Pure $\mathcal N=1$ $\mathrm{SU}(3)$ SYM has three trivially gapped vacua on $\mathbb{R}^4$, leading to three gapped vacua on $\mathbb{R}^3\times S^1$. The embedding of $G_\mathrm{ub} = \mathrm{SU}(3)$ in $G=\mathrm{SU}(3)$ is trivial ($I=1$), hence $\theta_\mathrm{IR}=\theta_\mathrm{UV}$. Therefore, $T$ is to be identified with $T_\mathrm{IR}$, which permutes the three confining vacua of pure $\mathcal N=1$ $\mathrm{SU}(3)$ SYM, both on $\mathbb{R}^4$ and on $\mathbb{R}^3\times S^1$.
    \end{itemize}

    These results are consistent with the action of the modular transformation $T^\mathrm{mod}$:
    \begin{equation}
        T^\mathrm{mod}\colon (\widetilde{\sigma}_i,a_i)\longmapsto (\widetilde{\sigma}_i+a_i,a_i)\ ,
    \end{equation} 
    on the Calogero--Moser diagrams corresponding to the $3d$ gapped vacua of the $\mathrm{SU}(3)$ $\mathcal N=1^*$ theory, cf.~\cref{fig:extremaCMA2totalSU3}.
    
    \item The three $\mathrm{PSU}(3)_i$ global variants are permuted by $T$: \begin{equation}\label{eq:Tmod}
        T\colon \mathrm{PSU}(3)_i \longmapsto \mathrm{PSU}(3)_{i+1}\ ,\quad i\in\mathbb{Z}_3\ .
    \end{equation}
    Therefore: 
    \begin{itemize}
        \item For all $i\in\mathbb{Z}_3$, the Higgs vacuum of $\mathrm{PSU}(3)_i$ $\mathcal{N}=1^*$ is mapped to the Higgs vacuum of $\mathrm{PSU}(3)_{i+1}$ $\mathcal{N}=1^*$ under $T$. This holds both on $\mathbb{R}^4$ and on $\mathbb{R}^3\times S^1$.
    
        \item In the trivial nilpotent orbit $\underline{\lambda} = 1^3$, the dynamics reduces to that of pure $\mathcal N=1$ $\mathrm{PSU}(3)_i$ SYM, which has three confining vacua on $\mathbb{R}^4$. Two of them are trivially gapped while the third supports a discrete magnetic $\mathbb{Z}_3$ gauge symmetry. These lead to five trivially gapped vacua on $\mathbb{R}^3\times S^1$. As in the previous case, one has $\theta_\mathrm{IR}=\theta_\mathrm{UV}$. Therefore, $T=T_\mathrm{IR}$ maps the $4d$ vacua of $\mathrm{PSU}(3)_i$ to the $4d$ vacua of $\mathrm{PSU}(3)_{i+1}$, in a way which preserves the low-energy dynamics (either trivial, or discrete $\mathbb{Z}_3$ gauge symmetry). This, in turn, induces the action of $T$ on the gapped vacua on $\mathbb{R}^3\times S^1$.
    \end{itemize} 

    These results are consistent with the action of the modular transformation $T^\mathrm{mod}$ of \cref{eq:Tmod}
    on the Calogero--Moser diagrams corresponding to the $3d$ gapped vacua of the $\mathrm{PSU}(3)_i$ $\mathcal N=1^*$ theories, cf.~\cref{fig:extremaCMA2totalPSU30,fig:extremaCMA2totalPSU31,fig:extremaCMA2totalPSU32} (keeping in mind that \cref{fig:extremaCMA2totalPSU30,fig:extremaCMA2totalPSU31,fig:extremaCMA2totalPSU32} are cyclically permuted by $T^\mathrm{mod} = T$).
\end{itemize}

In all cases, we have found that the identification $T=T^\mathrm{mod}$ is consistent with our expectations. This is what one expects considering the $\mathcal N=4$ origin of $\mathcal N=1^*$ theories.

\paragraph{Action of $S$.}

As for $T$, the $\mathcal N=4$ origin of $\mathcal N=1^*$ theories points towards identifying the duality generator $S$ with the modular transformation $S^\mathrm{mod}$ acting on Calogero--Moser diagrams. In general (for $\mathfrak g$ simply-laced), $S^\mathrm{mod}$ acts as:
\begin{equation}
    S^\mathrm{mod}\colon z_i=(\widetilde{\sigma}_i,a_i) \longmapsto \frac{z_i}{\tau}\ .
\end{equation}
When $\tau=i$, $S^\mathrm{mod}$ corresponds to a $-\pi/2$-rotation about the center of the cell. 

Interestingly, one finds that $S=S^\mathrm{mod}$ is not quite an involution: some Calogero--Moser diagrams are not fixed by the transformation: 
\begin{equation}
    S^2=(ST)^3=\mathcal C=-\mathrm{Id}\in\mathrm{SL}(2,\mathbb{Z})\ .
\end{equation}
Since the element $\mathcal C$ corresponds to charge conjugation, the isolated extrema which are not invariant under $\mathcal C$ correspond to $3d$ gapped vacua in which charge conjugation is spontaneously broken. The only vacua displaying this behavior are among those arising in families generated from a single $4d$ vacuum: non-trivial Wilson lines charged under the discrete $4d$ gauge symmetry and wrapping the compactification circle break charge conjugation $\mathcal C$.\footnote{The notion of charge conjugation that we refer to here is not the same that plays a role in the analysis of \cite{Kaidi:2022uux}, where it is defined in terms of the 1-form symmetry.} 

\begin{figure}
    \centering
    \includegraphics[width=0.9\linewidth]{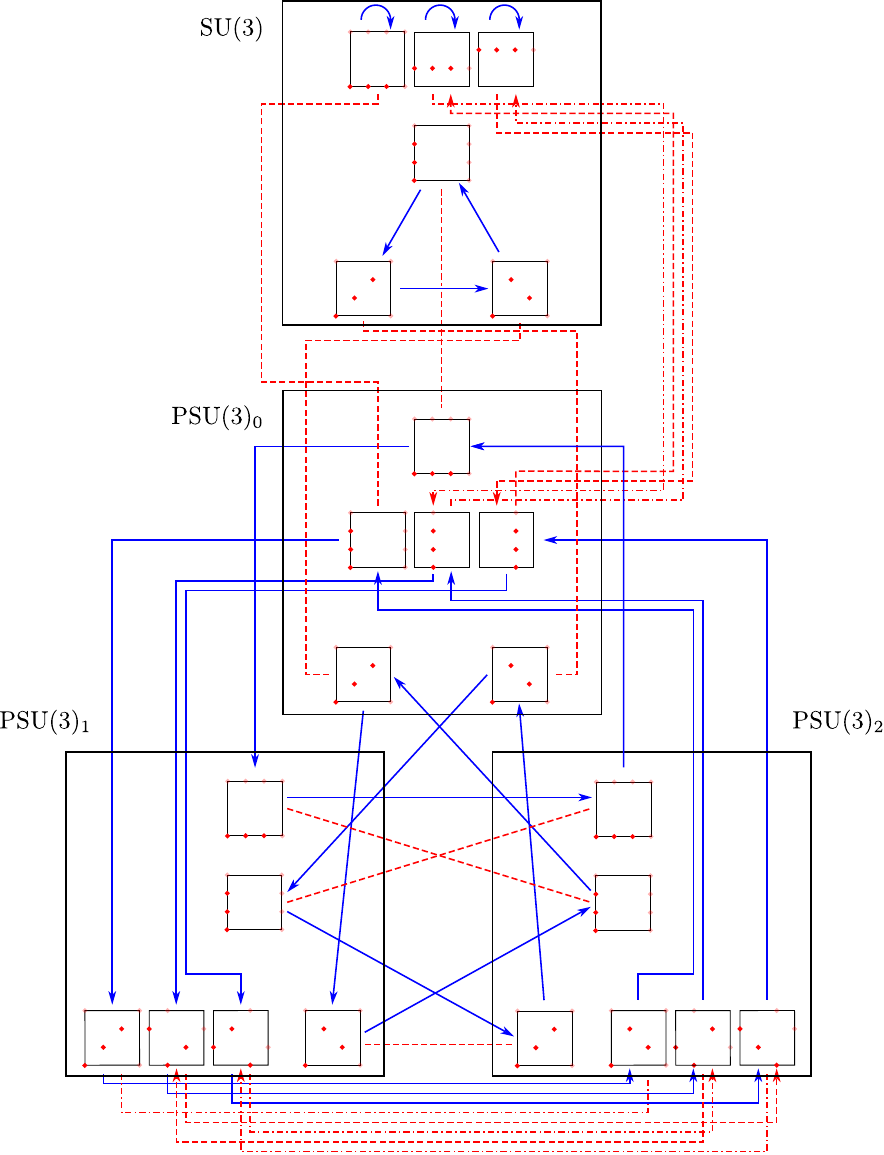}
    \caption{Duality web of $3d$ gapped vacua of $\mathfrak{su}(3)$ $\mathcal N=1^*$ theories. Dashed red arrows correspond to $S$ (we do not show the arrow when $S^2=1$) and plain blue arrows, to $T$. The corresponding disjoint orbits are shown in \cref{fig:su3_dualitygraph}.}
    \label{fig:dualityweb}
\end{figure}

\paragraph{Duality web.} The duality web of $3d$ gapped vacua of $\mathfrak{su}(3)$ $\mathcal N=1^*$ theories, computed from the action of $T=T^\mathrm{mod}$ and $S=S^\mathrm{mod}$ on Calogero--Moser diagrams at $\tau=i$, is shown in \cref{fig:dualityweb}. This graph refines that of global variants of $\mathfrak{su}(3)$ $\mathcal N=1^*$ theories (\cref{eq:dualityorbitsu(3)2occ}).

Gapped vacua of $\mathfrak{su}(3)$ $\mathcal N=1^*$ theories on $\mathbb{R}^3\times S^1$ split in three orbits under $\mathrm{SL}(2,\mathbb{Z})$ dualities, which are also depicted in \cref{fig:su3_dualitygraph}:
\begin{itemize}
    \item The first orbit contains all $3d$ gapped vacua arising from trivially gapped $4d$ vacua.
    \item The remaining two orbits contain the $3d$ gapped vacua arising from $4d$ gapped vacua supporting a discrete $\mathbb{Z}_3$ symmetry:
    \begin{itemize}
        \item One orbit consists of all such vacua in which charge conjugation $\mathcal C$ is unbroken,
        \item The other consists of all vacua in which charge conjugation $\mathcal C$ is spontaneously broken.
    \end{itemize}
\end{itemize}

\paragraph{General case.} Arguably the most important consequence of the correspondence between $\mathcal N=2^*$ or $\mathcal N=1^*$ theories and Calogero--Moser integrable systems as defined in \cref{def:globvarCM}, is that global variants of Calogero--Moser systems provide a useful handle on those theories even in non-Lagrangian regimes, when $\Im(\tau)$ is small. When $\Im(\tau)$ is large, one can study the dynamics of $\mathcal N=1^*$ theories via their Lagrangian. This allowed the investigation of the important relationship between $T$ and $T_\mathrm{IR}$ in \cref{subsec:TUVandTIR}.

How dualities relate vacua of $\mathcal N=1^*$ theories on $\mathbb{R}^3\times S^1$ can be studied by considering the action of $T^\mathrm{mod}$ and $S^\mathrm{mod}$ on Calogero--Moser diagrams, which are identified with the duality generators $T$ and $S$ of the UV $\mathcal N=1^*$ theory. In the sequel we will not distinguish $T^\mathrm{mod}$ and $S^\mathrm{mod}$ from $T$ and $S$ anymore.

Two subtleties arise when $\mathfrak{g}$ is non simply-laced: one deals with twisted Calogero--Moser systems, and the modular/duality generator $S$ acts as:
\begin{equation}
    S\colon z \longmapsto \frac{z}{\nu(\alpha_s)\tau}\ ,
\end{equation}
where $\nu(\alpha_s)$ is the index of \cref{eq:indexCM} for short roots of $\mathfrak{g}$. We will discuss these matters further in \cref{sec: so(5)}.

\subsection{\texorpdfstring{$\mathfrak{su}(6)$}{su(6)} theories}

We now turn to the study of global variants of $\mathfrak{su}(6)$ $\mathcal N=1^*$ theories and Calogero--Moser systems. This case introduces an additional complication as compared to $\mathfrak{su}(3)$, namely that the center $\mathbb{Z}_6$ of $\mathrm{SU}(6)$ admits non-trivial subgroups $\mathbb{Z}_2$ and $\mathbb{Z}_3$. However, since $2$ and $3$ are coprime, one has $\mathbb{Z}_6\cong \mathbb{Z}_2\times\mathbb{Z}_3$. The case of $\mathfrak{su}(4)$ theories, where $\mathbb{Z}_4$ is not isomorphic to $\mathbb{Z}_2\times\mathbb{Z}_2$, will be studied next. Both cases will serve as non-trivial tests of our findings, and inform on how more general cases can be approached.

We take as coroots of $\mathfrak{su}(6)$ the matrices $H_{\alpha_1},\dots,H_{\alpha_5}$, where $H_{\alpha_k}$ is the diagonal matrix with $1$ at entry $(k,k)$ and $-1$ at entry $(k+1,k+1)$, and 0 elsewhere, and as fundamental coweights, the matrices: 
\begin{equation}
    \begin{split}
        \chi_1 &= \frac{1}{6}\mathrm{Diag}(1,1,1,1,1,-5)\ , \quad 
        \chi_2 = \frac{1}{6}\mathrm{Diag}(2,2,2,2,-4,-4)\ , \\
        \chi_3 &= \frac{1}{6}\mathrm{Diag}(3,3,3,-3,-3,-3)\ ,\quad
        \chi_4 = \frac{1}{6}\mathrm{Diag}(4,4,-2,-2,-2,-2)\ ,\\
        \chi_5 &= \frac{1}{6}\mathrm{Diag}(5,-1,-1,-1,-1,-1)\ ,
    \end{split}
\end{equation}
where, for later convenience, we have used a different labeling of the coweights with respect to \cref{eq:coweigthssu3}, i.e.~ascending instead of descending first entry.

\begin{figure}
    \centering
    \includegraphics[width=\textwidth]{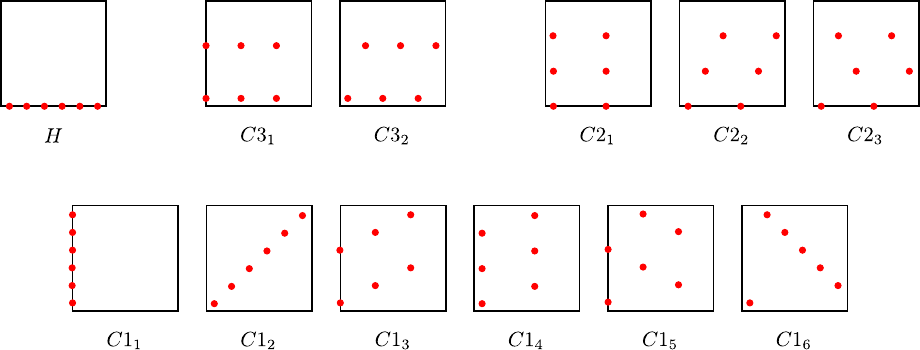}
    \caption{Isolated extrema of the $\mathfrak{su}(6)$ elliptic Calogero-Moser system. }
    \label{fig:extremaCMA5}
\end{figure}

The isolated extrema of the elliptic $\mathfrak{su}(6)$ Calogero--Moser system\footnote{That is, the isolated extrema of the elliptic Calogero--Moser system of type $A_5$ with horizontal (resp. vertical) periodicities elements of the weight lattice $\Lambda_w$ (resp. coweight lattice $\Gamma_w$) of $\mathfrak{su}(6)$.} are shown in \cref{fig:extremaCMA5}, in the convention where $\sum z_i=0$.

The count of $4d$ vacua for $\mathfrak{su}(6)$ $\N=1^*$ theories, shown in \cref{tab:su6unbrokenalgebras}, suggests a correspondence between the diagrams $H$, $C3_{1,2}$, $C2_{1,2,3}$, $C1_{1,\dots,6}$ in \cref{fig:extremaCMA5} and the partitions $6^1$, $3^2$, $2^3$, $1^6$; as we will see, this intuition is validated by carefully taking into account global variants of $\mathcal{N}=1^*$ theories and Calogero–Moser systems.

\begin{table}
\centering
\begin{tabular}{c|c|c}
    & $\mathfrak{g}_\mathrm{ub}$ & $\#$ $4d$ vacua \\[3pt]
    \hline
    $6^1$ & $\{0\}$ & 1 \\[3pt]
    $3^2$ & $\mathfrak{su}(2)$ & 2 \\[3pt]
    $2^3$ & $\mathfrak{su}(3)$ & 3 \\[3pt]
    $1^6$ & $\mathfrak{su}(6)$ & 6 
\end{tabular}
\caption{Massive partitions of $6$, $\mathfrak{g}_\mathrm{ub}$, and number of gapped $4d$ vacua.}
\label{tab:su6unbrokenalgebras}
\end{table}

\subsubsection{Unbroken subgroups}

\cref{tab:su6unbroken} shows the unbroken gauge group for each choice of UV gauge group $G$ such that $\mathrm{Lie}(G) = \mathfrak{su}(6)$ and for each massive partition $\underline{\lambda}$ of $6$, as obtained already in \cref{subsec:unbrokengaugegroups} from the general method discussed therein.

\begin{table}
\centering
\begin{tabular}{c|c|c|c|c}
    & $\mathrm{SU}(6)$ & $\mathrm{SU}(6)/\mathbb{Z}_2$ & $\mathrm{SU}(6)/\mathbb{Z}_3$ & $\mathrm{PSU}(6)$\\[3pt]
    \hline
    $6^1$ & $\mathbb{Z}_6$ & $\mathbb{Z}_3$ & $\mathbb{Z}_2$ & $\{\mathrm{id}\}$ \\[3pt]
    $3^2$ & $\mathbb{Z}_3\times\mathrm{SU}(2)$ & $\mathbb{Z}_3\times\mathrm{PSU}(2)$ & $\mathrm{SU}(2)$ & $\mathrm{PSU}(2)$ \\[3pt]
    $2^3$ & $\mathbb{Z}_2\times\mathrm{SU}(3)$ & $\mathrm{SU}(3)$ & $\mathbb{Z}_2\times\mathrm{PSU}(3)$ & $\mathrm{PSU}(3)$ \\[3pt]
    $1^6$ & $\mathrm{SU}(6)$ & $\mathrm{SU}(6)/\mathbb{Z}_2$ & $\mathrm{SU}(6)/\mathbb{Z}_3$ & $\mathrm{PSU}(6)$
\end{tabular}
\caption{Unbroken gauge groups for massive partitions in $\mathfrak{su}(6)$ $\mathcal{N}=1^*$ theories. }
\label{tab:su6unbroken}
\end{table}

The fact that the unbroken gauge groups always decompose as products of discrete and simple factors is a consequence of $2$ and $3$ being coprime, and in turn, of $6$ being square-free. This property makes the $\mathfrak{su}(6)$ case comparatively easier than, say, the $\mathfrak{su}(4)$ case.

\subsubsection{Emergent R-symmetries and \texorpdfstring{$\theta$}{theta} angles}

The relation between the UV theta angle $\theta$ and the IR one $\theta_\mathrm{IR}$ only depends on the embedding index $I$ of the unbroken gauge algebra $\mathfrak{g}_\mathrm{ub}$ in the UV gauge algebra $\mathfrak{g}$, which only depends on the row in \cref{tab:su6unbroken}. 

For the first row, $\mathfrak{g}_\mathrm{ub} = \{0\}$, hence the embedding index is not defined, and neither is $\theta_\mathrm{IR}$. In contrast, for the last row of \cref{tab:su6unbroken}, one has $\mathfrak{g}_\mathrm{ub} = \mathfrak{su}(6)$, embedding trivially in $\mathfrak g$, meaning that $I=1$ and $\theta=\theta_\mathrm{IR}$.

In the second row, one has $\mathfrak{g}_\mathrm{ub} = \mathfrak{su}(2)$, embedded in $\mathfrak{su}(6)$ as a $2\times 2$ matrix of $3\times 3$ blocks of the form $a\mathbf{1}_3$. This implies $I=3$, hence $\theta_\mathrm{IR}=3\theta_\mathrm{UV}$ (this case is the one discussed in \cref{subsec:TUVandTIR}).

In the third row of \cref{tab:su6unbroken}, one has $\mathfrak{g}_\mathrm{ub} = \mathfrak{su}(3)$, embedded in $\mathfrak{su}(6)$ as a $3\times 3$ matrix of $2\times 2$ blocks of the form $a\mathbf{1}_2$. This implies $I=2$, hence $\theta_\mathrm{IR}=2\theta_\mathrm{UV}$.

\subsubsection{\texorpdfstring{$\mathrm{SU}(6)$}{su(6)}}

We concentrate on the first column in \cref{tab:su6unbroken}. The unbroken gauge groups in every massive partition of $6$ as well as the corresponding numbers of $4d$ and $3d$ vacua are shown in \cref{tab:SU6unbroken}. 

\begin{table}
\centering
\begin{tabular}{c|c|c|c}
    & $\mathrm{G}_\mathrm{ub}$ & $\#$ $4d$ vacua & $\#$ $3d$ vacua \\[3pt]
    \hline
    $6^1$ & $\mathbb{Z}_6$ & 1 & $\mathbf{6}\times 1 = 6$ \\[3pt]
    $3^2$ & $\mathbb{Z}_3\times\mathrm{SU}(2)$ & 2 & $\mathbf{3}\times 2 = 6$ \\[3pt]
    $2^3$ & $\mathbb{Z}_2\times\mathrm{SU}(3)$ & 3 & $\mathbf{2}\times 3 = 6$ \\[3pt]
    $1^6$ & $\mathrm{SU}(6)$ & 6 & $\mathbf{1}\times 6 = 6$
\end{tabular}
\caption{Unbroken gauge groups and number of gapped vacua of the $\mathrm{SU}(6)$ $\N=1^*$ theory.}
\label{tab:SU6unbroken}
\end{table}

\begin{figure}
    \centering
    \includegraphics[width=\textwidth]{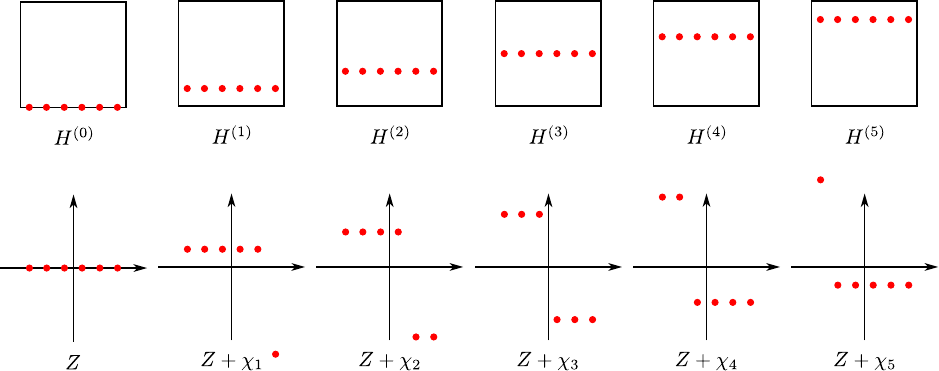}
    \caption{Inequivalent $3d$ Higgs vacua of $\mathrm{SU}(6)$ $\N=1^*$ theory.}
    \label{fig:HiggsvacuaSU6}
\end{figure}

Imposing as horizontal periodicity all translations by weights, and as vertical periodicity all translations by coroots, i.e. solving the equation
\begin{equation}
    E' + \sum_{i=1}^6 h_i \, \chi_i + \sum_{i=1}^6 v_i\, \tau H_{\alpha_i} \approx E \subset \mathbb{C}\ , 
\end{equation}
where $h_i, v_i \in\mathbb{Z}$, $i = 1,\dots,6$ (we recall that $\approx$ allows for Weyl transformations on the $z_i$), one finds that these diagrams have the expected multiplicity to match the count of $3d$ vacua.  For instance, the first row in \cref{tab:su6unbroken} (Higgs vacua) corresponds to the diagram $H$ in \cref{fig:extremaCMA5}. In fact, there are six inequivalent Calogero--Moser diagrams $H^{(k)}$ corresponding to $H$, $k=0,\dots,5$, shown in \cref{fig:HiggsvacuaSU6}. They are obtained from a configuration $Z$ corresponding to $H$ by a vertical shift by a coweight $\chi_i$, and cannot be mapped back to $H$ by an element of the coroot lattice.

Similarly, one finds three inequivalent $\mathrm{SU}(6)$ Calogero--Moser diagrams $C3_1^{(k)}$ and $C3_2^{(k)}$, $k=0,1,2$, corresponding to $C3_1$ and $C3_2$, two inequivalent $\mathrm{SU}(6)$ Calogero--Moser diagrams $C2_{1,2,3}^{(k)}$, $k=0,1$, for each $C2_{1,2,3}$, and only one for each $C1_{1,\dots,6}$. 

\begin{figure}
    \centering
    \includegraphics[width=\textwidth]{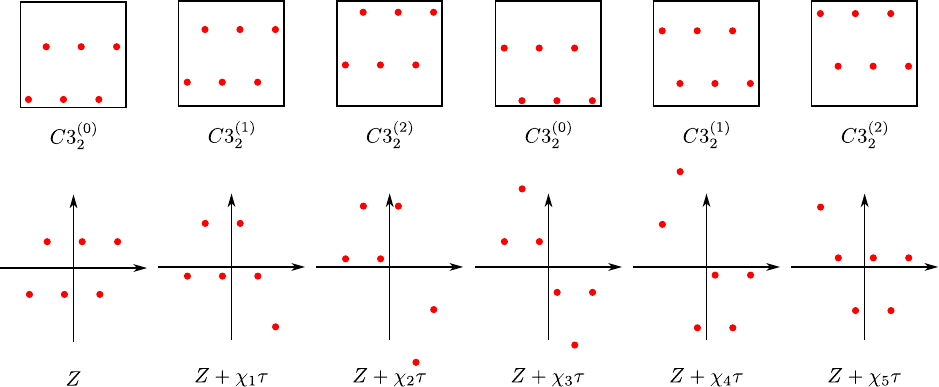}
    \caption{Inequivalent $C3_2^{(k)}$ vacua of $\mathrm{SU}(6)$ $\N=1^*$ theory.}
    \label{fig:C32kSU6}
\end{figure}

For instance, consider the diagram $C3_2^{(0)} = C3_2$ and a configuration $Z$ of six points in $\mathbb{C}$ corresponding to it and satisfying $\sum z_i = 0$, as depicted in \cref{fig:C32kSU6} (the points are ordered from left to right). 

One can verify that the configurations $Z$, $Z+\chi_1\tau$, and $Z+\chi_2\tau$ are mutually inequivalent modulo vertical translations by coroots and horizontal translations by coweights. In contrast, one has that $Z+\chi_3\tau$ is equivalent to $Z$. Indeed, this transformation can be undone by a vertical shift by $- (H_{\alpha_2}+H_{\alpha_3}+H_{\alpha_4})\in\Gamma_r$, followed by an horizontal shift by $(\chi_1-\chi_2+\chi_3-\chi_4+\chi_5) +(H_{\alpha_2}+H_{\alpha_3}+H_{\alpha_4})\in\Gamma_w$ and a permutation of the $z_i$. This shows that the fourth diagram in the top row of \cref{fig:C32kSU6} is equivalent to the first. Likewise, one finds that the second and fifth diagrams are equivalent, as well as the third and sixth. Here we should note that in the present case, it could seem somewhat contrived to show the equivalence of extrema on $\bC$ rather than on the torus. However we will see in the example based on $\mathfrak{so}(5)$ that this is the only way that clearly identifies equivalent and inequivalent extrema.

The transformation $T$ acts on the isolated extrema of the $\mathrm{SU}(6)$ Calogero--Moser systems in the following way:
\begin{equation}\label{eq:actofT}
    \begin{split}
        &H^{(i)} \longrightarrow H^{(i)}\ , \quad i=0,\dots,5\ ,\\
        &C3_1^{(i)} \longleftrightarrow C3_2^{(i)}\ , \quad i=0,1,2\ , \\
        &C2_1^{(i)} \longrightarrow C2_3^{(i)} \longrightarrow C2_2^{(i)} \longrightarrow C2_1^{(i)}\ , \quad i=0,1\ , \\
        &C1_1 \longrightarrow C1_2 \longrightarrow \dots \longrightarrow C1_6 \longrightarrow C1_1\ .
    \end{split}
\end{equation}
These equations are consistent with our proposal. In the partition $3^2$ one has $T=T_\mathrm{IR}^3$, which exchanges the two confining vacua of pure $\mathcal N=1$ $\mathrm{SU}(2)$ SYM, while in the partition $2^3$ on has $T=T_\mathrm{IR}^2$, which permutes the three confining vacua of pure $\mathcal N=1$ $\mathrm{SU}(3)$ SYM cyclically.

\subsubsection{\texorpdfstring{$\mathrm{PSU}(6)$}{PSU(6)}}

Let us now turn to the last column in \cref{tab:su6unbroken}. The unbroken gauge groups in every massive partition of $6$, as well as the corresponding numbers of gapped vacua on $\mathbb{R}^4$ and $\mathbb{R}^3\times S^1$, are shown in \cref{tab:PSU6unbroken}.

\begin{table}
\centering
\begin{tabular}{c|c|c|c}
    & $G_\mathrm{ub}$ & $\#$ $4d$ vacua & $\#$ $3d$ vacua \\[3pt]
    \hline
    $6^1$ & $\{1\}$ & 1 & $1$ \\[3pt]
    $3^2$ & $\mathrm{PSU}(2)$ & 2 & $3$ \\[3pt]
    $2^3$ & $\mathrm{PSU}(3)$ & 3 & $5$ \\[3pt]
    $1^6$ & $\mathrm{PSU}(6)$ & 6 & $15$
\end{tabular}
\caption{Unbroken gauge groups and number of gapped vacua of $\mathrm{PSU}(6)$ $\N=1^*$ theories.}
\label{tab:PSU6unbroken}
\end{table}

\paragraph{Global variant $\mathrm{PSU}(6)_0$.} The Lagrangian subgroup defining the global variant $\mathrm{PSU}(6)_0$ is:
\begin{equation}
    \mathcal L_{\mathrm{PSU}(6)_0} = \{(0,z)\,\mid\, z=0,\dots,5\} \subset \widehat{\mathbb{Z}}_6\times \mathbb{Z}_6\ .
\end{equation}
Therefore, gauged translations in the Calogero--Moser diagrams are horizontal translations by coroots and vertical translations by coweights. The multiplicities of the diagrams of \cref{fig:extremaCMA5} are obtained by solving
\begin{equation}
    E' + \sum_{i=1}^6 h_i \, H_{\alpha_i} + \sum_{i=1}^6 v_i\, \tau \chi_i \approx E \subset \mathbb{C}\ , 
\end{equation}
where $h_i, v_i \in\mathbb{Z}$, $i = 1,\dots,6$. Thus, this case is similar to that of the $\mathrm{SU}(6)$ theory, with horizontal and vertical directions exchanged (note that this corresponds to the action of $S$). We can thus readily deduce the ``multiplicities'' shown in \cref{tab:multipPSU(6)0}.

\begin{table}
\centering
\begin{tabular}{c||c|c|c|c|c|c|c|c|c|c|c|c}
    Diagram & $H$ & $C3_1$ & $C3_2$ & $C2_1$ & $C2_2$ & $C2_3$ & $C1_1$ & $C1_2$ & $C1_3$ & $C1_4$ & $C1_5$ & $C1_6$ \\[3pt]
    \hline
    Mult. & 1 & 2 & 1 & 3 & 1 & 1 & 6 & 1 & 2 & 3 & 2 & 1
\end{tabular}
\caption{Multiplicity of the diagrams of \cref{fig:extremaCMA5} for the global variant $\mathrm{PSU}(6)_0$.}
\label{tab:multipPSU(6)0}
\end{table}

These multiplicities are in accordance with the correspondence between the diagrams $C3_{1,2}$ of \cref{fig:extremaCMA5} and the $4d$ vacua of pure $\mathcal N=1$ $\mathrm{PSU}(2)_0$ SYM, where $C3_{1}$ is the vacuum with a discrete $\mathbb{Z}_2$ magnetic gauge symmetry. The three gapped vacua of pure $\mathcal N=1$ $\mathrm{PSU}(2)_0$ SYM on $\mathbb{R}^3\times S^1$ correspond to:
\begin{equation}
    C3_{1}^{(1)}\ , \quad C3_{1}^{(2)}\ , \quad C3_{2}\ .
\end{equation}

Similarly, the diagrams $C2_{1,2,3}$ correspond to the $4d$ vacua of pure $\mathcal N=1$ $\mathrm{PSU}(3)_0$ SYM on $S^1$, where $C2_1$ is the vacuum with a discrete $\mathbb{Z}_3$ gauge symmetry. The five gapped vacua of pure $\mathcal N=1$ $\mathrm{PSU}(3)_0$ SYM on $\mathbb{R}^3\times S^1$ correspond to: 
\begin{equation}
    C2_{1}^{(1)}\ , \quad C2_{1}^{(2)}\ , \quad C2_{1}^{(3)}\ , \quad C2_{2}\ , \quad C2_3\ .
\end{equation}

Lastly, the diagrams $C1_{1,\dots,6}$ correspond to the $4d$ vacua of pure $\mathcal N=1$ $\mathrm{PSU}(6)_0$ SYM. The $3d$ vacua correspond to: 
\begin{equation}
    C1_1^{(1),(2),(3),(4),(5),(6)}\ , \quad C1_2\ , \quad C1_3^{(1),(2)}\ , \quad C1_4^{(1),(2),(3)}\ , \quad C1_5^{(1),(2)}\ , \quad C1_6\ .\\
\end{equation}

\paragraph{Global variants $\mathrm{PSU}(6)_j$ $(j\neq 0)$.} 

The Lagrangian subgroup defining the global variant $\mathrm{PSU}(6)_j$, where $j=1,\dots,5$, is:
\begin{equation}
    \mathcal L_{\mathrm{PSU}(6)_j} = \{(jz,z)\,\mid\, z=0,\dots,5\} \subset \widehat{\mathbb{Z}}_6\times \mathbb{Z}_6\ .
\end{equation}
This means that gauged translations are those by elements of the form $\lambda+\chi\tau$, where $\lambda\in\Lambda_w$ and $\chi\in\Gamma_w$ must satisfy $(\phi_\Lambda(\lambda),\phi_\Gamma(\chi))\in\mathcal L$, in the notation of \cref{def:globvarCM}. Since the lift of $\mathcal L_{\mathrm{PSU}(6)_j}$ to $\mathbb{Z}\times\mathbb{Z}$ is generated by $(j,1)$ and $(6,0)$, and since the isomorphism given by the Killing form maps $\Lambda_w$ to $\Gamma_w$, the gauged translations in the $\mathrm{PSU}(6)_j$ Calogero--Moser system can be equivalently presented as follows: they are generated by all horizontal translations by elements of the coroot lattice, and all translations by elements of the coweight lattice in the diagonal direction $(j,1)$. This allows to compute the multiplicities of all diagrams of \cref{fig:extremaCMA5}, in the global variant $\mathrm{PSU}(6)_j$, by solving
\begin{equation}
    E' + \sum_{i=1}^6 h_i \, H_{\alpha_i} + \sum_{i=1}^6 d_i\, (j+\tau) \chi_i \approx E \subset \mathbb{C}\ , 
\end{equation}
where $h_i, d_i \in\mathbb{Z}$, $i = 1,\dots,6$. For instance, when $j=1$, one finds the multiplicities shown in \cref{tab:multipPSU(6)1}.

\begin{table}
\centering
\begin{tabular}{c||c|c|c|c|c|c|c|c|c|c|c|c}
    Diagram & $H$ & $C3_1$ & $C3_2$ & $C2_1$ & $C2_2$ & $C2_3$ & $C1_1$ & $C1_2$ & $C1_3$ & $C1_4$ & $C1_5$ & $C1_6$ \\[3pt]
    \hline
    Mult. & 1 & 1 & 2 & 1 & 1 & 3 & 1 & 6 & 1 & 2 & 3 & 2
\end{tabular}
\caption{Multiplicity of the diagrams of \cref{fig:extremaCMA5} for the global variant $\mathrm{PSU}(6)_1$.}
\label{tab:multipPSU(6)1}
\end{table}

Given the link between $T$ and $T_\mathrm{IR}$ in each massive partition of $6$, these periodicities are those one expects after acting with $T$ once on the diagrams of the global variant $\mathrm{PSU}(6)_0$. For instance, since $T=T_\mathrm{IR}^3$ in the partition $3^2$, $T$ maps the diagram $C3_1$ of the global variant $\mathrm{PSU}(6)_0$ to the diagram $C3_2$ of $\mathrm{PSU}(6)_1$. Since $T=T_\mathrm{IR}^2$ in the partition $2^3$, $T$ maps the diagram $C2_1$ of the global variant $\mathrm{PSU}(6)_0$ to the diagram $C2_3$ of $\mathrm{PSU}(6)_1$. In the partition $1^6$, $T = T_\mathrm{IR}$ maps each $C1_k$ of $\mathrm{PSU}(6)_0$ to $C1_{k+1}$ of $\mathrm{PSU}(6)_1$. Therefore, the multiplicities of the Calogero--Moser diagrams of \cref{fig:extremaCMA5} can easily be computed in any given global variant $\mathrm{PSU}(6)_j$, from those in the global variant $\mathrm{PSU}(6)_0$. This is a straightforward consequence of the construction, since the modular transformation $T$ leaves the horizontal direction invariant, and maps the diagonal direction $(j,1)$ to $(j+1,1)$.

\subsubsection{\texorpdfstring{$\mathrm{SU}(6)/\mathbb{Z}_2$}{SU(6)/Z2}}

The unbroken gauge groups in every massive partition of $6$, as well as the corresponding numbers of gapped vacua on $\mathbb{R}^4$ and $\mathbb{R}^3\times S^1$, are shown in \cref{tab:SU6/Z2unbroken}.

\begin{table}
\centering
\begin{tabular}{c|c|c|c}
    & $G_\mathrm{ub}$ & $\#$ $4d$ vacua & $\#$ $3d$ vacua \\[3pt]
    \hline
    $6^1$ & $\mathbb{Z}_3$ & 1 & $\mathbf{3}$ \\[3pt]
    $3^2$ & $\mathbb{Z}_3\times\mathrm{PSU}(2)$ & 2 & $\mathbf{3}\times 3=9$ \\[3pt]
    $2^3$ & $\mathrm{SU}(3)$ & 3 & $3$ \\[3pt]
    $1^6$ & $\mathrm{SU}(6)/\mathbb{Z}_2$ & 6 & $9$
\end{tabular}
\caption{Unbroken gauge groups and number of gapped vacua of $\mathrm{SU}(6)/\mathbb{Z}_2$ $\N=1^*$.}
\label{tab:SU6/Z2unbroken}
\end{table}

The multiplicities of the diagrams shown in \cref{fig:extremaCMA5} can be computed as before, for both global variants $(\mathrm{SU}(6)/\mathbb{Z}_2)_0$ and $(\mathrm{SU}(6)/\mathbb{Z}_2)_1$, defined by the following Lagrangian subgroups of $\widehat{\mathbb{Z}}_6\times\mathbb{Z}_6$:
\begin{equation}
    \begin{split}
        \mathcal L_{(\mathrm{SU}(6)/\mathbb{Z}_2)_0} &= \{(0,0),(0,3),(2,0),(2,3),(4,0),(4,3)\}\ ,\\
        \mathcal L_{(\mathrm{SU}(6)/\mathbb{Z}_2)_1} &= \{(0,0),(1,3),(2,0),(3,3),(4,0),(5,3)\}\ .
    \end{split}
\end{equation}

The lift of $\mathcal L_{(\mathrm{SU}(6)/\mathbb{Z}_2)_0}$ to $\mathbb{Z}\times\mathbb{Z}$ is generated by $(2,0)$ and $(0,3)$, thus the gauged translations in the $(\mathrm{SU}(6)/\mathbb{Z}_2)_0$ global variant of the Calogero--Moser system of type $A_5$ are all horizontal ones by coweights of $6$-ality $0,2$ or $4$ and all vertical ones by coweights of $6$-ality $0$ or $3$. Equivalently, since the lift of $\mathcal L_{(\mathrm{SU}(6)/\mathbb{Z}_2)_0}$ to $\mathbb{Z}\times\mathbb{Z}$ is generated by $(2,3)$ and $(4,3)$, one can solve the equation
\begin{equation}
    E' + \sum_{i=1}^6 d_i\,(2+3\tau) \chi_i + \sum_{i=1}^6 d_i'\, (4+3\tau) \chi_i \approx E \subset \mathbb{C}\ , 
\end{equation}
where $d_i, d_i' \in\mathbb{Z}$, $i = 1,\dots,6$. With such gauged translations, one finds for example that the diagrams $C3_1$ and $C3_2$ of \cref{fig:extremaCMA5} have multiplicity respectively six and three, which is in accordance with the fact that the intermediate gauge theory in the partition $3^2$ is pure $\mathcal N=1$ $\mathbb{Z}_3\times\mathrm{PSU}(2)_0$ SYM.

The lift of $\mathcal L_{(\mathrm{SU}(6)/\mathbb{Z}_2)_1}$ to $\mathbb{Z}\times\mathbb{Z}$ is generated by $(2,0)$ and $(1,3)$, thus the gauged translations in the $(\mathrm{SU}(6)/\mathbb{Z}_2)_1$ global variant of the Calogero--Moser system of type $A_5$ are all horizontal ones by coweights of $6$-ality $0,2$ or $4$ and the diagonals ones by any element of the coweight lattice, in the direction $(1,3)$. Equivalently, since the lift of $\mathcal L_{(\mathrm{SU}(6)/\mathbb{Z}_2)_0}$ to $\mathbb{Z}\times\mathbb{Z}$ is generated by $(0,6)$ and $(1,3)$, one can solve the equation
\begin{equation}
    E' + \sum_{i=1}^6 v_i\, H_{\alpha_i} + \sum_{i=1}^6 d_i\, (1+3\tau) \chi_i \approx E \subset \mathbb{C}\ , 
\end{equation}
where $v_i, d_i \in\mathbb{Z}$, $i = 1,\dots,6$. The diagrams $C3_1$ and $C3_2$ of \cref{fig:extremaCMA5} now have multiplicity three and six, which is in accordance with the intermediate gauge theory in the partition $3^2$ being pure $\mathcal N=1$ $\mathbb{Z}_3\times\mathrm{PSU}(2)_1$ SYM, and with the relation $T=T_\mathrm{IR}^3$ in this partition.

\subsubsection{\texorpdfstring{$\mathrm{SU}(6)/\mathbb{Z}_3$}{SU(6)/Z3}}

The unbroken gauge groups in every massive partition of $6$, as well as the corresponding numbers of gapped vacua on $\mathbb{R}^4$ and $\mathbb{R}^3\times S^1$, are shown in \cref{tab:SU6/Z3unbroken}.

\begin{table}
\centering
\begin{tabular}{c|c|c|c}
    & $G_\mathrm{ub}$ & $\#$ $4d$ vacua & $\#$ $3d$ vacua \\[3pt]
    \hline
    $6^1$ & $\mathbb{Z}_2$ & 1 & $\mathbf{2}$ \\[3pt]
    $3^2$ & $\mathrm{SU}(2)$ & 2 & $2$ \\[3pt]
    $2^3$ & $\mathbb{Z}_2\times\mathrm{PSU}(3)$ & 3 & $\mathbf{2}\times 5 = 10$ \\[3pt]
    $1^6$ & $\mathrm{SU}(6)/\mathbb{Z}_3$ & 6 & $10$
\end{tabular}
\caption{Unbroken gauge groups and number of gapped vacua of $\mathrm{SU}(6)/\mathbb{Z}_3$ $\N=1^*$.}
\label{tab:SU6/Z3unbroken}
\end{table}

The multiplicities of the diagrams shown in \cref{fig:extremaCMA5} can be computed as before, for the global variants $(\mathrm{SU}(6)/\mathbb{Z}_3)_0$, $(\mathrm{SU}(6)/\mathbb{Z}_3)_1$, and $(\mathrm{SU}(6)/\mathbb{Z}_3)_2$, for which the Lagrangian subgroups of $\widehat{\mathbb{Z}}_6\times\mathbb{Z}_6$ are respectively:
\begin{equation}
    \begin{split}
        \mathcal L_{(\mathrm{SU}(6)/\mathbb{Z}_3)_0} &= \{(0,0),(0,2),(0,4),(3,0),(3,2),(3,4)\}\ ,\\
        \mathcal L_{(\mathrm{SU}(6)/\mathbb{Z}_3)_1} &= \{(0,0),(1,2),(2,4),(3,0),(4,2),(5,4)\}\ ,\\
        \mathcal L_{(\mathrm{SU}(6)/\mathbb{Z}_3)_2} &= \{(0,0),(2,2),(4,4),(3,0),(5,2),(1,4)\}\ .
    \end{split}
\end{equation}

One can compute the multiplicities of the diagrams of \cref{fig:extremaCMA5} as in the previous cases, and again one finds a perfect agreement with the expected number of gapped vacua for $\mathrm{SU}(6)/\mathbb{Z}_3$ $\mathcal N=1^*$ theories on $\mathbb{R}^3\times S^1$.

\subsubsection{Dualities}

The duality web of $\mathfrak{su}(6)$ $\mathcal N=1^*$ theories is shown in \cref{eq:globvarsu6}, where dashed arrows represent the transformation $S$, and plain arrows, the transformation $T$.

\bea\label{eq:globvarsu6}
\begin{tikzcd}[column sep=-15pt]
& & & \mathrm{SU}(6) \arrow[loop above, "T"] \arrow[d,leftrightarrow,"S", dashed] & & & \\
& & & \mathrm{PSU}(6)_0 \arrow[drrr, "T"] & & & \\
\mathrm{PSU}(6)_5 \arrow[urrr, "T"] \arrow[rrrrrr,leftrightarrow,"S", dashed] & & & & & & \mathrm{PSU}(6)_5 \arrow[lddddd,"T"] \\
& & (\mathrm{SU}(6)/\mathbb{Z}_3)_1 \arrow[rd, "T"] & & (\mathrm{SU}(6)/\mathbb{Z}_3)_2 \arrow[ll,"T"] & & \\
& & & (\mathrm{SU}(6)/\mathbb{Z}_3)_0 \arrow[ru, "T"] \arrow[d,leftrightarrow,"S", dashed] & & & \\
& & & (\mathrm{SU}(6)/\mathbb{Z}_2)_0 \arrow[d,leftrightarrow,"T"] & & & \\
& & & (\mathrm{SU}(6)/\mathbb{Z}_2)_1 \arrow[d,leftrightarrow,"S",dashed] & & & \\
& \mathrm{PSU}(6)_4 \arrow[luuuuu,"T"] \arrow[uuuur,leftrightarrow,"S",dashed] & & \mathrm{PSU}(6)_3 \arrow[ll,"T"] & & \mathrm{PSU}(6)_2 \arrow[ll,"T"] \arrow[uuuul,leftrightarrow,"S",dashed] &
\end{tikzcd}
\eea

One could construct a duality diagram for the massive vacua of $\mathfrak{su}(6)$ $\mathcal N=1^*$ theories on $\mathbb{R}^3\times S^1$, refining \cref{eq:globvarsu6} in the same way that \cref{fig:dualityweb} refines \cref{eq:dualityorbitsu(3)2occ}. As in the case of $\mathfrak{su}(3)$ theories, one would find that the $S$ transformation is no longer always involutive, since compactification on $S^1$ can break charge conjugation invariance. For example, in the global variant $\mathrm{SU}(6)$, the transformation $S^2$ preserves $H^{(0)}$ and $H^{(3)}$, while exchanging $H^{(1)}$ with $H^{(5)}$ and $H^{(2)}$ with $H^{(4)}$, following the notation in \cref{fig:HiggsvacuaSU6}.

As anticipated in the discussion of the previous subsections, the duality web of $\mathfrak{su}(6)$ theories supports the notion of global variants of elliptic Calogero–Moser systems proposed in \cref{def:globvarCM}. Specifically, it acts on the lattices defining the gauged translations in each Calogero–Moser global variant exactly as depicted in \cref{eq:globvarsu6}. Consequently, this structure allows for an indirect computation of the multiplicities of isolated extrema in the $A_5$ elliptic Calogero–Moser system, provided the multiplicities are known in at least one global variant---for instance, $\mathrm{SU}(6)$. This follows from the fact that the duality web in \cref{eq:globvarsu6} is connected. However, this is not always the case: the lowest-rank example where the duality diagram of global variants is disconnected occurs in $\mathfrak{su}(4)$ $\N=1^*$ theories, which we now examine.

\subsection{\texorpdfstring{$\mathfrak{su}(4)$}{su(4)} theories}

The case of $\mathfrak{su}(4)$ $\N=1^*$ theories is similar to that of $\mathfrak{su}(6)$ $\N=1^*$ theories, but for the fact that $4$ is not square-free, and the subtleties this induces.

In the standard representation of the Lie algebra $\mathfrak{su}(4)$, the Cartan subalgebra of $\mathfrak{su}(4)$ consists of all $4\times 4$ traceless real diagonal matrices. Letting $L_i$ be the linear form such that $L_i ( \textrm{diag}(h_1,h_2,h_3,h_4)) = h_i$, one takes as positive simple roots $\alpha_1 = L_1 - L_2$, $\alpha_2 = L_2 - L_3$, and $\alpha_3 = L_3 - L_4$. The corresponding fundamental weights are $\lambda_1 = L_1$, $\lambda_2 = L_1 + L_2$, and $\lambda_3 = L_1+L_2+L_3 = -L_4$. The coroots associated to the simple roots $\alpha_1$, $\alpha_2$, and $\alpha_3$, are 
\begin{equation}
    H_{\alpha_1}=\begin{pmatrix}
    1 & 0 & 0 & 0 \\ 0 & -1 & 0 & 0 \\ 0 & 0 & 0 & 0 \\0 & 0 & 0 & 0 \end{pmatrix}, \quad H_{\alpha_2}=\begin{pmatrix}
    0 & 0 & 0 & 0 \\ 0 & 1 & 0 & 0 \\ 0 & 0 & -1  &0  \\0 & 0 & 0 & 0 \end{pmatrix}, \quad H_{\alpha_3}=\begin{pmatrix}
  0&0&0&0 \\0& 0 & 0 & 0 \\ 0 & 0 & 1 & 0 \\ 0& 0 & 0 & -1 \end{pmatrix},
\end{equation}
while the fundamental coweights read 
\begin{equation}
    \chi_1=\begin{pmatrix}
    \frac{3}{4} & 0 & 0 & 0 \\ 0 &  -\frac{1}{4} & 0 & 0 \\ 0 & 0 & -\frac{1}{4} & 0 \\0 & 0 & 0 & - \frac{1}{4} \end{pmatrix},  \quad \chi_2=\begin{pmatrix}
    \frac{1}{2} & 0 & 0 & 0 \\ 0 &  \frac{1}{2} & 0 & 0 \\ 0 & 0 & - \frac{1}{2}  &0  \\0 & 0 & 0 & - \frac{1}{2} \end{pmatrix}, \quad \chi_3=\begin{pmatrix}
  \frac{1}{4}&0&0&0 \\0& \frac{1}{4} & 0 & 0 \\ 0 & 0 & \frac{1}{4} & 0 \\ 0& 0 & 0 & -\frac{3}{4} \end{pmatrix}.
\end{equation}

\begin{figure}
    \centering
    \includegraphics[width=\textwidth]{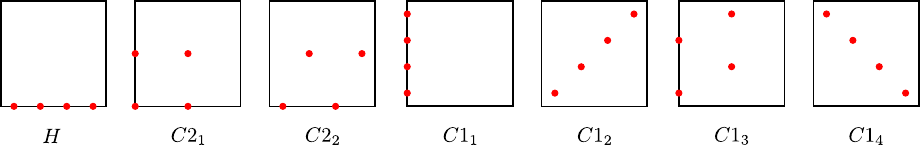}
    \caption{Isolated extrema of the $\mathfrak{su}(4)$ elliptic Calogero--Moser system}
    \label{fig:extremaCMA3}
\end{figure}

The isolated extrema of the elliptic $\mathfrak{su}(4)$ Calogero--Moser system\footnote{That is, the isolated extrema of the elliptic Calogero--Moser system of type $A_3$ with horizontal (resp. vertical) periodicities elements of the weight lattice $\Lambda_w$ (resp. coweight lattice $\Gamma_w$) of $\mathfrak{su}(4)$.} are shown in \cref{fig:extremaCMA3}, in the convention where $\sum z_i=0$.

\subsubsection{Unbroken subgroups}

Given the gauge group of an $\mathfrak{su}(4)$ $\mathcal N=1^*$ theory and the partition of $4$ corresponding to a massive vacuum of the theory, the unbroken gauge groups in these semi-classical vacua, obtained with the methods of \cref{subsec:unbrokengaugegroups}, are shown in \cref{tab:su4unbroken}.
\begin{table}
\centering
\begin{tabular}{c|c|c|c}
    & $\mathrm{SU}(4)$ & $\mathrm{SU}(4)/\mathbb{Z}_2$  & $\mathrm{PSU}(4)$\\[3pt]
    \hline
    $4^1$ & $\mathbb{Z}_4$ & $\mathbb{Z}_2$ & $\{\mathrm{id}\}$ \\[3pt]
    $2^2$ & $\mathrm{SU}(2)\rtimes \mathbb{Z}_2$ & $\mathbb{Z}_2\times\mathrm{PSU}(2)$  & $\mathrm{PSU}(2)$ \\[3pt]
    $1^4$ & $\mathrm{SU}(4)$ & $\mathrm{SU}(4)/\mathbb{Z}_2$  & $\mathrm{PSU}(4)$
\end{tabular}
\caption{Unbroken gauge groups in massive vacua of $\mathfrak{su}(4)$ $\mathcal{N}=1^*$ theories.}
\label{tab:su4unbroken}
\end{table}
The semi-direct product $\mathrm{SU}(2)\rtimes \mathbb{Z}_2$ in the second row is such that its center is $\mathbb{Z}_4$.

\subsubsection{UV and IR \texorpdfstring{$\theta$}{theta} angles}

As far as the relation between UV and IR theta angles is concerned, the only non-trivial case is the massive partition $2^2$ since there is no IR theta angle in the partition $4^1$, while the embedding $G_\mathrm{ub}<G$ is trivial in the partition $1^4$, yielding the equality $\theta_\mathrm{IR}= \theta_\mathrm{UV}$. 

In the massive partition $2^2$, the unbroken gauge algebra is $\mathfrak{g}_\mathrm{ub} = \mathfrak{su}(2)$. The embedding index of $\mathfrak{g}_\mathrm{ub}$ in $\mathfrak{su}(4)$ is 2, which implies that $\theta_\textrm{IR} = 2 \theta_\textrm{UV}$. Hence, the transformation $T$ acting on the CM diagrams corresponds to $\theta_\textrm{UV} \rightarrow \theta_\textrm{UV} + 2\pi$ or equivalently to $\theta_\textrm{IR} \rightarrow \theta_\textrm{IR} + 4\pi$. In pure $\mathcal{N}=1$ $\mathrm{SU}(2)$ gauge theory, $\theta_\textrm{IR} \rightarrow \theta_\textrm{IR} + 4\pi$ leaves the two vacua invariant. Thus, we expect to find Calogero--Moser diagrams invariant under $T=T_\mathrm{UV} = T^\mathrm{mod}$.

Conversely, the R-symmetry rotation $\theta_\textrm{IR} \rightarrow \theta_\textrm{IR} + 2\pi$ of the pure $\mathcal{N}=1$ $\mathrm{SU}(2)$ theory corresponds to a fractional shift of $\theta_\textrm{UV}$, which is not a symmetry in the UV theory. 

\subsubsection{Multiplicity of Calogero--Moser extrema}

Generalizing the cases of $\mathfrak{su}(3)$ and $\mathfrak{su}(6)$ theories discussed above, the multiplicity of the isolated extrema shown in \cref{fig:extremaCMA3} can be computed by considering equations of the form of \cref{eq:inequivalentCMSU3}. For the global variant $\mathrm{SU}(4)$ for which the vertical and horizontal gauged translations are those by coroots and coweights, respectively, the relevant equation is
\begin{equation}
    E' + h_1 \, \chi_1 + h_2\, \chi_2 + h_3\, \chi_3 + v_1\, \tau H_{\alpha_1} + v_2 \,\tau H_{\alpha_2} + v_3\, \tau H_{\alpha_3}   \approx E \subset \mathbb{C}\ , 
\end{equation}
where $h_a, v_a \in\mathbb{Z}$, $a = 1,2,3$. One finds that the diagram $H$ has multiplicity $4$, $C2_{1,2}$ multiplicity two, and $C1_{1,2,3,4}$ multiplicity one, as shown in the first row of \cref{tab:su4multiplicities}.

For the global variants $\mathrm{PSU}(4)_k$, $k=0,\dots,3$, the equation rather reads
\begin{equation}
    E' + h_1 \, H_{\alpha_1} + h_2\,H_{\alpha_2}  + h_3\, H_{\alpha_3} + v_1\, (k+\tau) \chi_1  + v_2 \,(k+\tau)  \chi_2 + v_3\, (k+\tau)  \chi_3    \approx E\ . 
\end{equation}
One finds the multiplicities in the last four rows of \cref{tab:su4multiplicities}.

\begin{table}
\centering
\begin{tabular}{c|c|c|c|c|c|c|c|c}
    & H & C2$_1$ & C2$_2$  & C1$_1$ & C1$_2$ & C1$_3$ & C1$_4$ & \textbf{Total} \\[3pt]
    \hline  
    $\mathrm{SU}(4)$ & 4 & 2 & 2 & 1 & 1 & 1 & 1 & 12 \\[3pt]
    $(\mathrm{SU}(4)/\Z_2)_0$ & 2 & 4 & 2 & 2 & 2 & 2 & 2 & 16\\[3pt]
    $(\mathrm{SU}(4)/\Z_2)_1$ & 2 & 2 & 4 & 1 & 1 & 1 & 1& 12 \\[3pt]
    $\mathrm{PSU}(4)_0$ & 1 & 2 & 1 & 4 & 1 & 2 & 1& 12 \\[3pt]
    $\mathrm{PSU}(4)_1$ & 1 & 2 & 1 & 1 & 4 & 1 & 2 & 12\\[3pt]
    $\mathrm{PSU}(4)_2$ & 1 & 2 & 1 & 2 & 1 & 4 & 1 & 12\\[3pt]
    $\mathrm{PSU}(4)_3$ & 1 & 2 & 1 & 1 & 2 & 1 & 4 & 12\\[3pt]
\end{tabular}
\caption{Multiplicities of the $\mathfrak{su}(4)$ CM extrema for each Lie group}
\label{tab:su4multiplicities}
\end{table}

Let us now turn to the global variants corresponding to the group $\mathrm{SU}(4)/\mathbb{Z}_2$. The corresponding Lagrangian subgroups of $\mathbb{Z}_4\times\mathbb{Z}_4$ are
\begin{equation}
    \begin{split}
        \mathcal{L}_{(\mathrm{SU}(4)/\mathbb{Z}_2)_0} &= \{(0,0),(2,0),(0,2),(2,2)\}\ ,\\
        \mathcal{L}_{(\mathrm{SU}(4)/\mathbb{Z}_2)_1} &= \{(0,0),(2,0),(1,2),(3,2)\}\ .
    \end{split}
\end{equation}
The lift of $\mathcal{L}_{(\mathrm{SU}(4)/\mathbb{Z}_2)_1}$ to $\mathbb{Z}\times\mathbb{Z}$ is generated by $(1,2)$ and $(0,4)$, therefore one finds the multiplicities of the diagrams in \cref{fig:extremaCMA3} by solving the equation
\begin{equation}
    E' + h_1 \,(1+2\tau) \chi_1 + h_2 \,(1+2\tau) \chi_2 + h_3 \,(1+2\tau) \chi_3 + v_1\, \tau H_{\alpha_1} + v_2 \,\tau H_{\alpha_2} + v_3\, \tau H_{\alpha_3}   \approx E\ . 
\end{equation}
The multiplicities are given in the third row of \cref{tab:su4multiplicities}.

The last global variant, $(\mathrm{SU}(4)/\mathbb{Z}_2)_0$, is of a somewhat different nature compared to the ones considered previously, notably because no vector in the lift of $\mathcal{L}_{(\mathrm{SU}(4)/\mathbb{Z}_2)_0}$ to $\mathbb{Z}\times\mathbb{Z}$, which is generated by $(2,0)$ and $(0,2)$, is primitive. Therefore, one has to consider the sublattice of the coweight lattice $\Gamma_w$ of $\mathfrak{su}(4)$, consisting of all elements of 4-ality divisible by two. The elements of this sublattice are the coweights of $\mathrm{SU}(4)/\mathbb{Z}_2$. One possible choice of generators for this sublattice is
\begin{equation}
    \{\chi_1+\chi_3,\chi_2,2\chi_3\}\ .
\end{equation}
Hence for $(\mathrm{SU}(4)/\Z_2)_0$, one needs to solve the equation 
\begin{equation}\label{eq:su4/Z20Eq}
    E' + h_1 \, (\chi_1 + \chi_3)  + h_2\,\chi_2 + h_3\, 2 \chi_3 + v_1\, \tau (\chi_1 + \chi_3)  + v_2 \,\tau  \chi_2 + v_3\, \tau 2 \chi_3 \approx E\ , 
\end{equation}
with integer coefficients, for every configuration $E'\subset\mathbb{C}$ obtained from $E$ by a translation $\chi_V+\chi_H\tau$, for every $\chi_V,\chi_H\in\Gamma_w$. The multiplicities obtained for all diagrams are shown in the second line of Tab.~\ref{tab:su4multiplicities}.

\begin{figure}
    \centering
    \includegraphics[width=\textwidth]{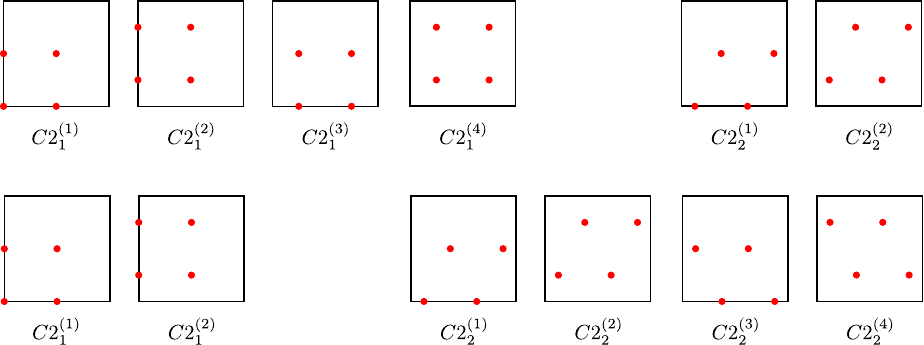}
    \caption{Isolated $C2_1$ and $C2_2$ extrema of the $(\mathrm{SU}(4)/\mathbb{Z}_2)_0$ (top) and $(\mathrm{SU}(4)/\mathbb{Z}_2)_1$ (bottom) Calogero--Moser system}
    \label{fig:extremaCMA3example}
\end{figure}

\Cref{tab:su4multiplicities} shows perfect agreement with the low-energy behavior of the unbroken gauge groups on $\mathbb{R}^3 \times S^1$ (cf. \cref{tab:su4unbroken}). An interesting case is the partition $2^2$ in $\mathrm{SU}(4)/\mathbb{Z}_2$ theories, where one can observe the consequences of the unbroken gauge group being $\mathbb{Z}_2 \times \mathrm{PSU}(2)$ rather than $\mathrm{SU}(2)$. The inequivalent isolated Calogero–Moser extrema corresponding to the diagrams $C2_1$ and $C2_2$ are shown in \cref{fig:extremaCMA3example} for $(\mathrm{SU}(4)/\mathbb{Z}_2)_0$ (top row) and $(\mathrm{SU}(4)/\mathbb{Z}_2)_1$ (bottom row).

The unbroken gauge group $G_\mathrm{ub}$ in the partition $2^2$, for the UV gauge group $\mathrm{SU}(4)/\mathbb{Z}_2$, satisfies $\mathrm{Lie}(G_\mathrm{ub}) = \mathfrak{su}(2)$ and has a center isomorphic to $\mathbb{Z}_2$. While $\mathrm{SU}(2)$ satisfies these constraints, if one had $G_\mathrm{ub} = \mathrm{SU}(2)$, the diagrams $C2_1$ and $C2_2$ in \cref{fig:extremaCMA3} would each have multiplicity two in either global variant $(\mathrm{SU}(4)/\mathbb{Z}_2)_0$ or $(\mathrm{SU}(4)/\mathbb{Z}_2)_1$. In contrast, the inequivalent isolated extrema shown in \cref{fig:extremaCMA3example} are consistent with $G_\mathrm{ub} = \mathbb{Z}_2 \times \mathrm{PSU}(2)$.

\subsubsection{Dualities}

The duality web of global variants of $\mathfrak{su}(4)$ $\N=1^*$ theories is given by 
\bea\label{eq:globalvarsu4}
\begin{tikzcd}
\raisebox{0pt}[0.5em][0pt]{$(\mathrm{SU}(4)/\bZ_2)_0 \quad\;\;$} \arrow[loop right, "{S,T}"]
\end{tikzcd} \\[-1.5em]
\begin{tikzcd}[row sep=tiny]
& & \mathrm{PSU}(4)_1 \arrow[rd, "T"] \arrow[dd, leftrightarrow, "S", dashed] \\
\mathrm{SU}(4) \arrow[loop below, "T"] \arrow[r, leftrightarrow, "S", dashed] & \mathrm{PSU}(4)_0 \arrow[ru, "T"]  & & \mathrm{PSU}(4)_2 \arrow[ld,"T"] \arrow[r, leftrightarrow, "S", dashed] & {(\mathrm{SU}(4)/\bZ_2)_1} \arrow[loop below, "T"] \\
& & \mathrm{PSU}(4)_3 \arrow[lu,"T"]
\end{tikzcd}
\eea
A novelty in the case of $\mathfrak{su}(4)$ theories, as compared to $\mathfrak{su}(3)$ and $\mathfrak{su}(6)$ theories, is that there exists a global variant, $(\mathrm{SU}(4)/\mathbb{Z}_2)_0$, which is fixed by $S$ and $T$, making it disconnected from the rest of the duality web.

This fully aligns with our findings regarding the relationship between the $3d$ vacua of $\mathfrak{su}(4)$ $\mathcal{N}=1^*$ theories and the global variants of $\mathfrak{su}(4)$ elliptic Calogero–Moser systems, where the global variant $(\mathrm{SU}(4)/\mathbb{Z}_2)_0$ is singled out with repect to the other by having a different number of 3d gapped vacua.

\begin{figure}
    \centering
    \includegraphics[width=0.92\textwidth]{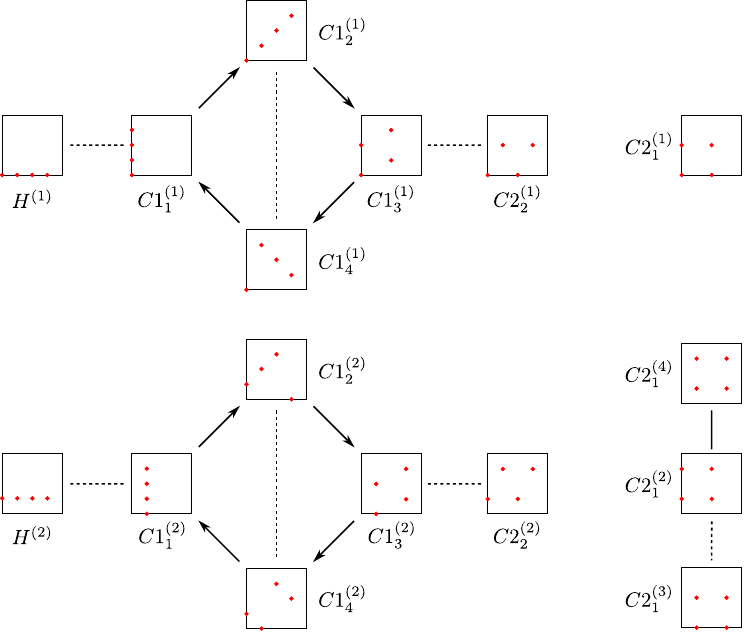}
    \caption{Duality web for the gapped vacua of $(\mathrm{SU}(4)/\Z_2)_0$ on $\mathbb{R}^3\times S^1$.}
    \label{fig:dualitywebsu4Z2}
\end{figure}

One could envision drawing a diagram similar to \cref{fig:dualityweb}; however, in the present case, one can restrict to the gapped $3d$ vacua for the global variant $(\mathrm{SU}(4)/\mathbb{Z}_2)_0$, which form their own duality orbits. This duality web of gapped $3d$ vacua of $(\mathrm{SU}(4)/\mathbb{Z}_2)_0$ refines \cref{eq:globalvarsu4} and is depicted in \cref{fig:dualitywebsu4Z2}. \Cref{fig:dualitywebsu4Z2} splits into four orbits under the duality group: two orbits of size 6 (each corresponding to the coset $\mathrm{PSL}(2,\mathbb{Z})/\Gamma_0(4)$), one orbit of size 3 (corresponding to $\mathrm{PSL}(2,\mathbb{Z})/\Gamma_0(2)$), and a singlet.

\subsection{Why naive extrema correspond to \texorpdfstring{$4d$}{4d} gapped vacua in type \texorpdfstring{$A$}{A}}

We mentioned at the end of \cref{Sec:N=1*andIRphases} that the discrepancy between the count of vacua for $\mathcal N=1^*$ theories on $\mathbb{R}^4$ and on $\mathbb{R}^3\times S^1$ has two origins.

On the one hand, a gapped vacuum on $\mathbb{R}^4$ may exhibit topological order in the form of a discrete gauge symmetry, which leads to a collection of gapped vacua on $\mathbb{R}^3\times S^1$. At the level of Calogero--Moser systems, this occurs because one considers periodicities associated with a specific global variant of the $\N=1^*$ theory rather than the standard periodicities of (twisted) elliptic Calogero--Moser systems, which instead correspond to the associated relative $\N=1^*$ theory (cf. \cref{subsec:generalization}). These gapped vacua on $\mathbb{R}^3\times S^1$ for a given $\N=1^*$ theory form equivalence classes labeled by the gapped vacua of the theory on $\mathbb{R}^4$: two gapped vacua on $\mathbb{R}^3\times S^1$ are equivalent if the corresponding isolated Calogero--Moser extrema differ only by horizontal translations by weights and vertical translations by coweights.

On the other hand, it sometimes happens that some gapless vacua of an $\N=1^*$ theory on $\mathbb{R}^4$ yield gapped vacua on $\R^3\times S^1$, typically when the unbroken gauge group in the classical $4d$ vacuum is a semi-direct product $\mathrm{SO}(2)\rtimes\mathbb{Z}_2$. Although the structure of the centralizer of $\mathfrak{sl}(2)$-triples can be determined in general using Sommers' generalization of the Bala--Carter theory \cite{Sommers:1998ago}, in type $A$ the results of \cref{subsec:unbrokengaugegroups} suffice to show that this phenomenon does not occur.

Therefore, all the gapped vacua of type $A$ $\N=1^*$ theories on $\mathbb{R}^3\times S^1$ fall into equivalence classes labeled by the gapped vacua on $\mathbb{R}^4$. These are precisely the isolated extrema of the ``naive'' Calogero--Moser systems, i.e.~those for which the horizontal periodicities form the lattice of weights and the vertical periodicities form the lattice of coweights. This correspondence explains the apparent match between the isolated extrema of the standard twisted elliptic Calogero--Moser systems of type $A$ and the gapped vacua on $\R^4$ of the corresponding $\N=1^*$ theories.

We emphasize that this result is a consequence of the specific structure of type $A$ algebras—more precisely, their theory of nilpotent orbits and $\mathfrak{sl}(2)$ triples. For other families of gauge groups, there often exist gapless $4d$ vacua leading to gapped $3d$ vacua, so one cannot expect a direct match between the isolated extrema of the standard twisted elliptic Calogero--Moser system and the $4d$ gapped vacua of the corresponding $\N=1^*$ theory. More precisely, in general, a standard twisted elliptic Calogero--Moser system has more isolated extrema than there are gapped $4d$ vacua in the corresponding $\N=1^*$ theory. We will illustrate a concrete example of this in \cref{sec: so(5)}, where we discuss $\mathfrak{so}(5)$ theories.

\subsection{Orbits of vacua}

The modular group acts simultaneously on global variants of $\mathfrak{su}(N)$ theories and on extrema of the Calogero-Moser potential. Our proposal thus gives rise to an intricate duality web, as illustrated for $N=3$ in \cref{fig:dualityweb}. In this subsection, we comment on generalizations of this web for arbitrary $N$. 

\paragraph{Count of vacua. } We first count how many vacua on $\mathbb{R}^3 \times S^1$ there are in total for algebra $\mathfrak{su}(N)$. This is equal to the sum over global variants of the number of vacua in each global variant. Since the number of vacua in a given global variant depends only on the $\mathrm{SL}(2,\mathbb{Z})$ orbit of this variant, we have 
\begin{equation}
    I^{\mathcal{N}=1^{\ast}}(N) = \sum\limits_{\textrm{Orbits of global variants}} \left[ \begin{tabular}{c}
 \textrm{Orbit } \\ \textrm{size}
    \end{tabular} \right]\times \left[ \begin{tabular}{c}
 \textrm{Number of vacua in one of  } \\ \textrm{the global variants of the orbit}
    \end{tabular} \right]
\end{equation}
Each term is known: 
\begin{itemize}
    \item The orbits of global variants are in one-to-one correspondence with divisors $\ell$ of $N$ such that $\ell^2$ divides $N$ \cite{Aharony:2013hda}. In addition, a distinguished global variant in the orbit labelled by $\ell$ is $(\mathrm{SU}(N)/\mathbb{Z}_{\ell})_0$. 
    \item The size of the orbit labelled by $\ell$ is the index of $\Gamma_0(N/\ell^2)$. This index is known as the Dedekind $\psi$-function: \begin{equation}
    \psi (n) = n \prod\limits_{p | n} \left( 1 + \frac{1}{p}  \right) = [\mathrm{SL}(2,\mathbb{Z}) : \Gamma_0 (n) ] \ . 
\end{equation}
\item The number of vacua on $\mathbb{R}^3 \times S^1$ in the global variant $(\mathrm{SU}(N)/\mathbb{Z}_{\ell})_0$ is \cite{Bourget:2016yhy}: 
\begin{equation}
    I^{\mathcal{N}=1^{\ast}} [(\mathrm{SU}(N)/\mathbb{Z}_{\ell})_0] = N \sum\limits_{d|N} \mathrm{gcd} \left( d,\ell ,  \frac{N}{d} , \frac{N}{\ell} \right) \ . 
\end{equation}
\end{itemize}
From this we conclude that the total number of vacua is\footnote{We conjecture, based on numerical coincidence, that this can be rewritten as 
\begin{equation}
 \frac{I^{\mathcal{N}=1^{\ast}}(N)}{N} = (\sigma_1 * \psi) (N) =  \sum\limits_{d | N} \sigma_1 (d) \psi (N/d)  
\end{equation} 
where 
\begin{equation}
    \sigma_1 (n) = \sum\limits_{d' | n} d' 
\end{equation}
is the sum of the divisors of $n$, and $*$ denotes the convolution product. } 
\begin{equation}
\label{eq:totalCountVacua}
    I^{\mathcal{N}=1^{\ast}}(N) = \sum\limits_{\ell \text{ such that } \ell^2 | N} \psi \left( \frac{N}{\ell ^2} \right)  I^{\mathcal{N}=1^{\ast}} [(\mathrm{SU}(N)/\mathbb{Z}_{\ell})_0]  \ . 
\end{equation}

\paragraph{Modular Action. }  
There is an action of the modular group on the $I^{\mathcal{N}=1^{\ast}}(N)$ vacua by permutation. This can be computed explicitly for low $N$. The duality graphs are shown in \cref{app:dualityGraphs}.\footnote{It is interesting to compare the structure and automorphisms of these duality graphs with the ``dessins'' corresponding to the cosets $\mathrm{PSL}(2,\mathbb{Z})/\Gamma_0(N)$ \cite{Tatitscheff:2018aht}.} 

We observe that many (but not all) orbits have sizes which are integer multiples of $\psi (N)$. So we record the size of orbits in terms of these multiples. The results for $1 \leq N \leq 12$ are shown in Table \ref{tab:orbit_sizes}, in which the last column lists non-isomorphic orbits and their multiplicities. Note that it can happen that non-isomorphic orbits can have the same cardinality. For instance, for $N=6$ we have $9$ orbits, with cardinality $72,48,36,36,24,24,24,12,12$. Two of the three size 24 orbits are isomorphic while the other one is not. Since $\psi (6)=12$, we indicate the orbit structure as $[6,4,3^2,2_{I},2_{II}^2,1^2]$.  

We observe that the largest orbit always has size $N \psi (N)$, and for $N$ prime, there seems to always be exactly 3 orbits, of size $N \psi (N)$, $(N-1) \psi (N)$ and $\psi (N)$.  It would be interesting to prove these statements and explore further the structure for non prime $N$. 

\begin{table}
\begin{equation*}
    \begin{array}{|c|c|c|c|c|} \hline 
  N &   I^{\mathcal{N}=1^{\ast}}(N) & \frac{I^{\mathcal{N}=1^{\ast}}(N)}{N}  & \# \textrm{Orbits}&\frac{\textrm{Orbit Sizes}}{\psi (N)}\\ \hline 
 1 & 1 & 1 & 1 &  [1] \\
 2 & 12 & 6 & 3& [2,1^2] \\
 3 & 24 & 8 & 3& [3,2,1] \\
 4 & 88 & 22 & 12& [4,2,1^8, \frac{1}{2}, \frac{1}{6}] \\
 5 & 60 & 12 & 3& [5,4,1] \\
 6 & 288 & 48 & 9& [6,4,3^2,2_{I},2_{II}^2,1^2]\\
 7 & 112 & 16 & 3& [7,6,1] \\
 8 & 528 & 66 & 28& [8,4,2_I^{6},2_{II}^{4},1_I,1_{II}^8,(\frac{1}{2})_I,(\frac{1}{2})_{II}^4,(\frac{1}{4})^2] \\
 9 & 369 & 41 & 14& [9,6,2^5,1^5,\frac{2}{3},\frac{1}{12}]\\
 10 & 720 & 72 & 9& [10,8,5^2,4^2,2,1^2] \\ 
 11 &  264 &  24 & 3&  [11,10,1]  \\ 
 12 & 2112  &  176 &36&  [12,8,6,4_I,4_{II},3^8,2^9,\frac{3}{2},1^8_I,1_{II},(\frac{1}{2})_I,(\frac{1}{2})_{II},\frac{1}{3},\frac{1}{6}]   \\ \hline
\end{array}
\end{equation*}
    \caption{For each $N$, we give the total number of vacua $I^{\mathcal{N}=1^{\ast}}(N)$ for $\mathfrak{su}(N)$ theories in all possible global forms on $\mathbb{R}^3 \times S^1$. This number is partitioned into the size of orbits under the action of $\mathrm{SL}(2,\mathbb{Z})$. This partition is given in terms of multiples of $\psi (N)$.  }
    \label{tab:orbit_sizes}
\end{table}

\section{Another case study: \texorpdfstring{$\mathfrak{so}(5)$}{so(5)} theories}\label{sec: so(5)}

In this section we present a comprehensive analysis of the low energy dynamics of $\cN=1^*$ theory with gauge algebra $B_2$, i.e.~$\mathfrak{so}(5)$. This is a special instance of a non-simply laced Lie algebra which is yet mapped to itself under Langlands duality, as a consequence of the exceptional isomorphism $\mathfrak{sp}(4) \cong \mathfrak{so}(5)$ (other examples include $G_2$ and $F_4$). Furthermore, the structure of vacua on $\bR^3\times S^1$ displays marked differences with respect to the general situation that occurs for theories based on type-$A$ algebras. Most notably, we observe the emergence of an additional gapped vacuum from compactification of a four dimensional gapless one. Previous studies of the $\mathfrak{so}(5)$ theories can be found in \cite{Bourget:2015cza, Bourget:2015upj}.

In the following, we will introduce some aspects concerning the Lie algebra $\mathfrak{so}(5)$ which will turn useful to study the solutions of the associated CM system. Then we will provide a description of the physics occurring in each type of vacua, together with some comments on their uplift to four dimensions.      

\subsection{(Co-)roots and (co-)weights of the \texorpdfstring{$\mathfrak{so}(5)$}{so(5)} Lie algebra}

Following the conventions of \cite{Fulton:1991rep}, the complex simple Lie algebra $\mathfrak{so}(5,\mathbb{C})$ is the Lie algebra of $5\times 5$ matrices with complex entries  
\begin{equation}\label{eq:so5def}
    \left\{\left.\begin{pmatrix}
        A & B & E \\
        C & D & F \\
        G & H & 0
    \end{pmatrix} \right\vert B^T = -B,\; C^T = -C,\; A=-D^T,\; E = -H^T,\; F=-G^T\right\}\ ,
\end{equation}
where $A,B,C,D$ are $(2\times 2)$ matrices, while $E,F$ and $G,H$ are respectively $(2\times 1)$ and $(1\times 2)$ matrices. This form is obtained via the requirement 
     \begin{equation}
 	X^T \cdot M + M \cdot X = 0 \ , \qquad \textrm{with} \qquad M = \begin{pmatrix}
 		0 & \mathrm{Id}_2 & 0 \\
 		\mathrm{Id}_2 & 0 & 0 \\
 		0 & 0 & 1
 	\end{pmatrix} \ . 
 \end{equation}
This particular choice of non-degenerate symmetric bilinear form of $\mathfrak{so}(5,\mathbb{C})$ is such that one can take as Cartan subalgebra $\fh$ the algebra of diagonal matrices satisfying the additional defining constraints of \cref{eq:so5def}.\footnote{The more standard choice of symmetric bilinear form on $\mathbb{C}^5$, i.e.~$M = \mathrm{Id}_5$, yields the standard presentation of $\mathfrak{so}(5,\mathbb{C})$ as the algebra of $5\times 5$ anti-symmetric matrices. In this case, one cannot take $\mathfrak{h}$ to consist of diagonal matrices.} We parametrize the Cartan subalgebra $\mathfrak{h}$ by traceless diagonal matrices of the form ${\rm diag}(a,b,-a,-b,0) = a h_1 + b h_2$.
The dual $\mathfrak{h}^*$ is generated by the linear forms $L_1$, $L_2$, satisfying $L_i(h_j) = \delta_{ij}$.
The roots of $\mathfrak{so}(5)$ are $\pm L_1$, $\pm L_2$, $\pm L_1\pm L_2$. The set of positive roots is taken as 
\begin{equation}
    \Delta^+(\mathfrak{so}(5)) = \{L_1,L_2,L_1+L_2,L_1-L_2\}\ ,
\end{equation} 
and the corresponding simple roots identified with $\alpha_1 = L_2$ (short) and $\alpha_2 = L_1-L_2$ (long). The norm on $\mathfrak{h}^*$ is induced by the Killing form and is normalized such that $|\alpha_2|^2 = 2 |\alpha_1|^2 = 2$. The Weyl group acting on the root system is of the form $ W_{\mathfrak{so}(5)} = S_2 \ltimes (\bZ_2)^2 \cong \mathrm{D}_4$.  In turn, the coroots associated to simple roots are 
\begin{equation}
	 H_{\alpha_1}=2 h_2 = \left(\begin{array}{ccccc}
		0 & 0 & 0 & 0 & 0 \\ 0 & 2 & 0 & 0 & 0\\ 0 & 0 & 0& 0 &0 \\
		0 & 0 & 0 & -2 & 0 \\ 0 & 0 & 0 & 0 & 0\end{array}\right)\ , \quad 
			H_{\alpha_2}=h_1 - h_2 = \left(\begin{array}{ccccc}
			1 & 0 & 0 & 0 & 0 \\ 0 & -1 & 0 & 0 & 0\\ 0 & 0 & -1 & 0 &0 \\
			0 & 0 & 0 & 1 & 0 \\ 0 & 0 & 0 & 0 & 0\end{array}\right) \ ,
\end{equation}  
and the fundamental weights, defined as $\lambda_i(H_{\alpha_j}) = \delta_{ij}$, given by 
\begin{equation}
	\lambda_1 =  \frac{1}{2} L_1 + \frac{1}{2} L_2  \ , \quad \lambda_2 =L_1\ . 
\end{equation}
On the other hand, the fundamental coweights are dual to simple roots, i.e.~$\alpha_i (\chi_j) = \delta_{ij}$, hence obtaining
\begin{equation}
	\chi_1=h_1 + h_2 = \left(\begin{array}{ccccc}
		1 & 0 & 0 & 0 & 0 \\ 0 & 1 & 0 & 0 & 0\\ 0 & 0 & -1& 0 &0 \\
		0 & 0 & 0 & -1 & 0 \\ 0 & 0 & 0 & 0 & 0\end{array}\right)\ , \quad \chi_2 =	h_1 = \left(\begin{array}{ccccc}
		1 & 0 & 0 & 0 & 0 \\ 0 & 0 & 0 & 0 & 0\\ 0 & 0 & -1 & 0 &0 \\
		0 & 0 & 0 & 0 & 0 \\ 0 & 0 & 0 & 0 & 0\end{array}\right)\ .
\end{equation}
We summarize here the lattices involved, regarded as lattices in $\bR^2$:
\begin{align}
&\Lambda_w = \mathbb{Z} \langle (1/2,1/2), (1,0)\rangle\  ,  &\Lambda_r =\mathbb{Z} \langle (0,1),(1,-1) \rangle\ , \\
&\Gamma_w = \mathbb{Z} \langle (1,1),(1,0) \rangle\  ,  &\Gamma_r =\mathbb{Z} \langle (0,2) (1,-1) \rangle\ .
\end{align}
These expressions make it manifest that the isomorphism $\mathfrak{h} \cong \mathfrak{h}^*$ provided by the Killing form does not map all roots to coroots, and all weights to coweights (this never happens for non-simply laced Lie algebras, see \cref{App:Lie}). 

\subsection{Solutions to the \texorpdfstring{$F$}{F}-term equations and duality}

As explained around \cref{eq: type B partitions}, the solutions to the $F$-term equations are classified in terms of certain partitions. For the case at hand, the partitions and the corresponding unbroken gauge groups for each global variant of $\mathfrak{so}(5)$ are summarized in the Table \ref{tab: so(5) partitions}, where we recall that $\mathrm{O}(2)\cong\mathrm{SO}(2)\rtimes \bZ_2$ and $\mathrm{Pin}(2)\cong\mathrm{Spin}(2)\rtimes \bZ_2$ is its double cover. 

\begin{table} 
\centering
\begin{tabular}{c|c|c|c}
$\empty  $   & $\mathrm{Spin}(5)$ & $\mathrm{SO}(5)_+$ & $\mathrm{SO}(5)_-$ \\[3pt]
    \hline
 $\{5 \}$ & $\bZ_2$   &  $\{1\}$ &  $\{1\}$  \\[3pt]
    
$ \{3,1^2\}$ & $\mathrm{Pin}(2)$ & $\mathrm{O}(2)$ & $\mathrm{O}(2)$\\[3pt]
    
 $\{2^2,1\}$ & $\bZ_2\times \mathrm{SU}(2)$ & $\mathrm{SU}(2)$ & $\mathrm{SU}(2)$ \\[3pt]
    
 $\{1^5\}$   & $\mathrm{Spin}(5)$ & $\mathrm{SO}(5)$ & $\mathrm{SO}(5)$
\end{tabular}
\caption{Preserved gauge groups for each partition associated to $\mathfrak{g}=\mathfrak{so}(5)$ according to the choice of global variant.}
\label{tab: so(5) partitions}
\end{table}

The unbroken gauge algebra in the partition $\{5\}$ is trivial, with the unbroken gauge group identified with the center of the UV gauge group. On the contrary, the partition $\{1^5\}$ preserves the gauge algebra $\mathfrak{so}(5)$ and the global form of the unbroken gauge group coincides with the one of the UV gauge group.

For the partition $\{2^2,1\}$, the methods of \cite{collingwood1993nilpotent} (cf.~\cref{App:Lie}) yield the following representative for the orbit of $\mathfrak{sl}(2)$-triples corresponding to this partition:
\begin{equation}
    H = \begin{pmatrix}
        1 & 0 & 0 & 0 & 0 \\
        0 & -1 & 0 & 0 & 0 \\
        0 & 0 & -1 & 0 & 0 \\
        0 & 0 & 0 & 1 & 0 \\
        0 & 0 & 0 & 0 & 0 \\
    \end{pmatrix}, \quad 
    X = \begin{pmatrix}
        0 & 1 & 0 & 0 & 0 \\
        0 & 0 & 0 & 0 & 0 \\
        0 & 0 & 0 & 0 & 0 \\
        0 & 0 & -1 & 0 & 0 \\
        0 & 0 & 0 & 0 & 0 \\
    \end{pmatrix},\quad
    Y = \begin{pmatrix}
        0 & 0 & 0 & 0 & 0 \\
        1 & 0 & 0 & 0 & 0 \\
        0 & 0 & 0 & -1 & 0 \\
        0 & 0 & 0 & 0 & 0 \\
        0 & 0 & 0 & 0 & 0 \\
    \end{pmatrix}.
\end{equation}
The centralizer of this $\mathfrak{sl}(2)$-triple in $\mathfrak{so}(5)$ consists of all matrices of the form
\begin{equation}\label{eq:SU2inSO5}
    M = \begin{pmatrix}
        a & 0 & 0 & b & 0 \\
        0 & a & -b & 0 & 0 \\
        0 & c & -a & 0 & 0 \\
        -c & 0 & 0 & -a & 0 \\
        0 & 0 & 0 & 0 & 0
    \end{pmatrix}\ ,
\end{equation}
where $a,b,c\in\mathbb{C}$. The unbroken gauge algebra in the partition $\{2^2,1\}$ is then isomorphic to $\mathfrak{su}(2)$. Correspondingly, the centralizer in $\mathrm{SO}(5)$ is $\mathrm{Sp}(2)\cong\mathrm{SU}(2)$, as follows from the Springer--Steinberg theorem (\cref{thm:SpringerSteinberg}). When $G_\mathrm{UV}=\mathrm{Spin}(5)$ instead, the short exact sequence
\begin{equation}
    1\longrightarrow \mathbb{Z}_2 \longrightarrow \mathrm{Spin}(5) \longrightarrow \mathrm{SO}(5) \longrightarrow 1
\end{equation}
implies that the unbroken gauge group $G_\mathrm{ub}$ is a double cover of $\mathrm{SU}(2)$:
\begin{equation}
    1\longrightarrow \mathbb{Z}_2 \longrightarrow G_\mathrm{ub} \longrightarrow \mathrm{SU}(2) \longrightarrow 1\ ,
\end{equation}
i.e.~$G_\mathrm{ub}\cong \mathbb{Z}_2\times\mathrm{SU}(2)$, because $\mathrm{SU}(2)$ is simply-connected. Therefore, we conclude that in both cases, there is an emergent $\bZ_2^{(1)}$ 1-form symmetry at low energies, associated to the center of $\mathrm{SU}(2)$. In the two gapped vacua that ensue, electrically charged particles under $\mathrm{SU}(2)$ are confined.\footnote{In contrast, electrically charged particles with respect to the UV gauge group are deconfined for all the global variants.} As we discuss later, this emergent set of charges is needed in order to characterize these particular massive vacua in terms of condensation of line operators. This phenomenon has no analog for type $A$ systems.

Finally, in the partition $\{3,1^2\}$, the unbroken gauge algebra is $\mathfrak{so}(2)$, hence there is a massless photon in $\bR^4$. The computation of $G_\mathrm{ub}$, depending on the $\mathfrak{so}(5)$ global variant at hand, can be achieved with the methods of \cite{collingwood1993nilpotent}. An $\mathfrak{sl}(2)$-triple corresponding to the partition $\{3,1^2\}$ can be chosen as: 
\begin{equation}
    H = \begin{pmatrix}
        2 & 0 & 0 & 0 & 0 \\
        0 & 0 & 0 & 0 & 0 \\
        0 & 0 & -2 & 0 & 0 \\
        0 & 0 & 0 & 0 & 0 \\
        0 & 0 & 0 & 0 & 0 \\
    \end{pmatrix}, \quad 
    X = \begin{pmatrix}
        0 & 0 & 0 & 0 & 1 \\
        0 & 0 & 0 & 0 & 0 \\
        0 & 0 & 0 & 0 & 0 \\
        0 & 0 & 0 & 0 & 0 \\
        0 & 0 & -1 & 0 & 0 \\
    \end{pmatrix}, \quad 
    Y = \begin{pmatrix}
        0 & 0 & 0 & 0 & 0 \\
        0 & 0 & 0 & 0 & 0 \\
        0 & 0 & 0 & 0 & -2 \\
        0 & 0 & 0 & 0 & 0 \\
        2 & 0 & 0 & 0 & 0 \\
    \end{pmatrix}.
\end{equation}
The stabilizer of this $\mathfrak{sl}_2$-triple in $\mathfrak{so}_5$ consists of all matrices of the form
\begin{equation}
    M = \begin{pmatrix}
        0 & 0 & 0 & 0 & 0 \\
        0 & a & 0 & 0 & 0 \\
        0 & 0 & 0 & 0 & 0 \\
        0 & 0 & 0 & -a & 0 \\
        0 & 0 & 0 & 0 & 0 
    \end{pmatrix},
\end{equation}
i.e. $\mathfrak{g}_\mathrm{ub}\simeq\mathfrak{so}(2)$. One readily finds that $G_\mathrm{ub}\simeq \mathrm{O}(2)\simeq\mathrm{SO}(2)\rtimes\mathbb{Z}_2$ when $G=\mathrm{SO}(5)$, while the centralizer in $\mathrm{Spin}(5)$ is the double cover $\mathrm{Pin}(2)$ of $\mathrm{O}(2)$. In general, gapless IR phases are not the main focus of this work. The reason to include it in our analysis is that it will give rise to a gapped vacuum in $\bR^3\times S^1$ by means of the mechanism described in section \ref{Sec:N=1*andIRphases}. Therefore, there should be a corresponding isolated extremum in the $\mathfrak{so}(5)$ CM systems, and we will show below that this is indeed the case.  

Due to the fact that the long and short roots of $\mathfrak{so}(5)$ have a ratio of the squared norms that is two, the electric-magnetic duality group acting on this class of theories is the Hecke group with elliptic elements of order four, with its order-two generator acting as\footnote{In general, $S$-duality reads $S_\nu : \tau \rightarrow - \frac{1}{\nu (\alpha_s)\tau}$, with $\nu(\alpha_s)$ defined in \cref{eq:indexCM}.}
\begin{equation}\label{eq: S2}
    S_2 \, : \, \tau \rightarrow - \frac{1}{2  \tau} \ .
\end{equation}

From the duality web of \cref{eq: so(5) S-duality}, we expect gapped vacua in the ${\rm Spin}(5)$ theory to be mapped to gapped vacua in the $\mathrm{SO}(5)_+$ theory, and vice-versa, under the action of $S_2$. Physically, this stems from the fact that $S_2$ exchanges the electric and magnetic line operators. Furthermore, the set of gapped vacua of the $\mathrm{SO}(5)_-$ global variant is fixed by $S_2$. On $\bR^3\times S^1$, these relations are realized by the modular $S_2$ transformation acting within the solutions of the CM system, as we will show below.

The action of $T$ is trivial at high energies, due to the absence of a mixed anomaly between the UV $\bZ_2^{(1)}$ 1-form symmetry and the shift $\theta_{UV}\to \theta_{UV}+2\pi$ \cite{Cordova:2019uob}.\footnote{Here it is important to clarify that we always take the spacetime manifold to be spin, as required by the presence of fermions in $\N=1^*$ theories.} In other words, theories based on the Lie algebra $\mathfrak{so}(5)$ have a trivial Witten effect. However, at low energies $T$ might act non-trivially as the dynamics effectively reduces to that of $\cN=1$ SYM theories, where there might be an emergent $R$-symmetry. In the gapped vacua of $\cN=1$ SYM, the emergent $R$-symmetry is always spontaneously broken by the condensation of the gaugino bilinear. As such, one expects the shift $\theta_\mathrm{IR}\to \theta_\mathrm{IR}+ 2\pi$ to shuffle these vacua. This is the case in the partitions $\{1^5\}$ and $\{2^2,1\}$, where the unbroken gauge algebras are $\mathfrak{so}(5)$ and $\mathfrak{su}(2)$, respectively. We will describe these two cases separately.

The $\cN=1$ $\mathfrak{g}=\mathfrak{so}(5)$ SYM theory arising at intermediate energies in the partition $\{1^5\}$ has a $\bZ_6$ R-symmetry, with its $\bZ_2$ subgroup acting as fermion number. Since this solution to the $F$-terms does not involve any Higgsing pattern, the embedding of the IR gauge group in the UV one is trivial, hence $\theta_\mathrm{UV}=\theta_\mathrm{IR}$ (see section \ref{subsec:TUVandTIR}). Flowing to the deep IR, the theory displays three gapped vacua characterized by a condensed gaugino bilinear, signaling the spontaneous breakdown of $\bZ_3 = \bZ_6/\bZ_2$. Due to the absence of Witten effect, the condensed line operator is the same  for each confining vacua, but $\theta_\mathrm{IR}\to \theta_\mathrm{IR}+2\pi$ induces a $\bZ_3$ phase shift on the gaugino condensate. Consequently, these three vacua must be permuted by the action of $T$. When studying the solutions of the CM system, we will explicitly verify the presence of solutions furnishing triplets under the action of the modular $T$-transformation. 

The case of the partition $\{2^2,1\}$ is slightly more involved. On the one hand, the emergent $R$-symmetry now corresponds to $\bZ_4$, with its $\bZ_2$ subgroup identified with $(-1)^F$. In addition, one can show that even if the Higgsging pattern is non-trivial, $\theta_\mathrm{UV}=\theta_\mathrm{IR}$ again holds. An explicit proof of this is provided in \cref{app: proof su(2) theta}. As before, the unbroken $\mathrm{SU}(2)$ gauge group introduces an emergent $\bZ_2^{(1)}$ 1-form symmetry, together with quantum numbers associated with the electric and magnetic charges of its line operators. Furthermore, there is a non-trivial Witten effect acting on these emergent charges when $\theta_\mathrm{IR}\to \theta_\mathrm{IR}+2\pi$. Pure $\cN=1$ $\mathrm{SU}(2)$ SYM theory develops two gapped vacua due to the $R$-symmetry breaking $\bZ_4\to \bZ_2$, which are exchanged by $2\pi$ shifts of $\theta_\mathrm{IR}$. In terms of the isolated extrema of the CM potential, we will see that this phenomenon manifests itself as the occurrence of a pair of solutions exchanged by the modular $T$-transformation. Lastly, note that in terms of the emergent 1-form symmetry charges, $T$ also trades the condensed monopole line operators for dyonic ones. This distinction is nevertheless not meaningful from the perspective of the high energy theory.   

We conclude this analysis with a comment about the four dimensional gapless vacuum arising from the partition $\{3,1^2\}$ in $\mathfrak{g}=\mathfrak{so}(5)$ $\cN=1^*$ theories. This is the only instance of a ground state featuring gapless degrees of freedom. Consequently, the action of duality should map it back to itself (up to a switch of global variant). Upon compactification on $\bR^3\times S^1$, this vacuum gives rise to an isolated extremum of the CM potential, invariant under both $S_2$ and $T$.

\subsection{CM system and its solutions}\label{Sec:so5CMsolutions}

We now turn to the analysis of the isolated minima of the CM potential associated to this theory or, equivalently, the gapped vacua in $\bR^3\times S^1$. Since $|\alpha_1|^2=1$ and $|\alpha_2|^2=2$, the potential of the twisted elliptic Calogero-Moser system of type $B_2$ reads:
\begin{equation}\label{eq:PotCMso5}
    V^{\mathrm{tw}}_{\mathfrak{so}(5)}(Z;\tau) = \wp(z_1 + z_2) +  \wp(z_1 - z_2) + \frac{1}{2} \left[ \wp_2(z_1) + \wp_2(z_2)  \right]\ , 
\end{equation}
where the twisted Weierstrass function is $\wp_2(z,\tau ) = \wp(z,\tau )  + \wp(z + \frac{1}{2},\tau )$. This twisted elliptic integrable system is indeed Langlands self-dual \cite{Bourget:2015cza}, with the modular $S_2$ transformation acting on $\tau$ as in \cref{eq: S2}, accompanied by $z_i\to z_i/2\tau$ and a shift that we will detail shortly. In addition, the potential $V^{\mathrm{tw}}_{\mathfrak{so}(5)}$ is invariant under permutations $z_1 \leftrightarrow z_2$ and sign flips $z_i \rightarrow -z_i$, which together correspond to the action of the Weyl group $ W_{\mathfrak{so}(5)} = \mathrm{D}_4$.

\begin{figure}
\captionsetup[subfloat]{labelformat=empty} 
\centering
\begin{minipage}{.32\linewidth}
    \centering
    \subfloat[][Extremum 1]{\label{fig:V1}\includegraphics[width=0.85\linewidth]{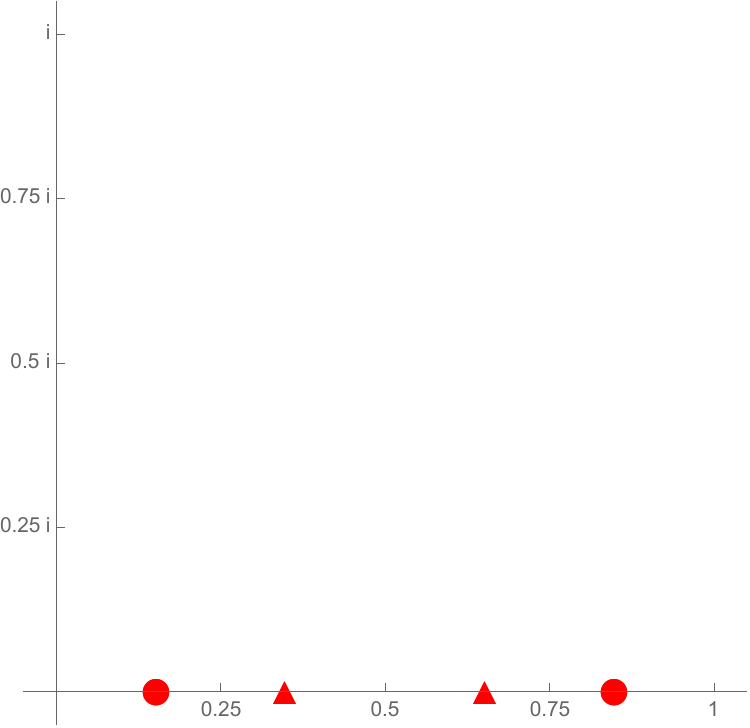}}
\end{minipage}
\begin{minipage}{.32\linewidth}
    \centering
    \subfloat[][Extremum 2]{\label{fig:V2}\includegraphics[width=0.85\linewidth]{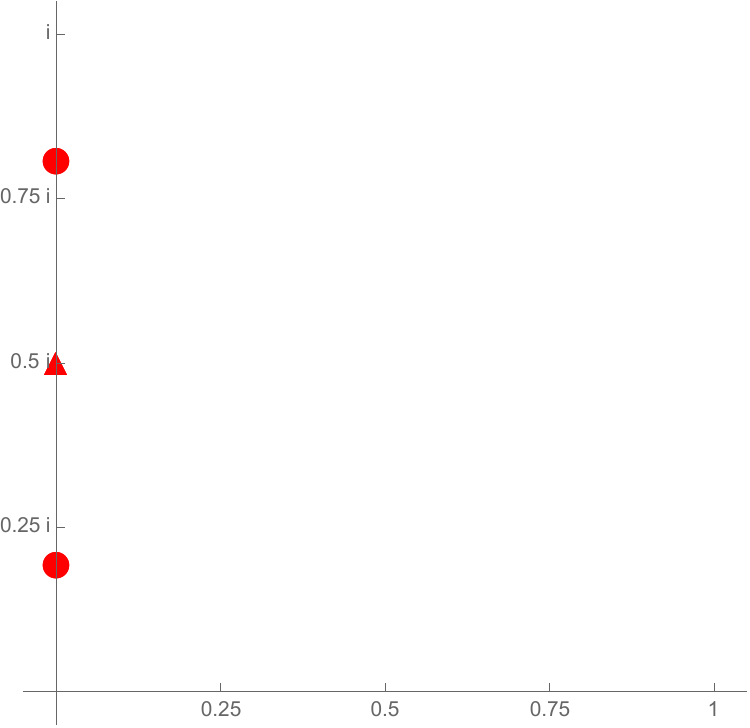}}
\end{minipage}
\begin{minipage}{.32\linewidth}
    \centering
    \subfloat[][Extremum 3]{\label{fig:V3}\includegraphics[width=0.85\linewidth]{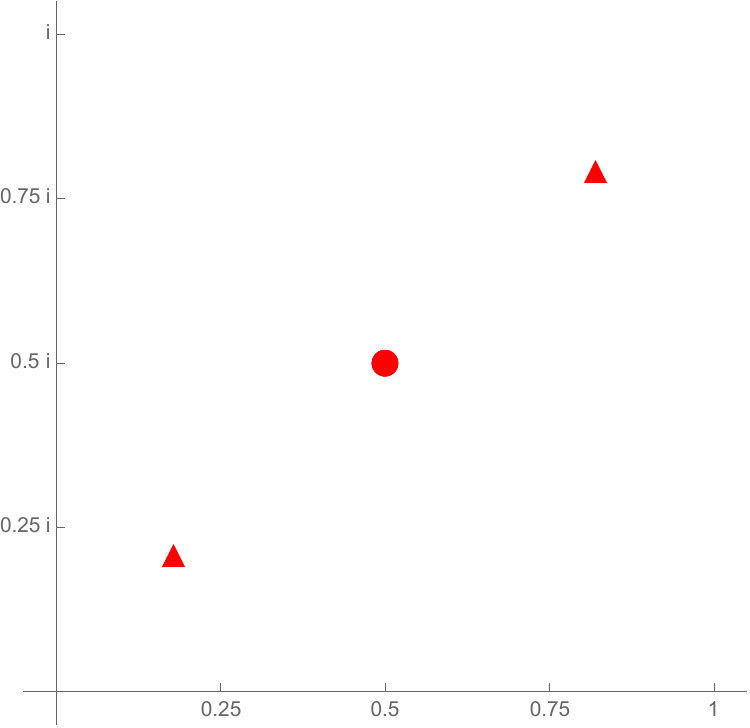}}
\end{minipage}
\par\medskip \medskip \medskip 
\begin{minipage}{.32\linewidth}
    \centering
    \subfloat[][Extremum 4]{\label{fig:V4}\includegraphics[width=0.85\linewidth]{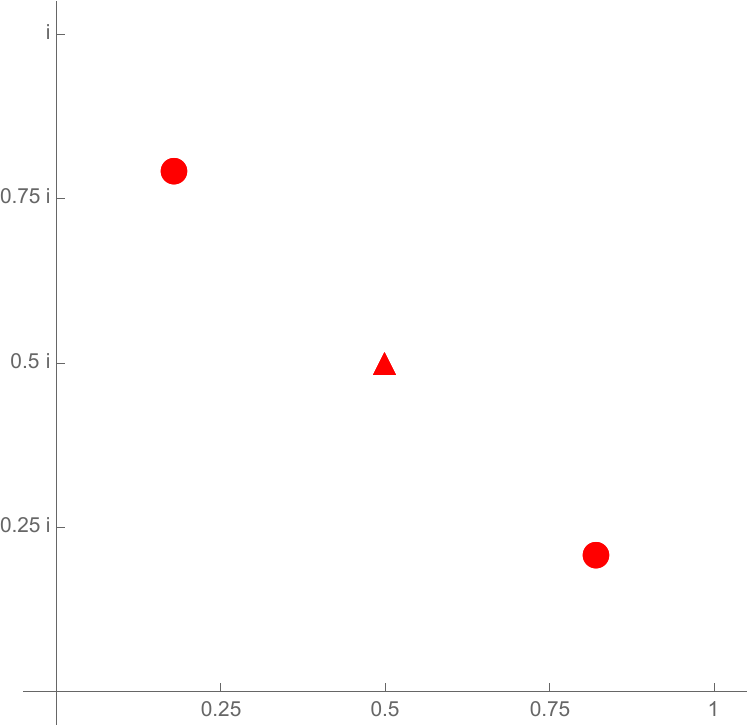}}
\end{minipage}
\begin{minipage}{.32\linewidth}
    \centering
    \subfloat[][Extremum 5]{\label{fig:V5}\includegraphics[width=0.85\linewidth]{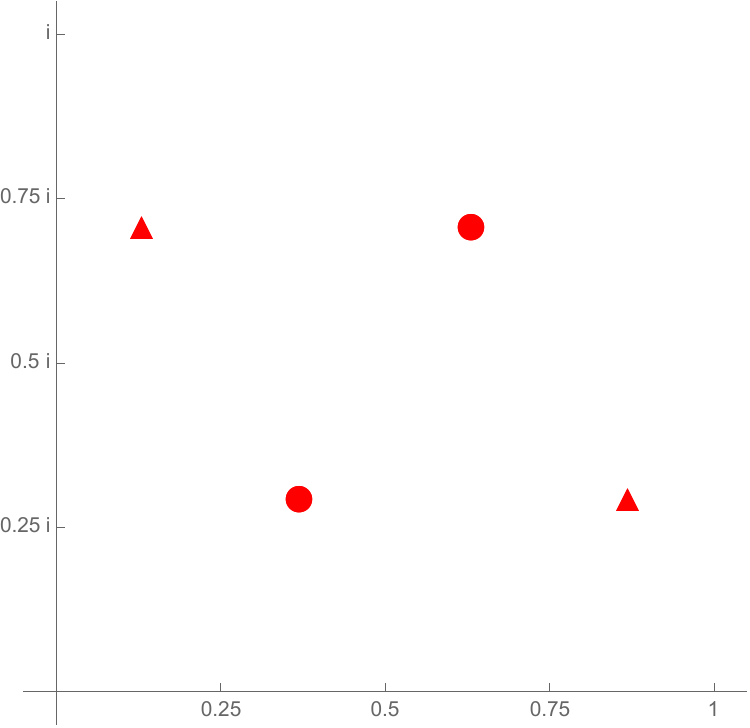}}
\end{minipage}
\begin{minipage}{.32\linewidth}
    \centering
    \subfloat[][Extremum 6]{\label{fig:V6}\includegraphics[width=0.85\linewidth]{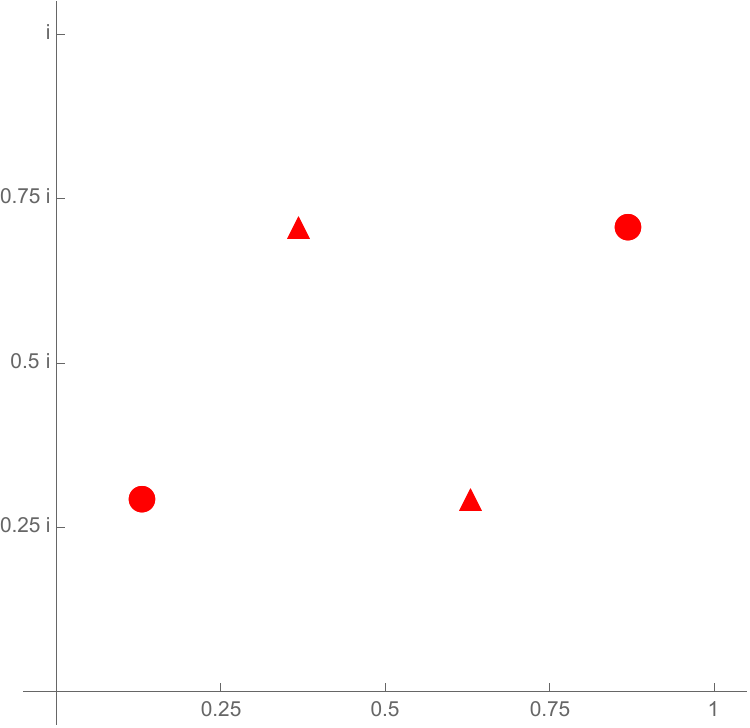}}
\end{minipage}
\par\medskip \medskip \medskip 
\centering
    \subfloat[][Extremum 7]{\label{fig:V7}\includegraphics[width=0.28\linewidth]{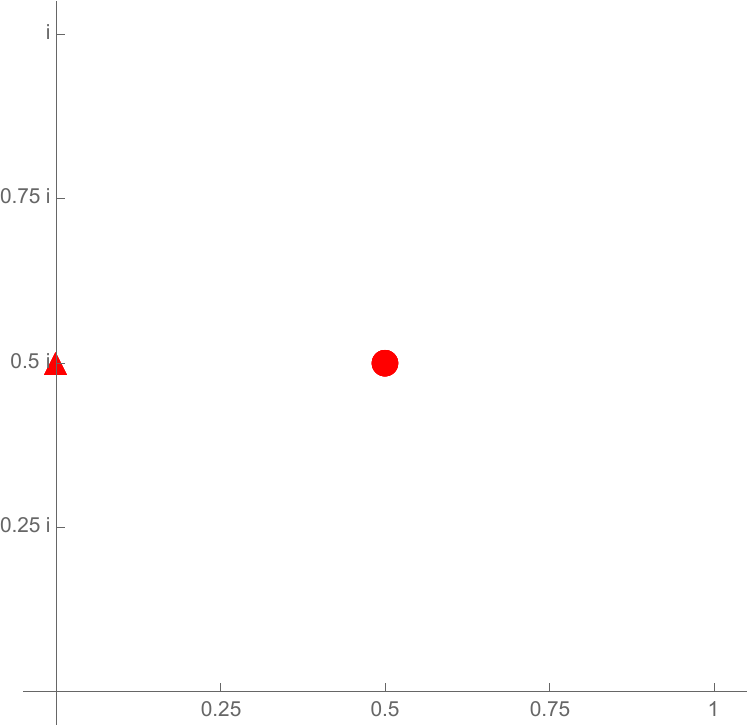}}
\caption{Isolated extrema of the $\mathfrak{so}(5)$ Calogero-Moser system at $\tau =i$. In each diagram, any two points represented with the same symbol are mapped one to another under sign flip, and any pair (circle, triangle) provides a representation of the extremum at hand.}
\label{fig:so5ExtremaAll}
\end{figure}

There are seven ``naive'' isolated minima of \cref{eq:PotCMso5} displayed in \cref{fig:so5ExtremaAll}. 
Note that to underscore the equivalence under the Weyl group transformations, for each extremum we display also the images under the sign flips, translated to the fundamental cell. In order to be consistent with the previous sections, these extrema are displayed at $\tau=i$.\footnote{Note, however, that $\tau=i$ is not the fixed point of $S_2$. The choice is made for the sake of simplicity, so that in this case again we display the extrema within a square fundamental cell.}


As emphasized throughout this article, the key distinction between the global variants associated to the Lie algebra arises by imposing specific periodicities for the potential. This yields the correct multiplicities in $\bR^3\times S^1$. The potential \cref{eq:PotCMso5} is manifestly invariant under (vertical) horizontal translations by (coweights) weights. However, this does not mean that any two configurations related by such a shift must be considered physically equivalent. Fixing the global variant of the CM system amounts to selecting which shifts become actual redundancies of the theory, while the others correspond to global symmetries rather than gauged ones.

The gauged translations corresponding to each global variant are:
\bea\label{eq: so(5) periodicities} \begin{array}{lcl}
\rho={\rm Spin}(5)\colon & E' \approx E+\lambda+H\tau \quad~ & \lambda\in \Lambda_w\ , \quad H\in \Gamma_r\ , \\
\rho=\mathrm{SO}(5)_+\colon & E'\approx E+\alpha+\chi \tau \quad~ & \alpha\in \Lambda_r\ , \quad \chi\in \Gamma_w\ , \\
\rho=\mathrm{SO}(5)_-\colon & E'\approx E+\widetilde\lambda+\widetilde\chi\tau \quad~ & \ \ \ (\widetilde\lambda,\widetilde\chi) \in \widetilde\Lambda\times \widetilde\Gamma\ ,
\end{array}
\eea 
with the sublattice $\widetilde\Lambda\times\widetilde\Gamma$ defined as
\be
\widetilde\Lambda\times\widetilde\Gamma =\mathbb{Z}\langle(\lambda_1,\chi_2)\rangle \oplus \Lambda_r\times\Gamma_r < \Lambda_w\times \Gamma_w\ ,
\ee
namely, the non-trivial generator $(\lambda_1,\chi_2)$ is such that $\left(\phi_\Lambda(\lambda_1),\phi_\Gamma(\chi_2)\right)=(1,1)$ as required by \cref{def:globvarCM}.
Crucially, the equivalences $\approx$ in \cref{eq: so(5) periodicities} must be understood up to the action of the Weyl group.

Taking this into account we find that, for $\rho={\rm Spin}(5)$, the configurations 1, 5, and 6 in \cref{fig:so5ExtremaAll} acquire multiplicity 2 because of the vertical shift $+ i \chi_2$. For these three extrema it is not possible to go back to the initial configuration by using the gauged translations of \cref{eq: so(5) periodicities}. For all the other extrema, the vertical shift by $\chi_2$ can be canceled by gauged translations. For instance, at $\tau = i$ the $V_3$ extremum coordinates are $E = (z_1^*,z_2^*) = \left( 0.5+ 0.5 i , 0.82 + 0.79i\right) $ and we have that 
\begin{align}
E + i\chi_2  = \left( 0.5+ 1.5 i , 0.82 + 0.79i\right) &\approx   \left( - 0.5 - 1.5 i , 0.82 + 0.79i\right)  \\
& = E - \lambda_2 -i H_{\alpha_1} - 2i H_{\alpha_2} \ ,  \nonumber
\end{align}
where we used a sign flip on the first coordinate.
The additional non-equivalent configurations with respect to extrema $1$,$5$ and $6$ will be denoted $1'$, $5'$, and $6'$, respectively. 
In this particular case, the new configurations formally coincide with the ones already displayed in \cref{fig:so5ExtremaAll} once brought to the fundamental cell, because $i\chi_2=(i,0)$. This example underlines the importance of computing equivalences between shifted configurations in the complex plane and not in the fundamental cell.\footnote{A similar subtlety arises in type $A$ theories when one chooses the $z_n=0$ convention to fix the global translation symmetry, as discussed in \cref{App:othergaugefixtypeA}.}
In conclusion, we find 10 non-equivalent solutions, corresponding to 10 physically inequivalent gapped vacua on $\bR^3\times S^1$.

\begin{figure}
\captionsetup[subfloat]{labelformat=empty} 
\centering
\begin{minipage}{.32\linewidth}
    \centering
    \subfloat[][Extremum 2]{\label{}\includegraphics[width=0.85\linewidth]{figures/so5V2.pdf}}
\end{minipage}
\begin{minipage}{.32\linewidth}
    \centering
    \subfloat[][Extremum 3]{\label{}\includegraphics[width=0.85\linewidth]{figures/so5V3.pdf}}
\end{minipage}
\begin{minipage}{.32\linewidth}
    \centering
    \subfloat[][Extremum 4]{\label{}\includegraphics[width=0.85\linewidth]{figures/so5V4.pdf}}
\end{minipage}
\par\medskip \medskip \medskip 
\begin{minipage}{.32\linewidth}
    \centering
    \subfloat[][Extremum $2'$]{\label{}\includegraphics[width=0.85\linewidth]{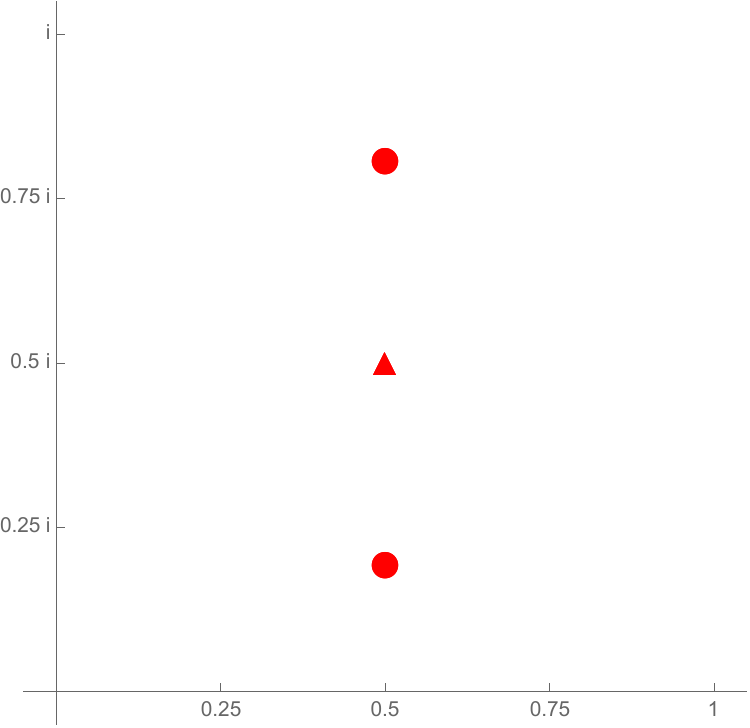}}
\end{minipage}
\begin{minipage}{.32\linewidth}
    \centering
    \subfloat[][Extremum $3'$]{\label{}\includegraphics[width=0.85\linewidth]{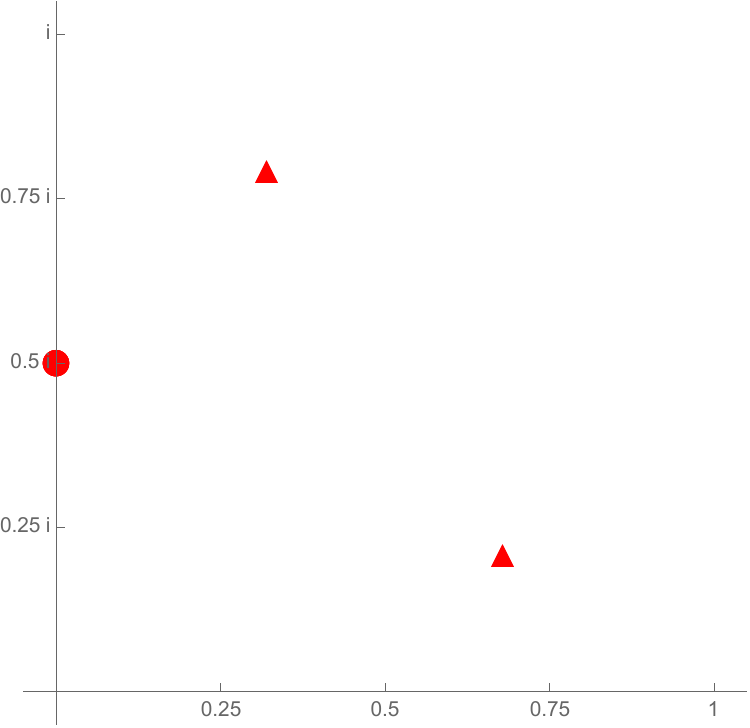}}
\end{minipage}
\begin{minipage}{.32\linewidth}
    \centering
    \subfloat[][Extremum $4'$]{\label{}\includegraphics[width=0.85\linewidth]{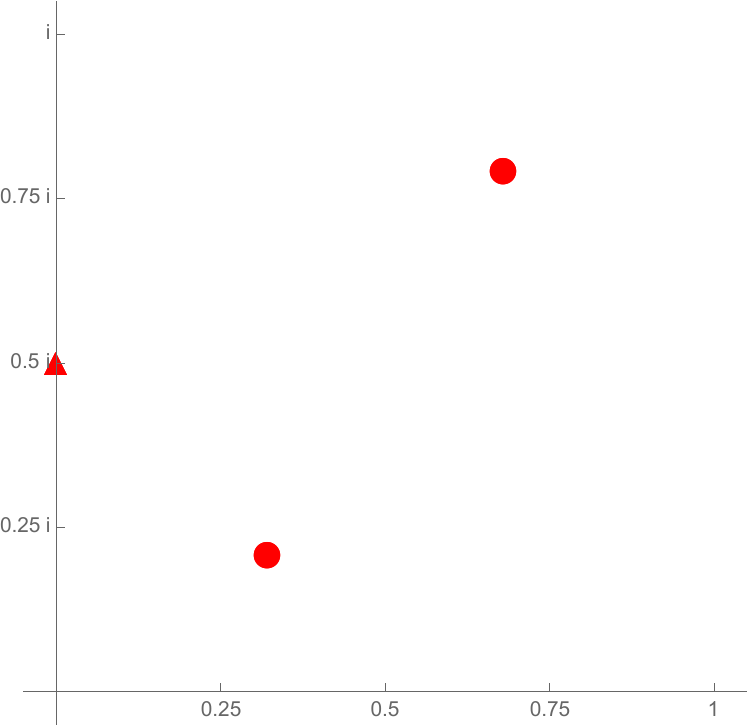}}
\end{minipage}

\caption{Duplicated isolated extrema of the $\mathrm{SO}(5)_+$ Calogero-Moser system at $\tau =i$}
\label{fig:SO5pduplic}
\end{figure}

Turning to $\rho=\mathrm{SO}(5)_+$, the two-fold multiplicity occurs now for the configurations 2, 3, and 4, which, under a shift by the fundamental weight $\lambda_1$, respectively lead to non-equivalent solutions $2'$, $3'$, and $4'$, as shown in \cref{fig:SO5pduplic}. This time, they project into different configurations in the fundamental cell. Again, this implies 10 physically distinct gapped phases on $\bR^3\times S^1$. 

Finally, implementing the periodicities associated to $\rho=\mathrm{SO}(5)_-$, one concludes that each configuration in \cref{fig:so5ExtremaAll} gives rise to exactly one vacuum. For instance, one can use the equivalence 
\begin{equation}\label{eq:shiftsequivso5-}
    E' \approx E + h_1 \alpha_1 + h_2 \alpha_2 + i v_1 H_{\alpha_2} + d_1 (\lambda_1 + i \chi_2)
\end{equation}
with integer horizontal, vertical and diagonal coefficients. Note that a vertical shift by $H_{\alpha_1}$ can be expressed as an integer linear combination of the shifts appearing in \cref{eq:shiftsequivso5-}, namely $i H_{\alpha_1} = 2(\lambda_1 + i \chi_2) - 2 H_{\alpha_1} -2 \alpha_1 - \alpha_2 $. Therefore, the $\mathrm{SO}(5)_-$ theory on $\bR^3\times S^1$ has 7 gapped ground states. All the above considerations are summarized in Table \ref{tab:so5multiplicities}.
\begin{table}
\centering
\begin{tabular}{c|c|c|c|c|c|c|c|c}
    & $V_1$ & $V_2$& $V_3$ & $V_4$ & $V_5$ & $V_6$ & $V_7$ & \textbf{Total} \\[3pt]
    \hline  
    $\mathrm{Spin}(5)$ & 2 & 1 & 1 & 1 & 2 & 2 & 1 & 10 \\[3pt]
    $\mathrm{SO}(5)_+$ & 1 & 2 & 2 & 2 & 1 & 1 & 1 & 10\\[3pt]
    $\mathrm{SO}(5)_-$ & 1 & 1 & 1 & 1 & 1 & 1 & 1& 7 
\end{tabular}
\caption{Multiplicities of the $\mathfrak{so}(5)$ CM extrema for each Lie group}
\label{tab:so5multiplicities}
\end{table}

All these vacua furnish a non-trivial representation of the duality group. This can be implemented by considering the modular properties of the potentials, as described in \cref{sec: N=1* and CM}, provided the transformation under $S$ duality (\cref{eq:SpotTransform}) is modified due to the presence of the twisted Weierstrass functions. Specifically, there is an additional shift in the potential \cite{Bourget:2015cza}, which in the case of $\mathfrak{so}(5)$ reads 
\begin{equation}
\frac{1}{2 \tau^2}  V_i \left(  \frac{-1}{2\tau} \right) =  V_{\sigma_{S_2} (i)} (\tau) + \frac{4\pi^2}{3} E_{2,2}(\tau)\ ,
\end{equation}
where we have used the linear combination of Eisenstein series $E_{2,2}(\tau)=E_{2}(\tau)-2E_{2}(2\tau)$.

As it will turn useful below, we list here explicitly the transformation of the potentials when evaluated at the fixed point of $S_2$, namely $\tau=i/\sqrt{2}$, for which one obtains
\begin{equation}\label{eq:S2fixedpoint}
V_i \left(\frac{i}{\sqrt{2}}\right) = - V_{\sigma_{S_2} (i)}\left(\frac{i}{\sqrt{2}}\right) -\frac{4\pi^2}{3} E_{2,2}\left(\frac{i}{\sqrt{2}}\right)\ . 
\end{equation}
Therefore, a singlet $V_j(\tau)$ under $S_2$ is characterized by:
\begin{equation}
    V_j \left( \frac{i}{\sqrt{2}} \right) = 
    \frac{-2\pi^2}{3} E_{2,2}\left(\frac{i}{\sqrt{2}}\right) \approx 8.45999\ . 
\end{equation} 
The seven different numerical values of the potentials evaluated at the fixed point $\tau=i/\sqrt{2}$ are listed in Table \ref{tab:so5PotsS2}. Recall that, depending on the global variants, some of these numerical values might correspond to multiple non-equivalent solutions. 
\begin{table}
\centering
\begin{tabular}{|c|c|}
    \hline 
    Extremum & Potential value at $\tau = \frac{i}{\sqrt{2}}$  \\
    \hline  
    $V_1$ & 90.0193 \\
    $V_2$ & $-73.0993$  \\
    $V_3$ & $11.1697 -13.8962\, i$ \\
    $V_4$ & $11.1697 + 13.8962\, i$ \\
    $V_5$ & $5.7503 + 13.8962\, i$ \\
    $V_6$ & $5.7503 -13.8962\, i$ \\
    $V_7$ & 8.45999 \\
    \hline 
\end{tabular}
\caption{Values of the $\mathfrak{so}(5)$ potentials at the $S_2$ fixed point.}
\label{tab:so5PotsS2}
\end{table}

\begin{figure}
    \centering
    \includegraphics[width=0.52\linewidth]{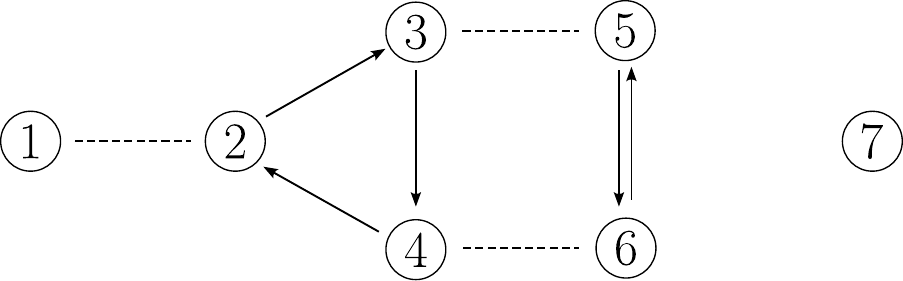}
    \caption{Actions of $S_2$ in dashed line and $T$ in solid line on the seven extrema $V_i(\tau)$ of the $\mathfrak{so}(5)$ twisted CM system. They coincide with the duality web of gapped vacua in $\bR^3 \times S^1$ for the $\mathrm{SO}(5)_-$ global variant.}
    \label{fig:so5Extr}
\end{figure}

\begin{figure}
    \centering
    \includegraphics[width=0.34\linewidth]{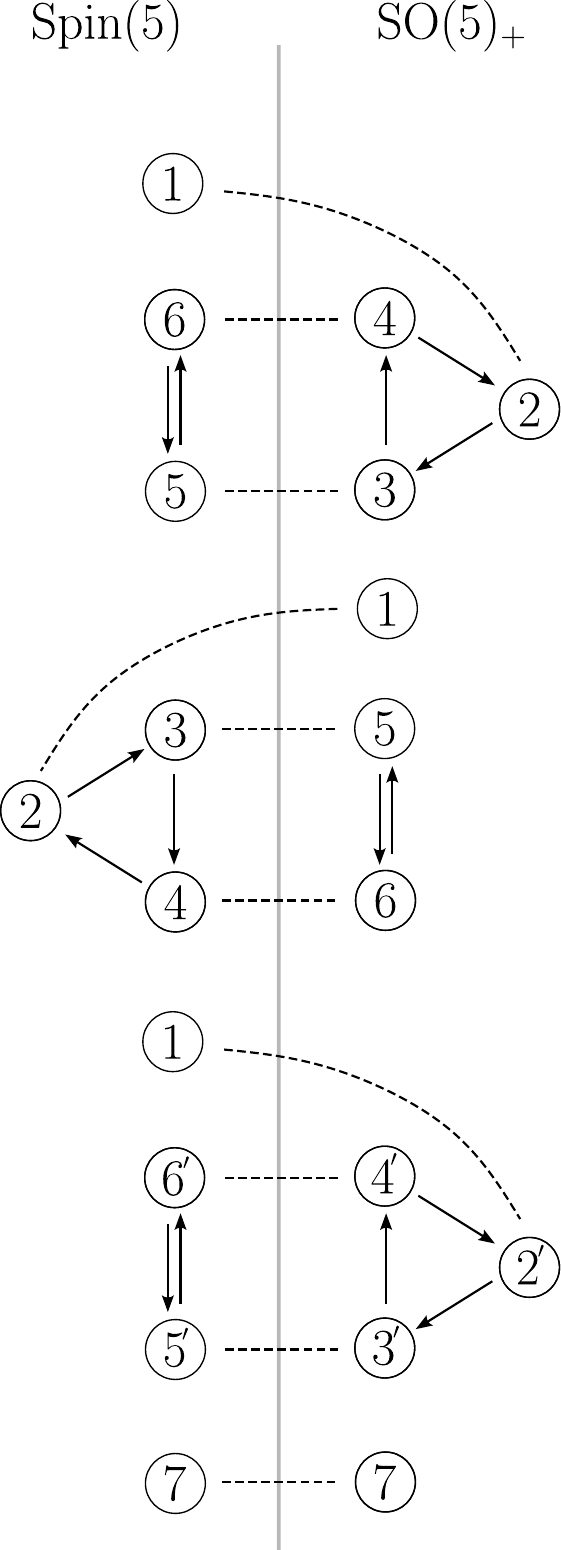}
    \caption{Duality web of gapped vacua in $\bR^3 \times S^1$ \cite[figure 4]{Bourget:2015upj}. There are 10 massive vacua for $\mathrm{Spin}(5)$ (on the left), as well as for $\mathrm{SO}(5)_+$ (on the right). The Langlands $S_2$-duality is depicted in dotted lines.}
    \label{fig:spin5so5}
\end{figure}

The simplest case corresponds to $\rho=\mathrm{SO}(5)_-$, where the action of duality does not change the global variant. In this case, the duality orbit is displayed in \cref{fig:so5Extr}. 
The more interesting case corresponds to the orbits involving the ground states of the theories with $\rho={\rm Spin}(5)$ and $\rho=\mathrm{SO}(5)_+$, which we display in \cref{fig:spin5so5}. Both of these duality webs were already obtained in \cite{Bourget:2015upj}.

Regardless of the global variant, one readily observes the occurrence of singlets, doublets and triplets under the action of $T$. 
Firstly, one can consistently identify the vacua associated to the configuration 1, which are singlets under $T$, as the Higgs vacua on each global variant, that is the ones associated to the partition $\{5\}$. Then, the doublets and triplets
are identified with the ground states featuring an emergent (spontaneously broken) $R$-symmetry, hence descending from the partitions $\{2^2,1\}$ (doublets) and $\{1^5\}$ (triplets). The different multiplicities of these orbits within each global variant stems from the presence of additional topological sectors (concretely a $\bZ_2$ gauge theory), associated to the spontaneous breakdown of the 1-form symmetry in $\bR^4$. We will comment further on this fact in the next subsection.  Finally, the configuration 7 in \cref{fig:so5ExtremaAll} corresponds to the partition $\{3,1^2\}$, as it is a singlet under all dualities.

\subsection{Duality symmetries and effective TQFT}

\begin{figure}
    \centering
    \includegraphics[width=0.52\linewidth]{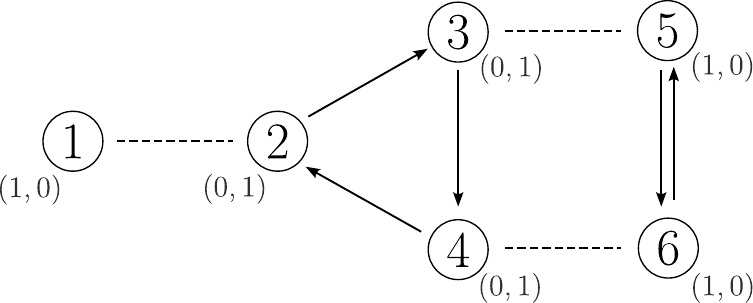}
    \caption{4d vacua of $\mathcal{N}=1^*$ $\mathfrak{so}(5)$ theories and line condensation. $S_2$ is depicted in dotted line, and $T$ in solid line acts trivially on the line operators.}
    \label{fig:so54dvac}
\end{figure}

We conclude this section with some remarks on the dynamics of the theory in $\bR^4$. In a sense, the following analysis is somewhat disconnected from the main theme of this article, but aligns with the one presented in \cite{Damia:2023ses} for $\cN=1^*$ theories with gauge algebras of type-$A$. The vacua in four dimensions can be readily obtained by ignoring the additional multiplicities and excluding the duality singlet, leading to the structure of six vacua depicted in \cref{fig:so54dvac}.

Let us first comment on the realization of duality symmetries. 
These theories enjoy the action of additional global symmetries associated to the $S_2$ self-duality of the theory at the self-dual coupling $\tau^*$, in this case $\tau^*=i/\sqrt{2}$. These symmetries are exact, in the sense that they are already present at the UV fixed point of the $\cN=1^*$ flow. Moreover, we consider a slightly modified version where the $S_2$-duality operation is composed with appropriate gaugings of the UV $\bZ_2^{(1)}$ 1-form symmetry in order to define maps among vacua within the same global variant. Note that the inclusion of this topological manipulation depends on the actual choice of global variant.\footnote{Strictly speaking, the definition of this symmetries also includes a particular phase rotation associated to the explicitly broken UV $R$-symmetry \cite{Damia:2023ses}. We will ignore this subtlety as it does not play any significant role within the present context.} In particular, a gauging of $\bZ_2^{(1)}$ is required when discussing either $\rho=\mathrm{Spin}(5)$ or $\rho=\mathrm{SO}(5)_+$, whereas it is not for the case of $\rho=\mathrm{SO}(5)_-$.

As with ordinary symmetries, duality symmetries can be either preserved or spontaneously broken in the deep IR of the $\cN=1^*$ theory. Inspection of \cref{tab:so5PotsS2} immediately instructs us that the $S_2$ duality symmetry is spontaneously broken on all six vacua. More precisely, they form three doublets related by the action of $S_2$, as it can be easily verified using the transformation rule \cref{eq:S2fixedpoint} for the potentials at the fixed point.

A further ingredient in the characterization of the low energy phases of $\cN=1^*$ SYM theory concerns the assignment of a condensed line operator in each massive vacua or, equivalently, a particular realization of the 1-form symmetry. This, in turns, prescribes a particular TQFT as the low energy description on top of each gapped ground state. The finite lattice of equivalence classes of line operators associated to the gauge algebra $\mathfrak{so}(5)$ is of the form $\bZ_2\times \bZ_2$ \cite{Aharony:2013hda}. Therefore, line operators are labeled by doublets $(e,m)$ with $e,m=0,1$. The global variant $\rho$ is determined by a choice of a Lagrangian subgroup within this lattice, which for the case at hand implies choosing a single non-trivial element $(e,m)\in \bZ_2\times \bZ_2$ as the class of genuine gauge invariant line operators, or, in other words, the assignment of which line operators are charged under the $\bZ_2^{(1)}$ 1-form symmetry for each global variant. For $\rho={\rm Spin}(5)$, this corresponds to the electric line $(1,0)$. On the other hand, the monopole $(0,1)$ and dyon $(1,1)$ are the genuine lines for $\rho=\mathrm{SO}(5)_+$ and $\rho=\mathrm{SO}(5)_-$, respectively. In turn, the action of $S_2$ exchanges the electric and monopole lines, while it leaves the dyonic one invariant.    

Taking into account all these considerations, together with the realization of the duality symmetry described above, we can unambiguously assign a pattern of line condensation as shown in \cref{fig:so54dvac}. According to their charges under $\bZ_2^{(1)}$, this may lead to a phase with either preserved or broken 1-form symmetry. This of course depends on the choice of global variant $\rho$. In this context, one finds an important deviation from the situation featured for type $A$ gauge algebras. For the latter case, the condensation furnishes a bijection between the equivalence classes of line operators and the various massive vacua. This no longer holds for other types of algebras, in particular for type $B$ or $D$. In a sense, this is an expected outcome rooted in the fact that the UV 1-form symmetry does not scale with the rank of the gauge group.

As an illustration of these facts, let us expand on the case of $\rho={\rm Spin}(5)$ where the genuine line operator $(1,0)$ is charged under $\bZ_2^{(1)}$. In the Higgs vacuum corresponding to the partition $\{5\}$ (node 1 in \cref{fig:so54dvac}), the line $(1,0)$ acquires a constant VEV signaling the spontaneous breaking of $\bZ_2^{(1)}$. This scenario is realized in four dimensions by a topological order described by a dynamical $\bZ_2$ 1-form connection, i.e.~a $\bZ_2$ gauge theory.  It is straightforward to extend these considerations for the remaining 5 vacua. Noticing that the massive ground state 2 is obtained by the action of the non-invertible self-duality symmetry, we assign the condensation of the line with unit magnetic charge $(0,1)$ to such a vacuum. As this line operator is non-genuine for this global variant, the $\bZ_2^{(1)}$ symmetry is preserved and this state is now described by a trivial $\bZ_2^{(1)}$ SPT.\footnote{The same conclusion can be obtained by considering the action of the topological manipulation, namely the gauging of $\bZ_2^{(1)}$, on the $\bZ_2$ gauge theory placed at the vacuum 1.}
Moreover, due to the absence of a non-trivial Witten effect for $\mathfrak{g}=\mathfrak{so}(5)$, this effective description extends without modification to the vacua 3 and 4. Of course, these form a triplet once the emergent $R$-symmetry is included in the analysis, labeled by the value of a fermion bilinear, but otherwise indistinguishable from the perspective of the $\bZ_2^{(1)}$ 1-form symmetry.   

\begin{figure}
\begin{minipage}{.5\linewidth}
\centering
\subfloat[][$\rho = \mathrm{Spin}(5)$]{\label{fig:spin5}\includegraphics[width=0.9\linewidth]{figures/Spin5.pdf}}
\end{minipage}%
\begin{minipage}{.5\linewidth}
\centering
\subfloat[][$\rho = \mathrm{SO}(5)_+$]{\label{fig:so5+}\includegraphics[width=0.9\linewidth]{figures/SO5p.pdf}}
\end{minipage}\par\medskip \medskip
\centering
\subfloat[][$\rho = \mathrm{SO}(5)_-$]{\label{fig:so5-}\includegraphics[width=0.45\linewidth]{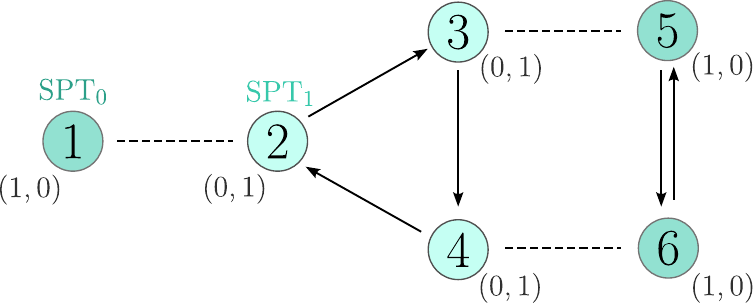}}
\caption{4d low-energy theories in the vacua of $\mathcal{N}=1^*$ $\mathfrak{so}(5)$ theories for the three global variants. The pink vacua are all described by a $\Z_2$ TQFT, while those in shades of teal are trivially gapped.}
\label{fig:so5VariantVac}
\end{figure}

The case of the vacua 5 and 6 is just a little bit different in a sense, as they may be distinguished by a 1-form symmetry charge, though an emergent one. On the one hand, it is consistent to assign the condensation of the genuine line $(1,0)$ to these phases. This stems from the action of $S_2$ duality. Relatedly, it also derives from a careful analysis of the multiplicities associated to the their associated solutions of the CM system, where we showed that these appear with double multiplicity for $\rho={\rm Spin}(5)$, consistently with the $\bZ_2$ factor within the centralizer group associated to the partition $\{2^2,1\}$ (see \cref{tab: so(5) partitions}). As explained before within this section, by including the emergent $\bZ_2^{(1)}$ 1-form symmetry charges, the vacua 5 and 6 might be further distinguished by the condensation of emergent magnetic and dyonic charges, hence related by the (emergent) Witten effect. We emphasize nevertheless that these lines have identical charge under the UV $\bZ_2^{(1)}$ center symmetry of $\mathrm{Spin}(5)$, hence this symmetry is spontaneously broken and both vacua host a four dimensional $\bZ_2$ gauge theory.

The pattern of line condensation displayed in \cref{fig:so54dvac} goes through without modifications for the remaining two global variants, but the effective topological field theories do change. Specifically, for $\rho=\mathrm{SO}(5)_+$, the magnetic 1-form symmetry acts on the genuine line $(0,1)$. Consequently, the 1-form symmetry is now spontaneously broken in the vacua labeled by the nodes 2, 3, and 4. Finally, since the line $(1,1)$ never condenses, the $\bZ_2^{(1)}$ symmetry is always preserved in the theory with $\rho=\mathrm{SO}(5)_-$. This discussion is summarized in \cref{fig:so5VariantVac}.\footnote{As a further consistency check, note that the scenarios proposed for $\rho=\mathrm{SO}(5)_\pm$ can be obtained from the one at $\rho={\rm Spin}(5)$ by gauging the $\bZ_2^{(1)}$ 1-form symmetry with or without discrete torsion. } As in the case of type $A$ theories studied above, this assignment of $\bZ_2$ gauge theories agrees perfectly with the occurrence of additional multiplicities on $\bR^3\times S^1$, for each global variant.

\section*{Acknowledgments}

We are grateful to Marco Fazzi, Louan Mol, Giovanni Rizi and Jan Troost for useful discussions and feedback. VT is thankful to Sibasish Banerjee and Raphael Senghaas for stimulating conversations. RA and RV are respectively a Research Director and a Research Fellow of the F.R.S.-FNRS (Belgium).  The research of RA and RV is funded through an ARC advanced project, and further supported by IISN-Belgium (convention 4.4503.15). VT is funded by the Deutsche Forschungsgemeinschaft (DFG, German Research Foundation) under Germany’s Excellence Strategy EXC 2181/1 -
390900948 (the Heidelberg STRUCTURES Excellence Cluster). The work of JAD is funded by the Spanish MCIN/AEI/10.13039/501100011033 grant PID2022-126224NB-C21.

\appendix

\section{Lie theory}\label{App:Lie}

\subsection{Lie algebras, weights and roots}

We refer to \cite{Fulton:1991rep} for details. Let $\mathfrak g$ be a complex\footnote{What follows applies \textit{mutatis mutandis} to compact Lie algebras.} simple Lie algebra, and let $\mathfrak h \subset \mathfrak g$ be a Cartan subalgebra. 

The elements of $\mathfrak{h}$ can be simultaneously diagonalized in the adjoint representation. This defines the finite set of \textit{roots} $\Delta(\mathfrak g)\subset\mathfrak{h}^*$ of $\mathfrak{g}$, as the collective eigenvalues of the elements of $\mathfrak h$ in the adjoint representation. A choice of generic hyperplane in $\mathfrak{h}^*$ defines a subset of \textit{positive roots} $\Delta^+(\mathfrak g)\subset \Delta(\mathfrak g)$. The extremal elements in $\Delta^+(\mathfrak g)$ are called \textit{simple roots}. The set $\Delta(\mathfrak g)$ generates the \textit{root lattice} $\Lambda_r\subset\mathfrak{h}^*$, of which the set of simple roots forms a basis.

A \textit{Killing form} on $\mathfrak g$ is an invariant symmetric bilinear form $(\cdot,\cdot)$ on $\mathfrak g$, i.e.~$([x,y],z) = (x,[y,z])$ for any $x,y,z\in\mathfrak g$. The algebra $\mathfrak g$ being simple, any such Killing form is a scalar multiple of $(x,y)_0 = \tr_\mathrm{adj}(xy)$, where $x,y\in\mathfrak{g}$ and $\tr_\mathrm{adj}$ denotes the trace in the adjoint representation. Moreover, Cartan's criterion implies that any Killing form on $\mathfrak{g}$ is non-degenerate. A Killing form induces an isomorphism between $\mathfrak h^*$ and $\mathfrak h$, together with a non-degenerate symmetric bilinear form on $\fh^*$, that we shall also denote $(\cdot, \cdot)$. To each element of $\fh^*$, we can associate a unique element of $\fh$ in the following way
 \begin{equation}\label{eq:isoKilling}
 	  \begin{array}{lcl}
 	  	\mathfrak{h}^* & \longrightarrow & \mathfrak{h}\\
 	  	\mu & \longmapsto & h_\mu 
 	  \end{array}\ ,
 \end{equation}
such that $(h_\mu,h) = \mu(h) , \, \forall  h   \in \fh$. Likewise, the induced bilinear form on $\fh^*$ is defined as $(\mu_1, \mu_2) = (h_{\mu_1} , h_{\mu_2})$. We denote by $|\cdot|$ the norm on $\mathfrak{h}^*$ induced by the Killing form.

The \textit{coroots} of $\mathfrak{g}$ are the elements $H_\alpha\in\mathfrak{h}$, $\alpha\in\Delta(\mathfrak g)$, such that for any $\beta\in\Delta(\mathfrak g)\subset \mathfrak{h}^*$:
\begin{equation}\label{eq:defcoroot}
	\beta(H_\alpha) = \frac{2(\beta,\alpha)}{(\alpha,\alpha)}\ .
\end{equation}
In other words, $\forall \alpha\in\Delta(\mathfrak g)$, $H_\alpha$ is the element of $\mathfrak h$ corresponding to $2\alpha/|\alpha|^2\in\mathfrak{h}^*$ under the isomorphism of \cref{eq:isoKilling}. For a root such that $|\alpha|^2=2$, $H_\alpha$ is canonically associated to $\alpha$ following the isomorphism of \cref{eq:isoKilling}, namely $H_\alpha = h_\alpha$.
Coroots generate the \textit{coroot lattice} $\Gamma_r\subset\mathfrak{h}$, and the set of coroots corresponding to simple roots forms a basis of $\Gamma_r$. The matrix with elements $C_{\alpha\beta} = \alpha(H_\beta)$ for all simple roots $\alpha,\beta$, is the \textit{Cartan matrix} of $\mathfrak{g}$.

The \textit{weight lattice} $\Lambda_w\subset\mathfrak{h}^*$ is the lattice dual to $\Gamma_r$, i.e.
\begin{equation}
	\Lambda_w = \left\{\lambda \in\mathfrak{h}^*\;\mid\; \forall \alpha\in\Delta(\mathfrak{g}),\; \lambda(H_\alpha) \in\mathbb{Z} \right\}\ .
\end{equation}
Elements of $\Lambda_w$ are called \textit{weights}. The elements of the basis of $\Lambda_w$ dual to the simple coroot basis of $\Gamma_r$, i.e.~the weights $\lambda_\alpha$ with $\alpha$ a simple root, such that for all simple root $\beta$, $\lambda_\alpha(H_\beta) = \delta_{\alpha\beta}$, are called \textit{fundamental weights}.

The \textit{coweight lattice} $\Gamma_w\subset\mathfrak{h}$ is the lattice dual to $\Lambda_r$, i.e.
\begin{equation}
	\Gamma_w = \left\{\chi \in\mathfrak{h}\;\mid\; \forall \alpha\in\Delta(\mathfrak{g}),\; \alpha(\chi) \in\mathbb{Z} \right\}\ .
\end{equation}
Elements of $\Gamma_w$ are called \textit{coweights}. The elements of the basis of $\Gamma_w$ dual to the simple root basis of $\Lambda_r$, i.e.~the coweights $\chi_\alpha$ with $\alpha$ a simple root, such that for all simple root $\beta$, $\beta(\chi_\alpha) = \delta_{\beta\alpha}$, are called \textit{fundamental coweights}.

Complex simple Lie algebras $\mathfrak g$ are classified by Dynkin diagrams. When the latter admit simple edges only, $\mathfrak{g}$ is said to be \textit{simply-laced}. This is the case for the $A$ and $D$ series, as well as the exceptional algebras $E_6,E_7$, and $E_8$. In a simply-laced algebra $\mathfrak g$, all the (simple) roots have the same length. The other algebras $B_n$, $C_n$, $F_4$, and $G_2$, are non simply-laced, and correspondingly, not all (simple) roots have the same length. However, note that there are exactly two different lengths in each of these cases; one speaks of \textit{long} and \textit{short} (simple) roots. When $\mathfrak g$ is simply-laced all roots are long.

Fixing the length of long roots fixes the normalization of the Killing form entirely. In the paper, we use the standard normalization in which \textit{long roots have squared length two}. Note that this implies, by \cref{eq:defcoroot}, that when $\mathfrak g$ is simply-laced the isomorphism of \cref{eq:isoKilling} induced by the Killing form identifies $\Gamma_r$ with $\Lambda_r$, and hence $\Gamma_w$ with $\Lambda_w$.

\subsection{Lie groups, characters, and cocharacters} 

For any complex (resp.~compact) simple Lie algebra $\mathfrak g$, there exists a unique complex (resp.~compact) connected and simply-connected Lie group $\widetilde{G}$ whose Lie algebra is $\mathfrak g$. Let $\mathcal Z$ be the center of $\widetilde{G}$. It is always a finite abelian group: 

\begin{table}
	\centering
	\begin{tabular}{c||c|c|c|c|c|c|c|c|c|c}
		Dynkin type & $A_N$ & $B_N$ & $C_N$ & $D_{2N}$ & $D_{2N+1}$ & $E_6$ & $E_7$ & $E_8$ & $F_4$ & $G_2$ \\[2pt]
		\hline 
		Center $\mathcal Z$ & $\mathbb{Z}_{N+1}$ & $\mathbb{Z}_2$ & $\mathbb{Z}_2$ & $\mathbb{Z}_2\times\mathbb{Z}_2$ & $\mathbb{Z}_4$ & $\mathbb{Z}_3$ & $\mathbb{Z}_2$ & $\{1\}$ & $\{1\}$ & $\{1\}$
	\end{tabular}
\end{table}

All the connected complex (resp.~compact) Lie groups with Lie algebra $\mathfrak g$ are quotients of $\widetilde{G}$ by a subgroup of $\mathcal Z$. The group $G_\mathrm{ad} = \widetilde{G}/\mathcal Z$ is the adjoint group corresponding to $\mathfrak g$.

The image of the exponential map $\exp_G:\mathfrak g \rightarrow G$ restricted to the Cartan subalgebra $\mathfrak h$ is a \textit{Cartan subgroup} $H$ of $G$. When $G=\widetilde{G}$, one has $\ker(\exp_G) = \Gamma_r$, and $\exp_G(\Gamma_w) \subset \mathcal Z$, hence $\Gamma_w/\Gamma_r \simeq \mathcal Z$. This isomorphism lifts to an abelian group morphisms $\phi_{\Gamma} : \Gamma_w \rightarrow \mathcal Z$ such that $\ker(\phi_{\Gamma})= \Gamma_r$. For $G$ any connected complex or compact Lie group with Lie algebra $\mathfrak g$, one defines $\Gamma^G = \ker(\exp_G)$, which is a sublattice of $\Gamma_w$ containing $\Gamma_r$, called the lattice of \textit{$G$-coweights}, or \textit{$G$-cocharacters}. The quotient $\Gamma_w/\Gamma^G$ is naturally isomorphic to the center of $G$, while $\Gamma^G/\Gamma_r\cong \pi_1(G)$.

If $G=\widetilde{G}/\Pi$ with $\Pi$ a subgroup of $\mathcal Z$, the center of $G$ is $Z(G) = \mathcal Z/\Pi$ and only the representations of $\mathfrak g$ in which $\Pi$ acts trivially are representations of $G$. The lattice of \textit{$G$-weights}, or \textit{$G$-characters}, is the corresponding sublattice $\Lambda^G$ of $\Lambda_w$. It contains $\Lambda_r$ as a sublattice, and $\Lambda^G/\Lambda_r\cong \widehat{Z(G)}$, where $\widehat{\cdot}$ is Pontryagin duality, while $\Lambda_w/\Lambda^G\cong \widehat{\pi_1(G)}\cong \Pi$. Each representation of $G$ indeed induces a representation of $Z(G)$. When $G=\widetilde{G}$, $\Lambda^G=\Lambda_w$ and $\Lambda_w/\Lambda_r\simeq\widehat{\mathcal Z}$, which lifts to an abelian group morphism $\phi_{\Lambda}\colon \Lambda_w \rightarrow \widehat{\mathcal Z}$ generalizing the notion of $N$-ality for representations of $\mathfrak{su}(N)$ or $\mathfrak{sl}(N,\mathbb{C})$. 

While a finite abelian group $A$ is not generally isomorphic to its dual $\widehat{A}$ in a natural way, the center $\mathcal Z$ of a simply-connected simple Lie group is naturally isomorphic to its dual. This follows from the pairing
\begin{equation}\label{eq:DSZpairing}
\begin{array}{rcl} \Gamma_w/\Gamma_r \times \Lambda_w/\Lambda_r & \longrightarrow & \mathbb{Q}/\mathbb{Z} \\ 
(\chi,\lambda) & \longmapsto & \lambda(\chi) 
\end{array}\ , 
\end{equation}
being perfect. \Cref{eq:DSZpairing} is obtained by restricting the natural pairing between $\mathfrak{h}$ and $\mathfrak{h}^*$ to the lattices $\Gamma_w$ and $\Lambda_w$, yielding a pairing valued in $\mathbb{Q}$, and in $\mathbb{Z}$ when further restricted to either $\Gamma_r$ or $\Lambda_r$. Note that the pairing in \cref{eq:DSZpairing} is precisely the one appearing in the non-abelian Dirac–Schwinger–Zwanziger quantization condition \cite{Kapustin:2006pk,Gaiotto:2010be}
\begin{equation}
    e^{2\pi i (\lambda_1(\chi_2)-\lambda_2(\chi_1))} = 1\ ,
\end{equation}
for two dyonic lines $(\lambda_1,\chi_1)$ and $(\lambda_2,\chi_2)$, with $\lambda_1,\lambda_2 \in \Lambda_w$ and $\chi_1,\chi_2 \in \Gamma_w$.

Given a connected simple complex (or compact) Lie group $G$ with character lattice $\Lambda_r \subset \Lambda^G \subset \Lambda_w$ and cocharacter lattice $\Gamma_r \subset \Gamma^G \subset \Gamma_w$, its Langlands dual is the connected simple complex (or compact) Lie group $G^L$ with character lattice $\Gamma_r \subset \Gamma^G \subset \Gamma_w$ and cocharacter lattice $\Lambda_r \subset \Lambda^G \subset \Lambda_w$. Langlands duality exchanges type $B_n$ and type $C_n$ groups, while preserving all other Dynkin types.

This is summarized in the following diagram, where dotted arrows denote lattice dualities, and groups above inclusions denote quotients:
\begin{center}
 \begin{tikzpicture}
 \node at (-3,0) {$\mathfrak{h}$};
 \node at (11,2) {$\mathfrak{h}^{\ast}$};
 \node at (-2,0) {$\supset$};
 \node at (2,0) {$\supseteq$};
 \node at (6,0) {$\supseteq$};
 \node at (2,2) {$\subseteq$};
 \node at (6,2) {$\subseteq$};
 \node at (10,2) {$\subset$};
 \node[align=center,text width = 2cm] (1) at (0,0) {Coweight \\ Lattice $\Gamma_w$};
 \node[align=center,text width = 2cm] (2) at (4,0) {Cocharacter \\ Lattice $\Gamma^{G}$};
 \node[align=center,text width = 2cm] (3) at (8,0) {Coroot \\ Lattice $\Gamma_r$};
 \node[align=center,text width = 2cm] (4) at (0,2) {Root \\ Lattice $\Lambda_r$};
 \node[align=center,text width = 2cm] (5) at (4,2) {Character \\ Lattice $\Lambda^{G}$};
 \node[align=center,text width = 2cm] (6) at (8,2) {Weight \\ Lattice $\Lambda_w$};
 \draw[<->,dotted] (1)--(4);
 \draw[<->,dotted] (2)--(5);
 \draw[<->,dotted] (3)--(6);
 \node at (2,2.5) {$\widehat{Z(G)}$};
 \node at (6,2.5) {$\widehat{\pi_1 (G)}$};
 \node at (2,0.5) {$Z(G)$};
 \node at (6,0.5) {$\pi_1 (G)$};
 \node (66) at (8,4) {$\widehat{\mathcal Z}$};
 \node (11) at (0,-2) {$\mathcal Z$};
 \draw[->] (6)--(66) node[midway,right] {$\phi_{\Lambda}$};
 \draw[->] (1)--(11) node[midway,left] {$\phi_{\Gamma}$};
 \end{tikzpicture}
\end{center} 

\subsection{Nilpotent orbits in complex simple Lie algebras}

Our main reference here is \cite{collingwood1993nilpotent}. 

Let $\mathfrak{g}$ be a complex simple Lie algebra. An element $X\in\mathfrak{g}$ is nilpotent if $\mathrm{ad}_X=[X,\cdot]$ is a nilpotent endomorphism of $\mathfrak g$. A nilpotent orbit in $\mathfrak g$ is a subset of $\mathfrak g$ of the form
\begin{equation}
	\mathcal O_X = \{g\cdot X g^{-1}\,\mid\, g\in G_\mathrm{ad}\}\ ,
\end{equation}
where $X$ is nilpotent. In words, a nilpotent orbit in $\mathfrak g$ is the orbit of a nilpotent element in $\mathfrak g$, under the adjoint action of $G_\mathrm{ad}$.

An $\mathfrak{sl}(2)$-triple in $\mathfrak g$ is a triple $(H,X,Y)$ of elements of $\mathfrak g$ satisfying the commutation relations
\begin{equation}
	[H,X] = 2X\ , \quad [H,Y] = -2Y\ , \quad [X,Y] = H\ .
\end{equation}
The elements $H,X$, and $Y$, are called the neutral, nilpositive and nilnegative elements of the $\mathfrak{sl}(2)$-triple $(H,X,Y)$, respectively. One defines the orbits of $\mathfrak{sl}(2)$-triples in $\mathfrak g$ under the adjoint action of $G_\mathrm{ad}$ as before.

By the Jacobson--Morozov lemma, every nilpotent element $X\in\mathfrak g$ is the nilpositive element of an $\mathfrak{sl}(2)$-triple in $\mathfrak g$. In fact, the set of nilpotent orbits in $\mathfrak g$ is in bijection with the set of orbits of $\mathfrak{sl}(2)$-triples.

\paragraph{Orbits of $\mathfrak{sl}_2$-triples in type $A$ algebras.} 

Let $G$ be of type $A$, i.e. $\mathfrak{g}_\mathbb{C} = \mathfrak{sl}(N,\mathbb{C})$. A distinguished orbit of $\mathfrak{sl}(2)$-triples in $\mathfrak{g}_\mathbb{C}$ is that of so-called principle $\mathfrak{sl}(2)$-triples. A representative in this orbit in $\mathfrak{sl}(N,\mathbb{C})$ is the triple $(X_N,Y_N,H_N)$, where:
\begin{equation}\label{eq:principalnil}
    X_N = \begin{pmatrix}
        0 & 1 & 0 & 0 & \ldots & 0 \\
        0 & 0 & 1 & 0 & \ldots & 0 \\
        \vdots &  & \ddots & \ddots & & \\
        0 & 0 & \ldots & 0 & 1 & 0 \\
        0 & 0 & \ldots & 0 & 0 & 1 \\
        0 & 0 & \ldots & 0 & 0 & 0 \\
    \end{pmatrix},\quad
    Y_N = \begin{pmatrix}
        0 & 0 & 0  & \ldots & 0 \\
        N-1 & 0 & 0 & \ldots & 0 \\
        0 & 2(N-2) & 0 & \ldots & 0 \\
        \vdots & 0 & \ddots & \ddots & \\
        0 & \ldots & 0 & N-1 & 0 \\
    \end{pmatrix},
\end{equation}
where the only non-zero entries of $Y_N$ are $(Y_N)_{i+1,i} = i(N-i)$, and:
\begin{equation}\label{eq:principalnilCartan}
    H_N = [X_N,Y_N] = \begin{pmatrix}
        N-1 & 0 & 0 & \ldots & 0 & 0 \\
        0 & N-3 & 0 & \ldots & 0 & 0 \\
        0 & 0 & N-5 & \ldots & 0 & 0 \\
        \vdots & \vdots & & \ddots & \vdots \\
        0 & 0 & 0 & \ldots & 3-N & 0 \\
        0 & 0 & 0 & \ldots & 0 & 1-N
    \end{pmatrix}.
\end{equation}

General $\mathfrak{sl}(2)$-triples in $\mathfrak{sl}(N,\mathbb{C})$ are in one-to-one correspondence with partitions of $N$, defined in \cref{subsec:unbrokengaugegroups}. A representative $(X_{\underline{\lambda}}, Y_{\underline{\lambda}}, H_{\underline{\lambda}})$ in the orbit of $\mathfrak{sl}(2)$-triples corresponding to some partition $\underline{\lambda}=\{\lambda_1^{\mu_1}\dots\lambda_k^{\mu_k}\}$ is given by:
\begin{equation}
    X_{\underline{\lambda}} = \mathrm{Diag}(X_{\lambda_1},\dots,X_{\lambda_1},X_{\lambda_2},\dots,X_{\lambda_n})\ ,  
\end{equation}
where $X_{\lambda_1}$ appears $\mu_1$ times, $X_{\lambda_2}$, $\mu_2$ times, etc, and where the blocks $X_{\lambda_i}$ are the $\lambda_i\times\lambda_i$ matrices defined in \cref{eq:principalnil}, together with $X_1:=0\in\mathbb{C}$. The elements $Y_{\underline{\lambda}}$ and $H_{\underline{\lambda}}$ are constructed similarly, as block-diagonal matrices with blocks the matrices of \cref{eq:principalnil,eq:principalnilCartan}.

\paragraph{Orbits of $\mathfrak{sl}(2)$-triples in $\mathfrak{so}(2N+1,\mathbb{C})$.}

General orbits of $\mathfrak{sl}(2)$-triples (or equivalently nilpotent orbits) in $\mathfrak{so}(2N+1,\mathbb{C})$ are in one-to-one correspondence with $B$-type partitions of $2N+1$, i.e.~partitions of $2N+1$ where every even factor appears an even number of times. In other words, a partition $\underline{\lambda} = \{\lambda_1^{\mu_1}\dots\lambda_k^{\mu_k}\}$ of $N$ is a $B$-type partition if for every $\lambda_i$ even, $\mu_i$ is even.

A general recipe to obtain standard representatives of $\mathfrak{sl}(2)$-triples in $\mathrm{so}(2N+1,\mathbb{C})$ from $B$-type partitions of $2N+1$ is provided in \cite[Chap. 5]{collingwood1993nilpotent}, to which we refer for more details. As for type $A$ algebras, one first breaks the partition $\underline{\lambda}$ into ``chunks'' of prescribed form, specifically pairs $\{\lambda_i,\lambda_i\}$ of equal part, pairs $\{\lambda_j,\lambda_k\}$ of unequal odd parts, and a unique chunk $\{\lambda_l\}$ consisting of a single odd $\lambda_l$. Then, each chunk $\mathcal C$ is assigned a triple $(X_{\mathcal C},Y_{\mathcal C},H_{\mathcal C})$, and the $\mathfrak{sl}(2)$-triple $(X_{\underline{\lambda}},Y_{\underline{\lambda}},H_{\underline{\lambda}})$ corresponding to $\underline{\lambda}$ is obtained by summing the elementary triples $(X_{\mathcal C},Y_{\mathcal C},H_{\mathcal C})$ over all chunks.

\paragraph{Springer--Steinberg theorem}

Given a element $X\in\mathfrak{g}$, its centralizer in $\mathfrak{g}$ is the subalgebra $\{Y\in\mathfrak{g}\,\mid\,[Y,X]=0\}$. It is the Lie algebra of the centralizer in $G_\mathrm{ad}$, i.e.~the subgroup $\{g\in G_\mathrm{ad}\,\mid\, g\cdot X\cdot g^{-1} = X\}$.

Let now $\mathfrak{s}=(X,Y,H)$ be an $\mathfrak{sl}(2)$-triple in $\mathfrak{g}$. The centralizer of $\mathfrak{s}$ is the subalgebra
\begin{equation}
    \mathfrak{g}^{\mathfrak{s}} = \left\{Z\in\mathfrak{g}~\mid~[Z,V] = 0,~\forall~V\in\mathrm{Span}(X,Y,H)\right\}\subset\mathfrak{g}\ .
\end{equation}
Let also $\widetilde{G}^\mathfrak{s}$ and $G_\mathrm{ad}^\mathfrak{s}$ be the centralizers of $\mathfrak{s}$ in $\widetilde{G}$ and $G_\mathrm{ad}$. The form of $\widetilde{G}^\mathfrak{s}$ and $G_\mathrm{ad}^\mathfrak{s}$ is provided by the Springer--Steinberg theorem \cite[Thm. 6.1.3]{collingwood1993nilpotent}. The restriction to the cases of our interest, namely type $A$ and $B$ classical Lie algebras, is the following.

\begin{theorem}[Springer--Steinberg]\label{thm:SpringerSteinberg}
    Let $\mathfrak{g}$ be of type $A$, i.e. $\mathfrak{g} = \mathfrak{su}(N)$, let $\underline{\lambda}=\{\lambda_1^{\mu_1} \dots\lambda_k^{\mu_k}\}$ be a partition of $N$, and $\mathfrak{s}_{\underline{\lambda}}$ an $\mathfrak{sl}(2)$-triple in the orbit corresponding to $\underline{\lambda}$. Then:
    \begin{equation}
        \widetilde{G}^{\mathfrak{s}_{\underline{\lambda}}} = S\left(\prod_i \mathrm{GL}(\mu_i)^{\lambda_i}_\Delta\right) \quad \mathrm{and}\quad G^{\mathfrak{s}_{\underline{\lambda}}}_\mathrm{ad} = \widetilde{G}^{\mathfrak{s}_{\underline{\lambda}}} / Z(\mathrm{SL}(N))\ ,
    \end{equation}
    where $\mathrm{GL}(\mu_i)^{\lambda_i}_\Delta$ denotes the diagonal copy of $\mathrm{GL}(\mu_i)$ inside the direct product $\mathrm{GL}(\mu_i)^{\lambda_i}$, and where for any $H$ a matrix group, $S(H)$ is the subgroup of matrices of determinant 1.
    
    Let $\mathfrak{g}$ be of type $B$, i.e.~$\mathfrak{g} = \mathfrak{so}(2N+1)$, let $\underline{\lambda}=\{\lambda_1^{\mu_1} \dots\lambda_k^{\mu_k}\}$ be a type $B$ partition of $2N+1$, and $\mathfrak{s}_{\underline{\lambda}}$ an $\mathfrak{sl}(2)$-triple in the orbit corresponding to $\underline{\lambda}$. Then:
    \begin{equation}
        G^{\mathfrak{s}_{\underline{\lambda}}}_\mathrm{ad} \cong \prod_{i,~\lambda_i~\mathrm{even}} \mathrm{Sp}(\mu_i) \times  S\left( \prod_{i,~\lambda_i~\mathrm{odd}} \mathrm{O}(\mu_i)\right) \ . 
    \end{equation}
    while $\widetilde{G}^{\mathfrak{s}_{\underline{\lambda}}}$ is a double cover of $G^{\mathfrak{s}_{\underline{\lambda}}}_\mathrm{ad}$ induced by the covering $\mathrm{Spin}(2N+1)\rightarrow\mathrm{SO}(2N+1)$.
\end{theorem}

\section{Other gauge fixing in type \texorpdfstring{$A$}{A}}\label{App:othergaugefixtypeA}

An important aspect of \cref{Sec:so5CMsolutions} is that certain solutions of the $G = \mathrm{Spin}(5)$ Calogero--Moser system exhibit non-trivial multiplicity. Notably, these solutions cannot be interpreted as distinct configurations of points on the elliptic curve $E_\tau$. This highlights the importance of analyzing gauge translations in Calogero--Moser systems at the level of point configurations in the complex plane rather than on the elliptic curve itself.

This phenomenon can also arise in type $A$, as a consequence of the fact that type $A$ root systems of rank $N-1$ are typically embedded in a particular hyperplane of an $N$-dimensional real vector space—namely, the hyperplane consisting of vectors whose coordinates sum to zero. This leads to an additional translation symmetry:
\begin{equation}
(z_1, \dots, z_N) \rightarrow (z_1+\omega, \dots, z_N+\omega)\ ,
\end{equation}
which is absent, for instance, in type $B$ Calogero--Moser potentials. A similar property holds for the standard representations of type $E$ and $G$ root systems.

This is an aspect we have largely glossed over in \cref{sec:GlobvarCM,Sec:typeA}, since the choice we made to fix this translation symmetry---namely, imposing $\sum z_i = 0$---prevents the appearance of Calogero–Moser diagrams with non-trivial multiplicities. 

However, an alternative way to fix the additional translation symmetry is to freeze one of the $z_i$, for instance by setting $z_N = 0$. This choice is, in fact, more common in the literature than imposing $\sum z_i = 0$. To maintain the condition $z_N = 0$, shifts by coroots and coweights must be combined with a global translation of the $z_i$, implying that these shifts are implemented as translations of the $z_i$, for $i=1,\dots,N-1$, by integer values only. For instance, for $N=3$, one obtains:
\begin{equation}\label{eq:otherconvention}
    \begin{split}
        \widetilde H_{\alpha_1}\colon &(z_1,z_2,0) \longmapsto (z_1+1,z_2-1,0)\ ,\\
        \widetilde H_{\alpha_2}\colon &(z_1,z_2,0) \longmapsto (z_1+1,z_2+2,0)\ ,\\
        \widetilde \chi_1\colon &(z_1,z_2,0) \longmapsto (z_1+1,z_2,0)\ ,\\
        \widetilde \chi_2\colon &(z_1,z_2,0) \longmapsto (z_1+1,z_2+1,0)\ ,
    \end{split}
\end{equation}
for horizontal shifts, and a similar list of transformations for vertical ones.

When translations by arbitrary coweights correspond to shifts of the $z_i$'s by integers, potentially inequivalent extrema always project to the same configuration of points on the torus, and thus must be counted with multiplicity. The interpretation of these diagrams in terms of vacua of the $\mathrm{SU}(N)$ $\mathcal{N}=1^*$ theory remains unchanged.

\section{Infrared \texorpdfstring{$\theta$}{theta}-angle for \texorpdfstring{$\mathfrak{so}(5)$}{so(5)} theories}\label{app: proof su(2) theta}

The only non-trivial case, as far as the relationship between $\theta_\mathrm{UV}$ and $\theta_\mathrm{IR}$ is concerned, is that of the partition $\{2^2,1\}$, where the dynamics at intermediate scales reduces to that of pure $\mathfrak{su}(2)$ $\N=1$ SYM theory. 

There is an important subtlety  concerning the normalization of the theta terms in $\mathfrak{so}(5)$ and $\mathfrak{su}(2)$. Recall that the normalization of \cref{eq:thetatermYM}, 
\begin{equation}
    \frac{\theta}{16\pi^2 h^\vee} \int \tr_\mathrm{adj}(F\wedge F)\ ,
\end{equation}
ensures that the periodicity of $\theta$ is exactly $2\pi$ when the gauge group $G$ is simply-connected. When $\mathfrak{g}=\mathfrak{so}(n)$ ($n\geq 5$), the traces in the adjoint and fundamental representations satisfy $\tr_\mathrm{adj}(F\wedge F) = h^\vee \tr_\mathrm{FO}(F\wedge F)$, where $\mathrm{FO}$ denotes the vector representation of $\mathfrak{so}(n)$. Therefore, the theta term in theories based on the gauge algebra $\mathfrak{so}(n)$ can be written as
\begin{equation}
    \frac{\theta_\mathrm{UV}}{16\pi^2} \int \tr_\mathrm{FO}(F\wedge F)\ .
\end{equation}

In contrast, when $\mathfrak{g}=\mathfrak{su}(n)$ ($n\geq 2$), one has $\tr_\mathrm{adj}(F\wedge F) = 2 h^\vee \tr_\mathrm{FU}(F\wedge F)$, where $\mathrm{FU}$ denotes the fundamental representation of $\mathfrak{su}(n)$. Therefore, the theta term in the $\mathfrak{su}(n)$ low energy theory reads
\begin{equation}
    \frac{\theta_\mathrm{IR}}{8\pi^2} \int \tr_\mathrm{FU}(F\wedge F)\ .
\end{equation}

The representation $\mathbf{r}$ of $G^\mathrm{ub} = \mathrm{SU}(2) < \mathrm{SO}(5)$ induced by the fundamental of $\mathrm{SO}(5)$, cf. \cref{eq:SU2inSO5}, is not the fundamental. The relation between $\theta_\mathrm{UV}$ and $\theta_\mathrm{IR}$, in the partition $\{2^2,1\}$, only depends on the Dynkin index of $\mathbf{r}$. Let $H,X,Y$ be the standard Cartan--Weyl generators of $\mathfrak{su}(2)$:
\begin{equation}
    [H,X] = 2X\ , \quad [H,Y] = -2Y\ , \quad [X,Y] = H\ .
\end{equation}
In the fundamental representation: 
\begin{equation}
    H = \begin{pmatrix}
        1 & 0 \\ 0 & -1
    \end{pmatrix}, \quad X = \begin{pmatrix}
        0 & 1 \\ 0 & 0
    \end{pmatrix}, \quad Y = \begin{pmatrix}
        0 & 0 \\ 1 & 0
    \end{pmatrix},
\end{equation}
and $\tr(H^2) = 2$ and $\tr(XY)=1$. In the representation of \cref{eq:SU2inSO5} instead, one has:
\begin{equation}
    h = \begin{pmatrix}
        1 & 0 & 0 & 0 & 0 \\
        0 & 1 & 0 & 0 & 0 \\ 
        0 & 0 & -1 & 0 & 0 \\
        0 & 0 & 0 & -1 & 0 \\
        0 & 0 & 0 & 0 & 0
    \end{pmatrix}, \quad x = \begin{pmatrix}
        0 & 0 & 0 & 1 & 0 \\
        0 & 0 & -1 & 0 & 0 \\ 
        0 & 0 & 0 & 0 & 0 \\
        0 & 0 & 0 & 0 & 0 \\
        0 & 0 & 0 & 0 & 0
    \end{pmatrix}, \quad y = \begin{pmatrix}
        0 & 0 & 0 & 0 & 0 \\
        0 & 0 & 0 & 0 & 0 \\ 
        0 & -1 & 0 & 0 & 0 \\
        1 & 0 & 0 & 0 & 0 \\
        0 & 0 & 0 & 0 & 0
    \end{pmatrix},
\end{equation}
and $\tr(h^2) = 4$ and $\tr(xy) = 2$. Therefore, the Dynkin index of the representation of \cref{eq:SU2inSO5} is two times that of $\mathrm{FU}$, hence:
\begin{equation}
    \frac{\theta_\mathrm{UV}}{16\pi^2} \int \tr_\mathrm{FO}(F^\mathrm{ub}\wedge F^\mathrm{ub}) = \frac{\theta_\mathrm{UV}}{8\pi^2} \int \tr_\mathrm{FU}(F^\mathrm{ub}\wedge F^\mathrm{ub})\ ,
\end{equation}
leading to the identification $\theta_\mathrm{UV} = \theta_\mathrm{IR}$.

\section{Duality graphs}
\label{app:dualityGraphs}

\begin{figure}
    \centering
    \includegraphics[width=0.5\linewidth]{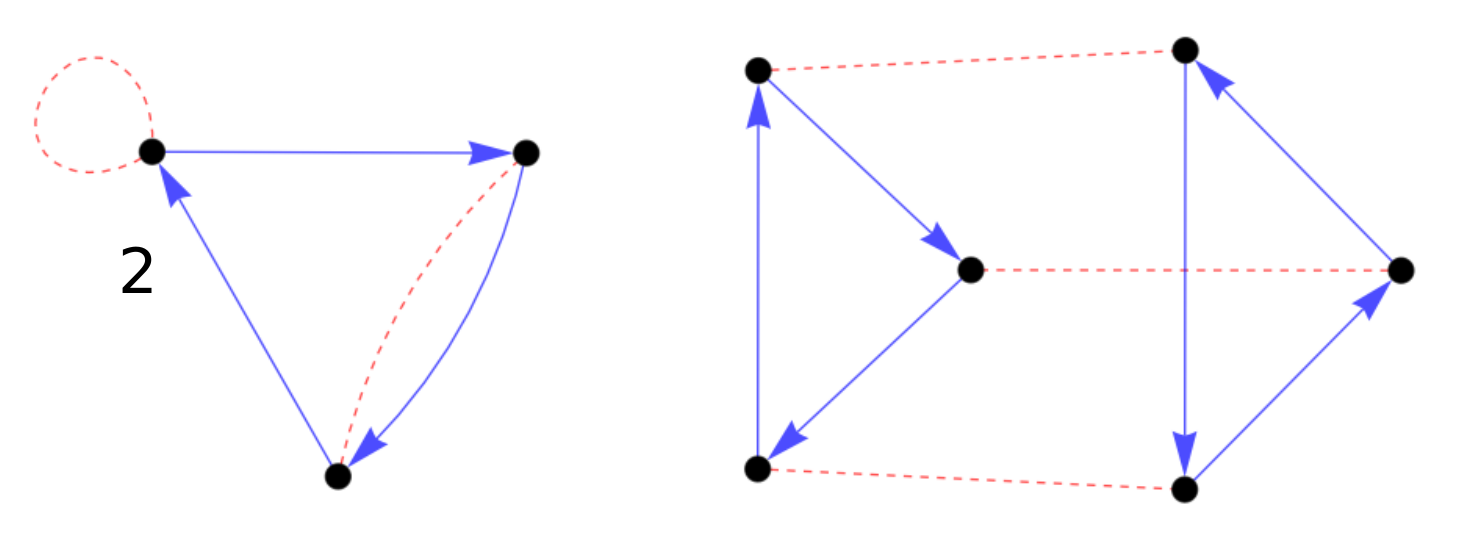}
    \caption{Duality graph for $\mathfrak{su}(2)$ vacua. }
    \label{fig:su2_dualitygraph}
\end{figure}

\begin{figure}
    \centering
    \includegraphics[width=0.5\linewidth]{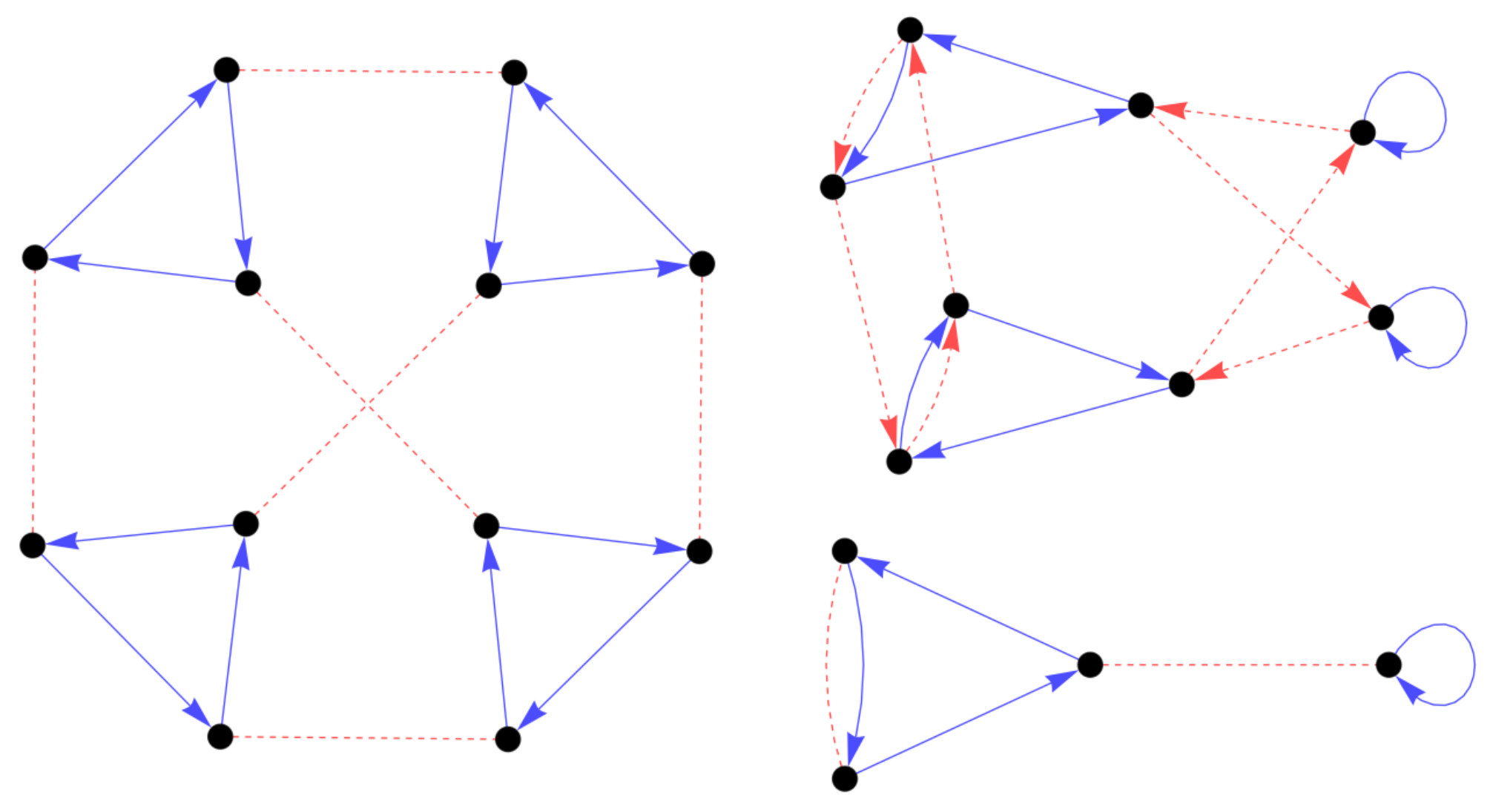}
    \caption{Duality graph for $\mathfrak{su}(3)$ vacua. }
    \label{fig:su3_dualitygraph}
\end{figure}

\begin{figure}
    \centering
    \includegraphics[width=0.7\linewidth]{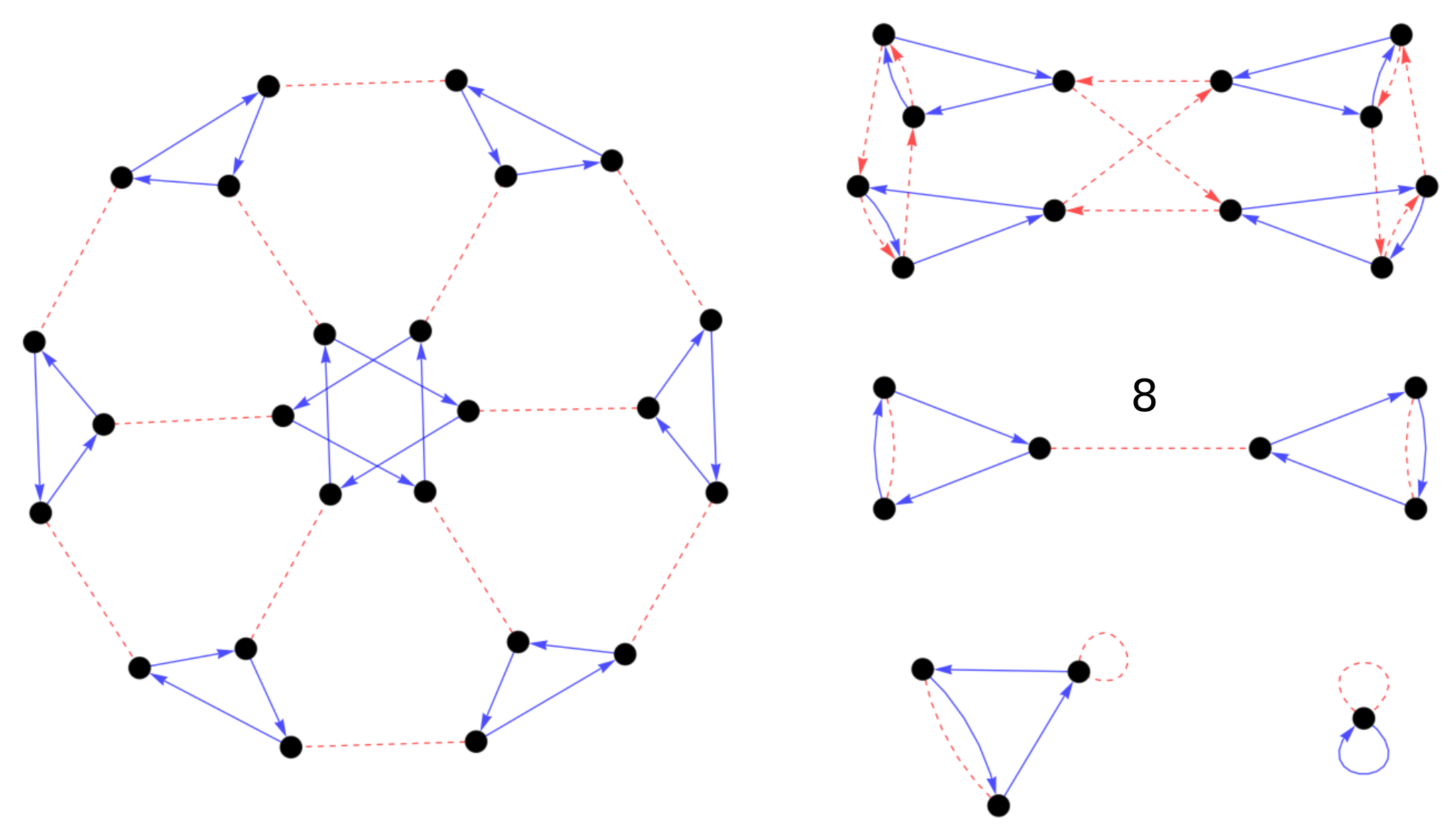}
    \caption{Duality graph for $\mathfrak{su}(4)$ vacua. }
    \label{fig:su4_dualitygraph}
\end{figure}

\begin{figure}
    \centering
    \includegraphics[width=0.7\linewidth]{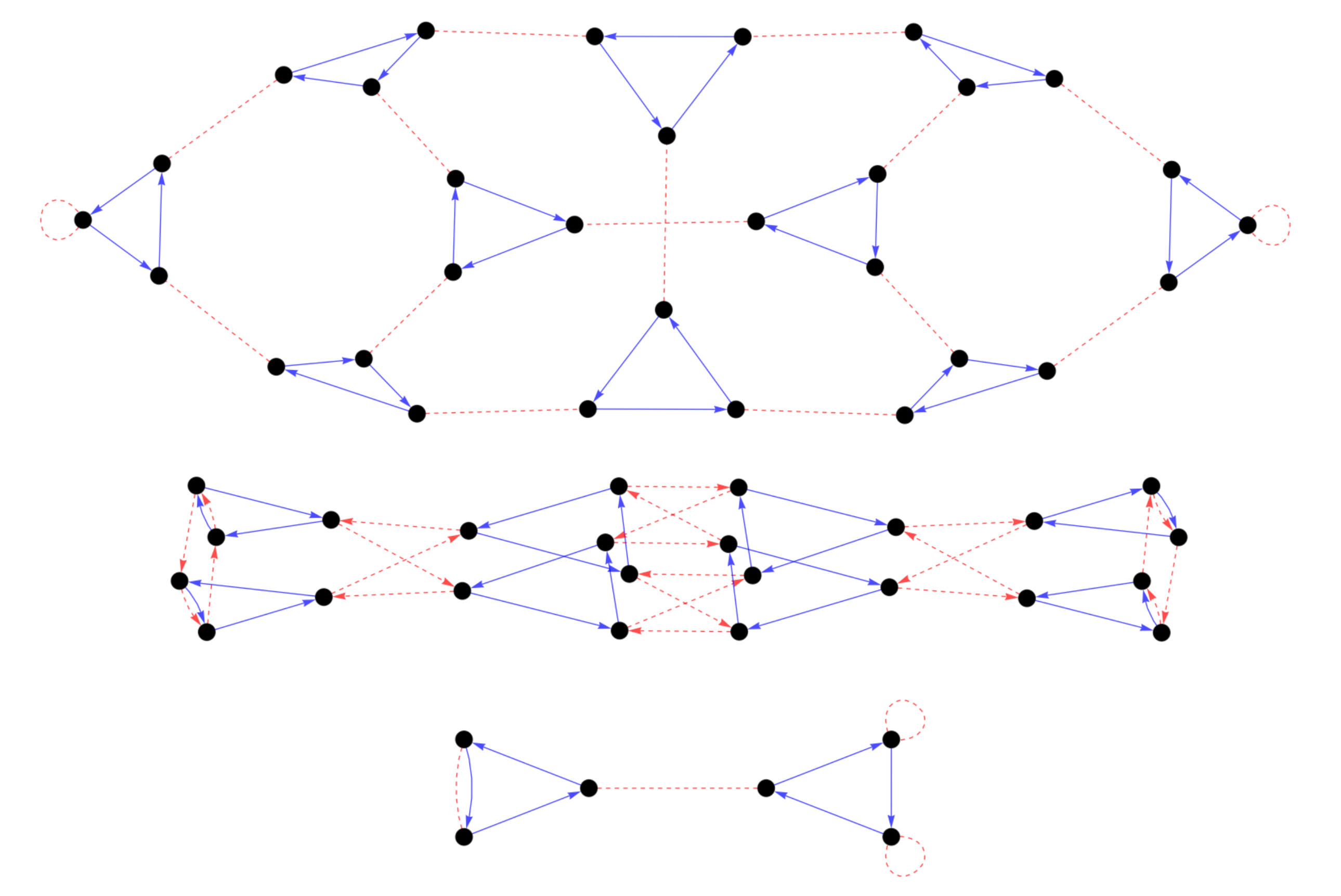}
    \caption{Duality graph for $\mathfrak{su}(5)$ vacua. }
    \label{fig:su5_dualitygraph}
\end{figure}

\begin{figure}
    \centering
    \includegraphics[width=1\linewidth]{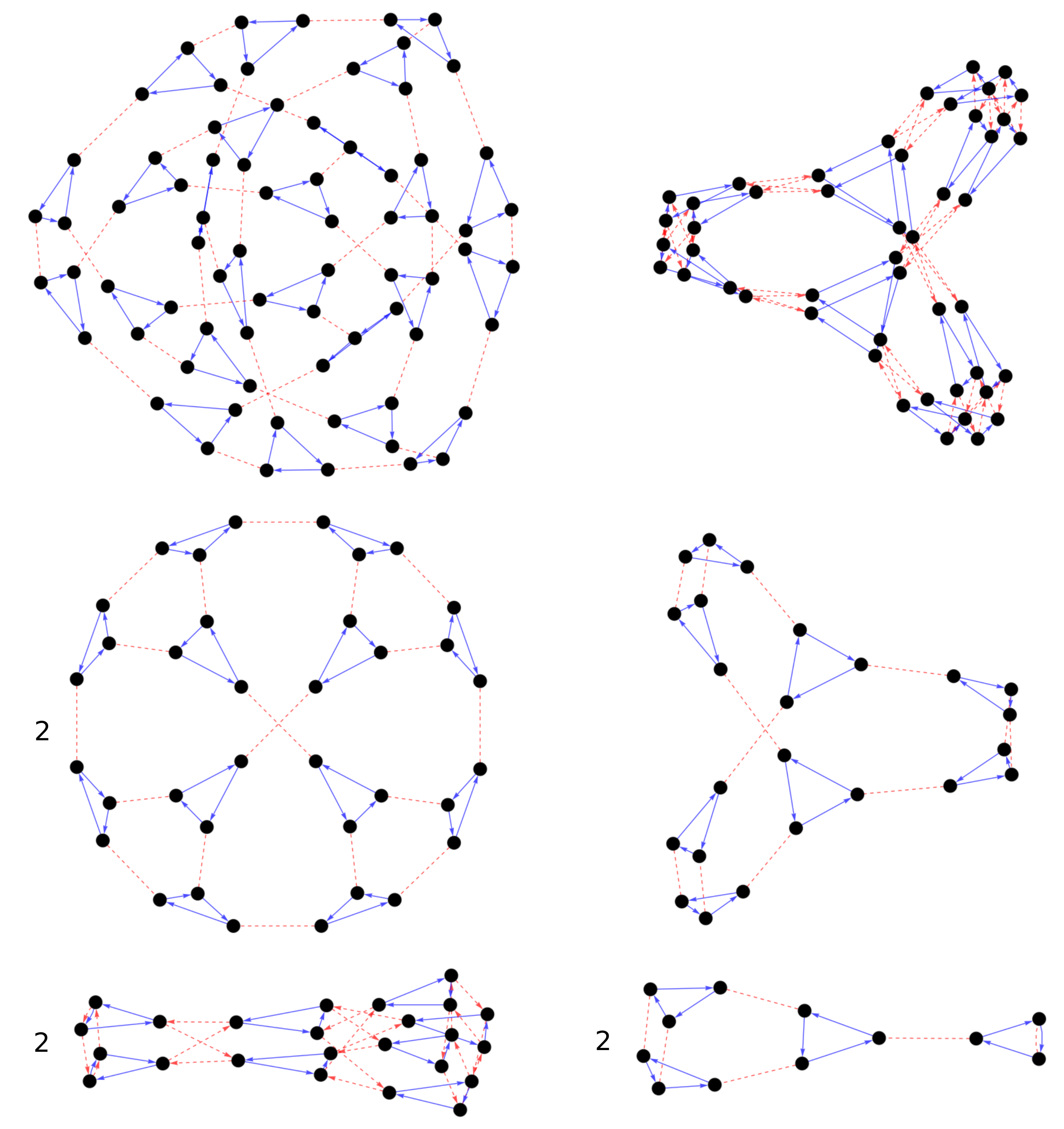}
    \caption{Duality graph for $\mathfrak{su}(6)$ vacua. }
    \label{fig:su6_dualitygraph}
\end{figure}

\begin{figure}
    \centering
    \includegraphics[width=1\linewidth]{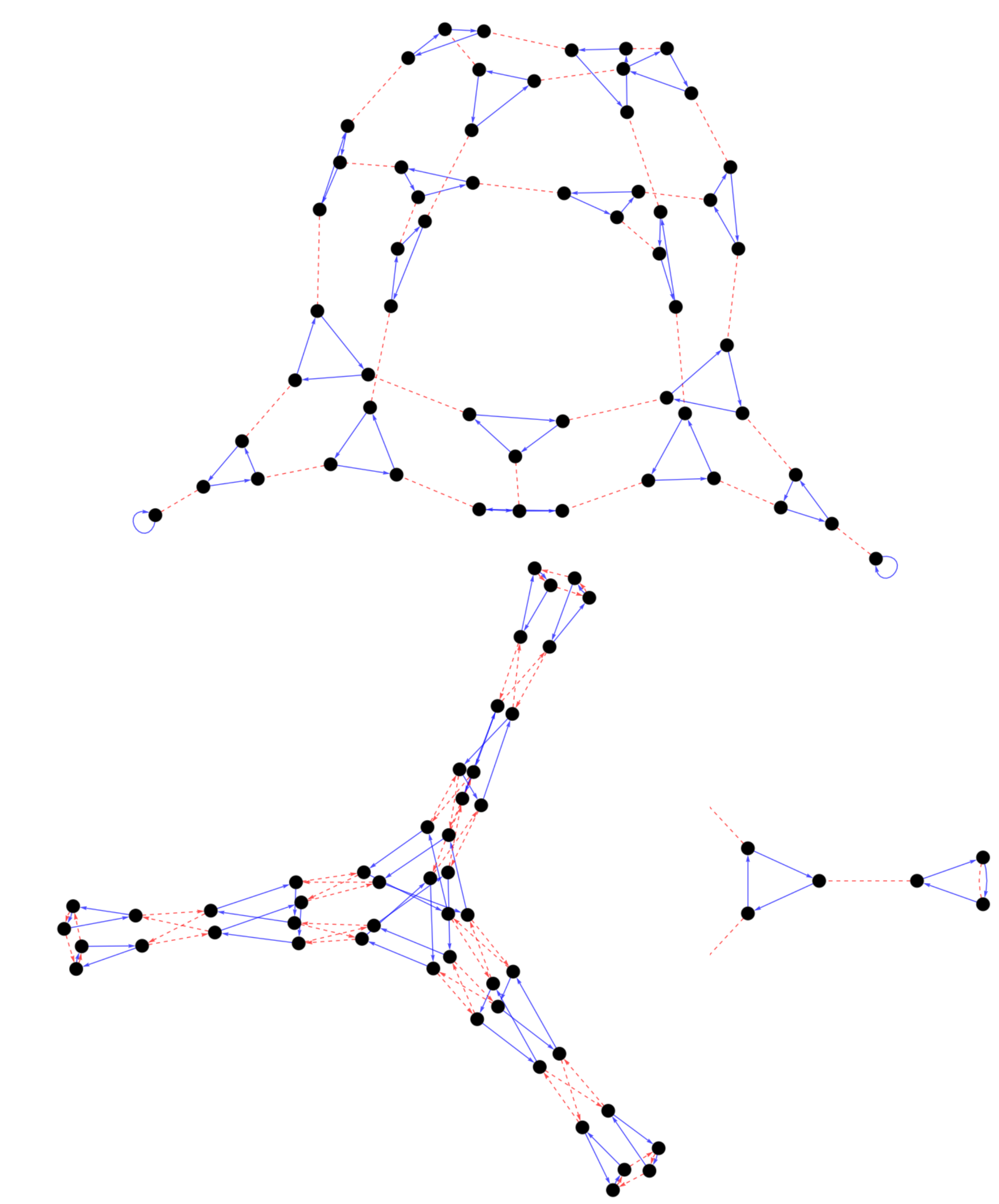}
    \caption{Duality graph for $\mathfrak{su}(7)$ vacua. }
    \label{fig:su7_dualitygraph}
\end{figure}

\begin{figure}
    \centering
    \includegraphics[width=1\linewidth]{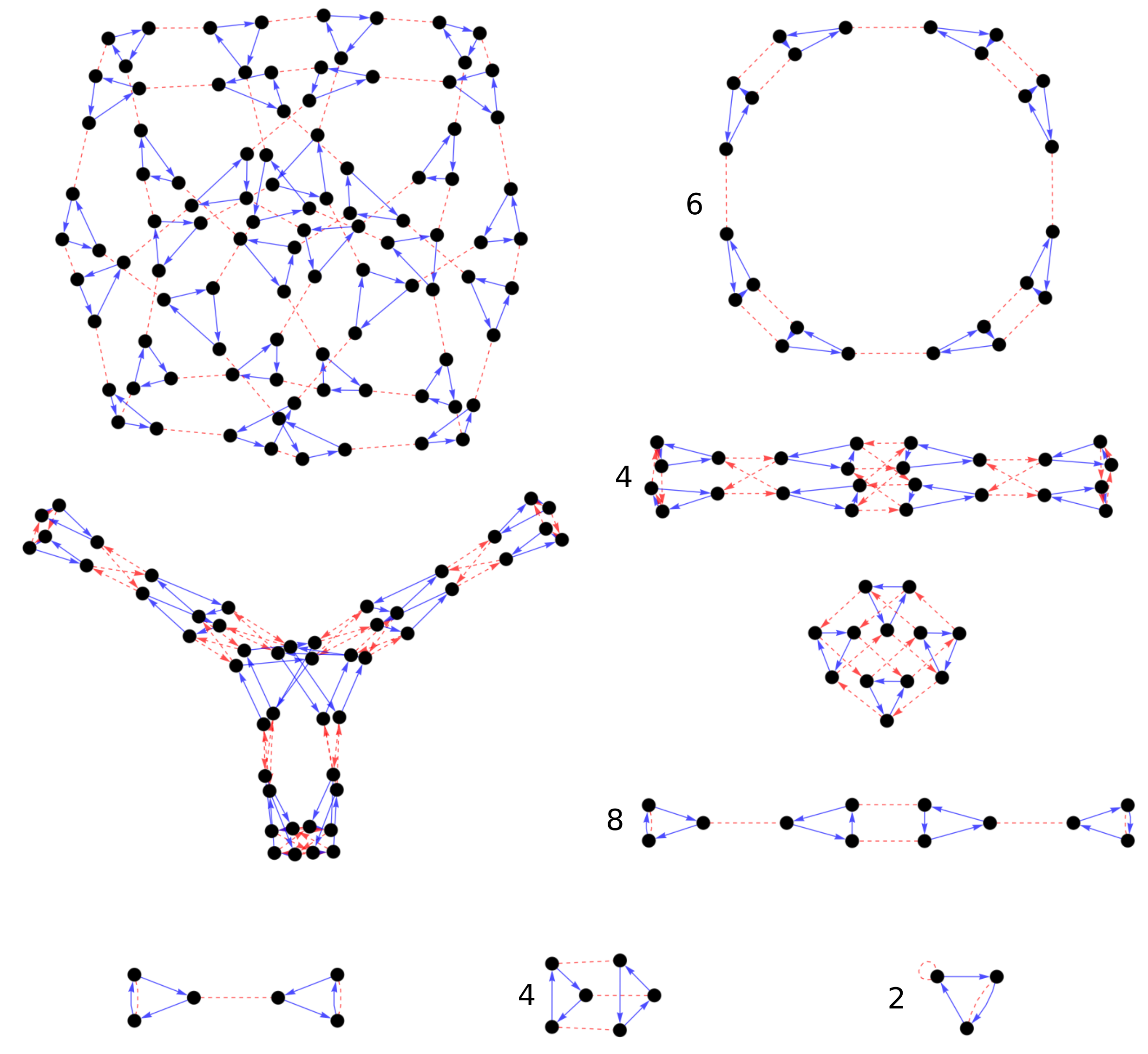}
    \caption{Duality graph for $\mathfrak{su}(8)$ vacua. }
    \label{fig:su8_dualitygraph}
\end{figure}

\Cref{fig:su2_dualitygraph,fig:su3_dualitygraph,fig:su4_dualitygraph,fig:su5_dualitygraph,fig:su6_dualitygraph,fig:su7_dualitygraph,fig:su8_dualitygraph} show the duality graphs of the orbits of $\mathfrak{su}(N)$ $\mathcal{N}=1^{\ast}$ theories on $\mathbb{R}^3 \times S^1$ for all global structures. Each black dot represents a vacuum in a global structure. Blue arrows indicate the action of the order 3 element $ST$, while dashed red arrows indicate the action of $S$. When $S^2 = 1$, we only draw one line with no arrow. The duality graphs are made up of a number of connected components, some of which are isomorphic. When two or more components are isomorphic, we represent only one of them and add a multiplicity number next to it. 

\clearpage
\bibliographystyle{JHEP}
\bibliography{ref}

\providecommand{\href}[2]{#2}\begingroup\raggedright\begin{thebibliography}{10}

\bibitem{Aharony:2013hda}
O.~Aharony, N.~Seiberg and Y.~Tachikawa, \emph{{Reading between the lines of four-dimensional gauge theories}}, \href{https://doi.org/10.1007/JHEP08(2013)115}{\emph{JHEP} {\bfseries 08} (2013) 115} [\href{https://arxiv.org/abs/1305.0318}{{\ttfamily 1305.0318}}].

\bibitem{Wilson:1974sk}
K.G.~Wilson, \emph{{Confinement of Quarks}}, \href{https://doi.org/10.1103/PhysRevD.10.2445}{\emph{Phys. Rev. D} {\bfseries 10} (1974) 2445}.

\bibitem{tHooft:1977nqb}
G.~'t~Hooft, \emph{{On the Phase Transition Towards Permanent Quark Confinement}}, \href{https://doi.org/10.1016/0550-3213(78)90153-0}{\emph{Nucl. Phys. B} {\bfseries 138} (1978) 1}.

\bibitem{Gaiotto:2014kfa}
D.~Gaiotto, A.~Kapustin, N.~Seiberg and B.~Willett, \emph{{Generalized Global Symmetries}}, \href{https://doi.org/10.1007/JHEP02(2015)172}{\emph{JHEP} {\bfseries 02} (2015) 172} [\href{https://arxiv.org/abs/1412.5148}{{\ttfamily 1412.5148}}].

\bibitem{Witten:1979ey}
E.~Witten, \emph{{Dyons of Charge e theta/2 pi}}, \href{https://doi.org/10.1016/0370-2693(79)90838-4}{\emph{Phys. Lett. B} {\bfseries 86} (1979) 283}.

\bibitem{Donagi:1995cf}
R.~Donagi and E.~Witten, \emph{{Supersymmetric Yang-Mills theory and integrable systems}}, \href{https://doi.org/10.1016/0550-3213(95)00609-5}{\emph{Nucl. Phys. B} {\bfseries 460} (1996) 299} [\href{https://arxiv.org/abs/hep-th/9510101}{{\ttfamily hep-th/9510101}}].

\bibitem{Olshanetsky:1981dk}
M.A.~Olshanetsky and A.M.~Perelomov, \emph{{Classical integrable finite dimensional systems related to Lie algebras}}, \href{https://doi.org/10.1016/0370-1573(81)90023-5}{\emph{Phys. Rept.} {\bfseries 71} (1981) 313}.

\bibitem{Olshanetsky:1983wh}
M.A.~Olshanetsky and A.M.~Perelomov, \emph{{Quantum Integrable Systems Related to Lie Algebras}}, \href{https://doi.org/10.1016/0370-1573(83)90018-2}{\emph{Phys. Rept.} {\bfseries 94} (1983) 313}.

\bibitem{DHoker:1999yni}
E.~D'Hoker and D.H.~Phong, \emph{{Lectures on supersymmetric Yang-Mills theory and integrable systems}},  in \emph{{9th CRM Summer School: Theoretical Physics at the End of the 20th Century}}, pp.~1--125, 12, 1999 [\href{https://arxiv.org/abs/hep-th/9912271}{{\ttfamily hep-th/9912271}}].

\bibitem{Kumar:2001iu}
S.P.~Kumar and J.~Troost, \emph{{Geometric construction of elliptic integrable systems and N=1* superpotentials}}, \href{https://doi.org/10.1088/1126-6708/2002/01/020}{\emph{JHEP} {\bfseries 01} (2002) 020} [\href{https://arxiv.org/abs/hep-th/0112109}{{\ttfamily hep-th/0112109}}].

\bibitem{Dorey:1999sj}
N.~Dorey, \emph{{An Elliptic superpotential for softly broken N=4 supersymmetric Yang-Mills theory}}, \href{https://doi.org/10.1088/1126-6708/1999/07/021}{\emph{JHEP} {\bfseries 07} (1999) 021} [\href{https://arxiv.org/abs/hep-th/9906011}{{\ttfamily hep-th/9906011}}].

\bibitem{Choi:2021kmx}
Y.~Choi, C.~Cordova, P.-S.~Hsin, H.T.~Lam and S.-H.~Shao, \emph{{Noninvertible duality defects in 3+1 dimensions}}, \href{https://doi.org/10.1103/PhysRevD.105.125016}{\emph{Phys. Rev. D} {\bfseries 105} (2022) 125016} [\href{https://arxiv.org/abs/2111.01139}{{\ttfamily 2111.01139}}].

\bibitem{Kaidi:2021xfk}
J.~Kaidi, K.~Ohmori and Y.~Zheng, \emph{{Kramers-Wannier-like Duality Defects in (3+1)D Gauge Theories}}, \href{https://doi.org/10.1103/PhysRevLett.128.111601}{\emph{Phys. Rev. Lett.} {\bfseries 128} (2022) 111601} [\href{https://arxiv.org/abs/2111.01141}{{\ttfamily 2111.01141}}].

\bibitem{ArabiArdehali:2019orz}
A.~Arabi~Ardehali, J.~Hong and J.T.~Liu, \emph{{Asymptotic growth of the 4d $ \mathcal{N} $ = 4 index and partially deconfined phases}}, \href{https://doi.org/10.1007/JHEP07(2020)073}{\emph{JHEP} {\bfseries 07} (2020) 073} [\href{https://arxiv.org/abs/1912.04169}{{\ttfamily 1912.04169}}].

\bibitem{Benini:2021ano}
F.~Benini and G.~Rizi, \emph{{Superconformal index of low-rank gauge theories via the Bethe Ansatz}}, \href{https://doi.org/10.1007/JHEP05(2021)061}{\emph{JHEP} {\bfseries 05} (2021) 061} [\href{https://arxiv.org/abs/2102.03638}{{\ttfamily 2102.03638}}].

\bibitem{Closset:2017bse}
C.~Closset, H.~Kim and B.~Willett, \emph{{$ \mathcal{N} $ = 1 supersymmetric indices and the four-dimensional A-model}}, \href{https://doi.org/10.1007/JHEP08(2017)090}{\emph{JHEP} {\bfseries 08} (2017) 090} [\href{https://arxiv.org/abs/1707.05774}{{\ttfamily 1707.05774}}].

\bibitem{Nekrasov:2009uh}
N.A.~Nekrasov and S.L.~Shatashvili, \emph{{Supersymmetric vacua and Bethe ansatz}}, \href{https://doi.org/10.1016/j.nuclphysbps.2009.07.047}{\emph{Nucl. Phys. B Proc. Suppl.} {\bfseries 192-193} (2009) 91} [\href{https://arxiv.org/abs/0901.4744}{{\ttfamily 0901.4744}}].

\bibitem{Closset:2024sle}
C.~Closset, E.~Furrer and O.~Khlaif, \emph{{One-form symmetries and the 3d $\mathcal{N}=2$ $A$-model: Topologically twisted indices and CS theories}}, \href{https://doi.org/10.21468/SciPostPhys.18.2.066}{\emph{SciPost Phys.} {\bfseries 18} (2025) 066} [\href{https://arxiv.org/abs/2405.18141}{{\ttfamily 2405.18141}}].

\bibitem{Gorsky:1995zq}
A.~Gorsky, I.~Krichever, A.~Marshakov, A.~Mironov and A.~Morozov, \emph{{Integrability and Seiberg-Witten exact solution}}, \href{https://doi.org/10.1016/0370-2693(95)00723-X}{\emph{Phys. Lett. B} {\bfseries 355} (1995) 466} [\href{https://arxiv.org/abs/hep-th/9505035}{{\ttfamily hep-th/9505035}}].

\bibitem{Martinec:1995by}
E.J.~Martinec and N.P.~Warner, \emph{{Integrable systems and supersymmetric gauge theory}}, \href{https://doi.org/10.1016/0550-3213(95)00588-9}{\emph{Nucl. Phys. B} {\bfseries 459} (1996) 97} [\href{https://arxiv.org/abs/hep-th/9509161}{{\ttfamily hep-th/9509161}}].

\bibitem{Seiberg:1996nz}
N.~Seiberg and E.~Witten, \emph{{Gauge dynamics and compactification to three-dimensions}},  in \emph{{Conference on the Mathematical Beauty of Physics (In Memory of C. Itzykson)}}, pp.~333--366, 6, 1996 [\href{https://arxiv.org/abs/hep-th/9607163}{{\ttfamily hep-th/9607163}}].

\bibitem{Kapustin:1998xn}
A.~Kapustin, \emph{{Solution of N=2 gauge theories via compactification to three-dimensions}}, \href{https://doi.org/10.1016/S0550-3213(98)00520-3}{\emph{Nucl. Phys. B} {\bfseries 534} (1998) 531} [\href{https://arxiv.org/abs/hep-th/9804069}{{\ttfamily hep-th/9804069}}].

\bibitem{DHoker:1998zuv}
E.~D'Hoker and D.H.~Phong, \emph{{Calogero-Moser Lax pairs with spectral parameter for general Lie algebras}}, \href{https://doi.org/10.1016/S0550-3213(98)00568-9}{\emph{Nucl. Phys. B} {\bfseries 530} (1998) 537} [\href{https://arxiv.org/abs/hep-th/9804124}{{\ttfamily hep-th/9804124}}].

\bibitem{DHoker:1998xad}
E.~D'Hoker and D.H.~Phong, \emph{{Spectral curves for superYang-Mills with adjoint hypermultiplet for general Lie algebras}}, \href{https://doi.org/10.1016/S0550-3213(98)00630-0}{\emph{Nucl. Phys. B} {\bfseries 534} (1998) 697} [\href{https://arxiv.org/abs/hep-th/9804126}{{\ttfamily hep-th/9804126}}].

\bibitem{Kapustin:2005py}
A.~Kapustin, \emph{{Wilson-'t Hooft operators in four-dimensional gauge theories and S-duality}}, \href{https://doi.org/10.1103/PhysRevD.74.025005}{\emph{Phys. Rev. D} {\bfseries 74} (2006) 025005} [\href{https://arxiv.org/abs/hep-th/0501015}{{\ttfamily hep-th/0501015}}].

\bibitem{Gaiotto:2010be}
D.~Gaiotto, G.W.~Moore and A.~Neitzke, \emph{{Framed BPS States}}, \href{https://doi.org/10.4310/ATMP.2013.v17.n2.a1}{\emph{Adv. Theor. Math. Phys.} {\bfseries 17} (2013) 241} [\href{https://arxiv.org/abs/1006.0146}{{\ttfamily 1006.0146}}].

\bibitem{Kapustin:2006pk}
A.~Kapustin and E.~Witten, \emph{{Electric-Magnetic Duality And The Geometric Langlands Program}}, \href{https://doi.org/10.4310/CNTP.2007.v1.n1.a1}{\emph{Commun. Num. Theor. Phys.} {\bfseries 1} (2007) 1} [\href{https://arxiv.org/abs/hep-th/0604151}{{\ttfamily hep-th/0604151}}].

\bibitem{Kaidi:2022uux}
J.~Kaidi, G.~Zafrir and Y.~Zheng, \emph{{Non-invertible symmetries of $ \mathcal{N} $ = 4 SYM and twisted compactification}}, \href{https://doi.org/10.1007/JHEP08(2022)053}{\emph{JHEP} {\bfseries 08} (2022) 053} [\href{https://arxiv.org/abs/2205.01104}{{\ttfamily 2205.01104}}].

\bibitem{Choi:2022zal}
Y.~Choi, C.~Cordova, P.-S.~Hsin, H.T.~Lam and S.-H.~Shao, \emph{{Non-invertible Condensation, Duality, and Triality Defects in 3+1 Dimensions}},  \href{https://arxiv.org/abs/2204.09025}{{\ttfamily 2204.09025}}.

\bibitem{Argyres:1999xu}
P.C.~Argyres, K.A.~Intriligator, R.G.~Leigh and M.J.~Strassler, \emph{{On inherited duality in N=1 d = 4 supersymmetric gauge theories}}, \href{https://doi.org/10.1088/1126-6708/2000/04/029}{\emph{JHEP} {\bfseries 04} (2000) 029} [\href{https://arxiv.org/abs/hep-th/9910250}{{\ttfamily hep-th/9910250}}].

\bibitem{Damia:2023ses}
J.A.~Damia, R.~Argurio, F.~Benini, S.~Benvenuti, C.~Copetti and L.~Tizzano, \emph{{Non-invertible symmetries along 4d RG flows}}, \href{https://doi.org/10.1007/JHEP02(2024)084}{\emph{JHEP} {\bfseries 02} (2024) 084} [\href{https://arxiv.org/abs/2305.17084}{{\ttfamily 2305.17084}}].

\bibitem{Intriligator:1998ig}
K.A.~Intriligator, \emph{{Bonus symmetries of N=4 superYang-Mills correlation functions via AdS duality}}, \href{https://doi.org/10.1016/S0550-3213(99)00242-4}{\emph{Nucl. Phys. B} {\bfseries 551} (1999) 575} [\href{https://arxiv.org/abs/hep-th/9811047}{{\ttfamily hep-th/9811047}}].

\bibitem{Polchinski:2000uf}
J.~Polchinski and M.J.~Strassler, \emph{{The String dual of a confining four-dimensional gauge theory}},  \href{https://arxiv.org/abs/hep-th/0003136}{{\ttfamily hep-th/0003136}}.

\bibitem{Bourget:2015lua}
A.~Bourget and J.~Troost, \emph{{Counting the Massive Vacua of N=1* Super Yang-Mills Theory}}, \href{https://doi.org/10.1007/JHEP08(2015)106}{\emph{JHEP} {\bfseries 08} (2015) 106} [\href{https://arxiv.org/abs/1506.03222}{{\ttfamily 1506.03222}}].

\bibitem{Bourget:2015cza}
A.~Bourget and J.~Troost, \emph{{Duality and modularity in elliptic integrable systems and vacua of $ \mathcal{N}={1}^{\ast } $ gauge theories}}, \href{https://doi.org/10.1007/JHEP04(2015)128}{\emph{JHEP} {\bfseries 04} (2015) 128} [\href{https://arxiv.org/abs/1501.05074}{{\ttfamily 1501.05074}}].

\bibitem{Heidenreich:2021xpr}
B.~Heidenreich, J.~McNamara, M.~Montero, M.~Reece, T.~Rudelius and I.~Valenzuela, \emph{{Non-invertible global symmetries and completeness of the spectrum}}, \href{https://doi.org/10.1007/JHEP09(2021)203}{\emph{JHEP} {\bfseries 09} (2021) 203} [\href{https://arxiv.org/abs/2104.07036}{{\ttfamily 2104.07036}}].

\bibitem{Antinucci:2022eat}
A.~Antinucci, G.~Galati and G.~Rizi, \emph{{On continuous 2-category symmetries and Yang-Mills theory}}, \href{https://doi.org/10.1007/JHEP12(2022)061}{\emph{JHEP} {\bfseries 12} (2022) 061} [\href{https://arxiv.org/abs/2206.05646}{{\ttfamily 2206.05646}}].

\bibitem{Damia:2023gtc}
J.A.~Damia, R.~Argurio and S.~Chaudhuri, \emph{{When the moduli space is an orbifold: spontaneous breaking of continuous non-invertible symmetries}}, \href{https://doi.org/10.1007/JHEP03(2024)042}{\emph{JHEP} {\bfseries 03} (2024) 042} [\href{https://arxiv.org/abs/2309.06491}{{\ttfamily 2309.06491}}].

\bibitem{Seiberg:1994rs}
N.~Seiberg and E.~Witten, \emph{{Electric - magnetic duality, monopole condensation, and confinement in N=2 supersymmetric Yang-Mills theory}}, \href{https://doi.org/10.1016/0550-3213(94)90124-4}{\emph{Nucl. Phys. B} {\bfseries 426} (1994) 19} [\href{https://arxiv.org/abs/hep-th/9407087}{{\ttfamily hep-th/9407087}}].

\bibitem{Bourget:2015upj}
A.~Bourget and J.~Troost, \emph{{On the $ \mathcal{N}={1}^{\ast } $ gauge theory on a circle and elliptic integrable systems}}, \href{https://doi.org/10.1007/JHEP01(2016)097}{\emph{JHEP} {\bfseries 01} (2016) 097} [\href{https://arxiv.org/abs/1511.03116}{{\ttfamily 1511.03116}}].

\bibitem{Bourget:2017goy}
A.~Bourget and J.~Troost, \emph{{Permutations of Massive Vacua}}, \href{https://doi.org/10.1007/JHEP05(2017)042}{\emph{JHEP} {\bfseries 05} (2017) 042} [\href{https://arxiv.org/abs/1702.02102}{{\ttfamily 1702.02102}}].

\bibitem{Bourget:2016yhy}
A.~Bourget and J.~Troost, \emph{{The Arithmetic of Supersymmetric Vacua}}, \href{https://doi.org/10.1007/JHEP07(2016)036}{\emph{JHEP} {\bfseries 07} (2016) 036} [\href{https://arxiv.org/abs/1606.01022}{{\ttfamily 1606.01022}}].

\bibitem{Sommers:1998ago}
E.~Sommers, \emph{A generalization of the {Bala}-{Carter} theorem for nilpotent orbits}, \href{https://doi.org/10.1155/S107379289800035X}{\emph{Int. Math. Res. Not.} {\bfseries 1998} (1998) 539}.

\bibitem{Dorey:2001qj}
N.~Dorey, T.J.~Hollowood and S.P.~Kumar, \emph{{An Exact elliptic superpotential for N=1* deformations of finite N=2 gauge theories}}, \href{https://doi.org/10.1016/S0550-3213(01)00647-2}{\emph{Nucl. Phys. B} {\bfseries 624} (2002) 95} [\href{https://arxiv.org/abs/hep-th/0108221}{{\ttfamily hep-th/0108221}}].

\bibitem{Argurio:2024kdr}
R.~Argurio, A.~Collinucci, S.~Mancani, S.~Meynet, L.~Mol and V.~Tatitscheff, \emph{{Inherited non-invertible duality symmetries in quiver SCFTs}}, \href{https://doi.org/10.1007/JHEP06(2025)227}{\emph{JHEP} {\bfseries 06} (2025) 227} [\href{https://arxiv.org/abs/2409.03694}{{\ttfamily 2409.03694}}].

\bibitem{Gaiotto:2009we}
D.~Gaiotto, \emph{{N=2 dualities}}, \href{https://doi.org/10.1007/JHEP08(2012)034}{\emph{JHEP} {\bfseries 08} (2012) 034} [\href{https://arxiv.org/abs/0904.2715}{{\ttfamily 0904.2715}}].

\bibitem{collingwood1993nilpotent}
D.H.~Collingwood and W.M.~McGovern, \emph{{Nilpotent orbits in semisimple Lie algebra: an introduction}}, CRC Press (1993).

\bibitem{Nekrasov:2004vw}
N.~Nekrasov and S.~Shadchin, \emph{{ABCD of instantons}}, \href{https://doi.org/10.1007/s00220-004-1189-1}{\emph{Commun. Math. Phys.} {\bfseries 252} (2004) 359} [\href{https://arxiv.org/abs/hep-th/0404225}{{\ttfamily hep-th/0404225}}].

\bibitem{Tatitscheff:2018aht}
V.~Tatitscheff, Y.-H.~He and J.~McKay, \emph{{Cusps, Congruence Groups and Monstrous Dessins}}, \href{https://doi.org/10.1016/j.indag.2020.09.005}{\emph{Indag. Math.} {\bfseries 31} (2020) 1015} [\href{https://arxiv.org/abs/1812.11752}{{\ttfamily 1812.11752}}].

\bibitem{Fulton:1991rep}
W.~Fulton and J.~Harris, \emph{Representation theory. {A} first course}, vol.~129 of \emph{Grad. Texts Math.}, New York etc.: Springer-Verlag (1991).

\bibitem{Cordova:2019uob}
C.~C\'ordova, D.S.~Freed, H.T.~Lam and N.~Seiberg, \emph{{Anomalies in the Space of Coupling Constants and Their Dynamical Applications II}}, \href{https://doi.org/10.21468/SciPostPhys.8.1.002}{\emph{SciPost Phys.} {\bfseries 8} (2020) 002} [\href{https://arxiv.org/abs/1905.13361}{{\ttfamily 1905.13361}}].

\end{thebibliography}\endgroup

\end{document}